\documentclass[10pt,aps,prb,amsmath,twocolumn,amssymb,floatfix,showpacs,nofootinbib,longbibliography,superscriptaddress]{revtex4-2}

\usepackage[utf8]{inputenc}

\usepackage{amssymb,amsfonts,amsmath} 
\usepackage{graphicx,epsfig,psfrag}
\usepackage{color}
\usepackage{natbib}
\usepackage{url}
\usepackage[breaklinks=true]{hyperref}
\usepackage{mathtools}
\usepackage{subfigure}
\usepackage{physics}
\usepackage{lipsum}
\usepackage{comment}
\usepackage{centernot}
\usepackage[normalem]{ulem}
\usepackage[dvipsnames]{xcolor}
\hypersetup{
        colorlinks = true,
        citecolor = blue
}

\usepackage{orcidlink}

\mathchardef\mhyphen="2D 

\begin{document}

\title{Optimized quantum state transfer \\ in a quasiperiodic ultracold atomic gas}

\author{Andreas Völkering}
\affiliation{Department of Physics, Stockholm University, SE-106 91 Stockholm, Sweden}
\author{Arnob Kumar Ghosh\,\orcidlink{0000-0003-0990-8341}}
\affiliation{Department of Physics and Astronomy, Uppsala University, Box 516, 75120 Uppsala, Sweden}
\author{Patric Holmvall\,\orcidlink{0000-0002-1866-2788}}
\affiliation{Department of Physics and Astronomy, Uppsala University, Box 516, 75120 Uppsala, Sweden}
\author{Paolo Molignini\,\orcidlink{0000-0001-6294-3416}}
\thanks{paolo.s.molignini@jyu.fi}
\affiliation{Department of Physics, Stockholm University, SE-106 91 Stockholm, Sweden}
\affiliation{Department of Physics and Nanoscience Center, University of Jyväskylä, P.O. Box 35 (YFL), University of Jyväskylä, FI-40014 Jyväskylä, Finland}

\begin{abstract}
Ultracold atomic systems offer a highly controllable platform for investigating quantum state transfer through the precise dynamical manipulation of system parameters.
While quantized Thouless pumping has been extensively explored in these systems, adiabatic edge-to-edge transfer of localized quantum states remains largely unexplored.
Here, we consider a one-dimensional ultracold atomic gas confined in a bichromatic optical lattice realizing a quasiperiodic Aubry--Andr\'e--Harper system.
We use its edge-localized winding states to implement quantum state transfer between opposite boundaries.
Starting from the instantaneous spectral properties of the corresponding tight-binding model, we construct locally adiabatic protocols and extend the approach to higher protocol orders.
We then simulate their dynamics under the full continuum bichromatic-lattice Hamiltonian.
Our results reveal a tradeoff between edge localization and the minimum spectral gap: strongly localized states require longer transfer times, but can benefit substantially from higher protocol orders.
We are also able to capture the main fidelity trends and coherent oscillations over a broad parameter regime using an effective two-level Landau--Zener description.
Our results provide practical guidelines for selecting experimentally accessible parameters and tailoring quantum transfer protocols in quasiperiodic ultracold atomic systems.
\end{abstract}
\maketitle

\section{Introduction}
\label{sec:introduction}

Ultracold atomic systems provide a versatile platform for simulating complex condensed matter systems that are often difficult to realize in material systems~\cite{JAKSCHAP2005, Lewenstein2007, CiracNP2012, BlochNP2012, ChristianScience2017, LangenNP2024}. 
In particular, they have emerged as a promising platform for quantum information science~\cite{HollandScience2023, YichengScience2023}, where robust quantum-state transfer remains a key challenge.

\begin{figure}[t!]
    \centering
    \includegraphics[width=0.45\textwidth]{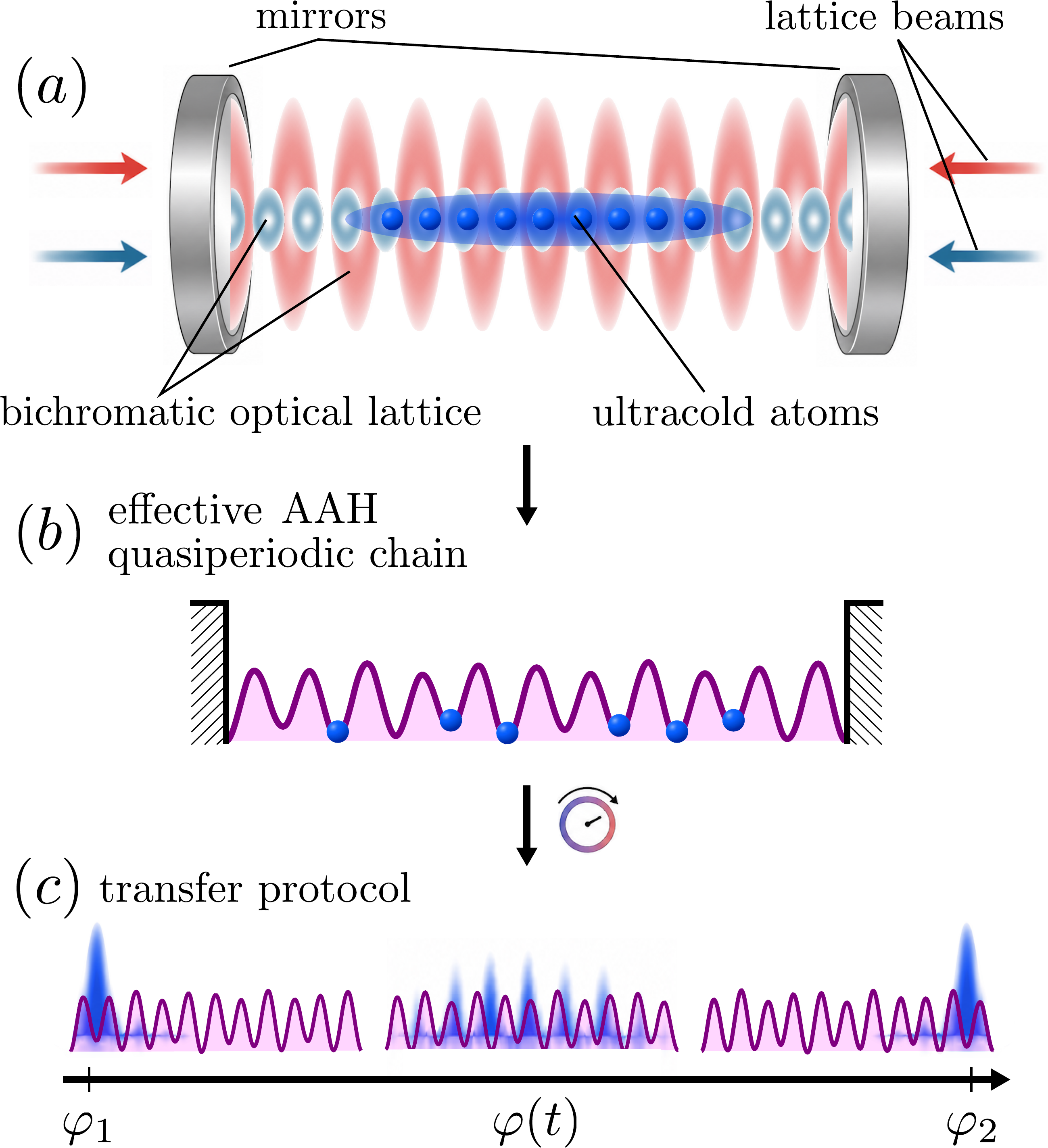}
    \caption{\textbf{Phason-based state transfer in an AAH quasiperiodic chain realized with ultracold atoms.} 
    (a) A gas of ultracold atoms is loaded into a bichromatic optical lattice formed by two superimposed standing waves with incommensurate wavelengths.
    (b) In the tight-binding regime, the resulting quasiperiodic potential is described by an Aubry--Andr\'e--Harper (AAH) model.
    (c) Varying the phason angle $\varphi$ adiabatically shifts the quasiperiodic potential and pumps edge-localized states across the system, exhibiting quantum state transfer between opposite boundaries.}
    \label{fig:schematics}
\end{figure}

Quantum pumps offer a natural mechanism for transferring quantum states and establishing communication between distant elements of a quantum system~\cite{LonghiPRB2019, ghosh2025quantum}.
In particular, adiabatic topological pumping provides a disorder-resilient route for transporting quantum states~\cite{CitroNRP2023}.
This can take the form of quantized charge pumping, as in a Thouless pump
~\cite{ThoulessPRB1983,PhysRevLett.129.053201,CitroNRP2023}, or the coherent transfer of a localized quantum state between distant regions of a system~\cite{LonghiPRB2019}.
Setups include dimerized Su–Schrieffer–Heeger models~\cite{LangNQI2017, MeiPRA2018, LonghiPRB2019, LonghiLandauZenerAQT2019, ZhengLiNaPRA2020, QiLuPRA2020, DAngelisPRR2020, PalaiodimopoulosPRA2021, CaoPRA2021, YuanAPLPh2021, QiPRRTR2021, WangPRA2022, ZhengLiNaYiPRApp2022, Liu:2022, WangDaWeiPRA2023, ZhaonppjQuantumInf2023, Zurita2023fastquantumtransfer, RomeroPRApp2024, HanJinXuanPRApp2024, TianPRB2024, Fernandez2024, WangDaWeiPRA2024, ZhengCJP2025}, chiral edge states in quantum spin liquids~\cite{YaoNatComm2013, DlaskaQSI2017}, and even in quasiperiodic systems~\cite{Kraus:2012-2, VerbinPump2015, SinghPRA2015, ghosh2025quantum}. 
Classical analogs of such transfer protocols have also been realized experimentally in photonic waveguide arrays based on both the SSH model~\cite{Liu:2022, ChaohuaPRA2023} and Fibonacci quasicrystals~\cite{Kraus:2012-2, VerbinPump2015, ghosh2026FCwaveguide}.

Quantized Thouless pumping has already been demonstrated experimentally in ultracold bosonic~\cite{LohseNP2016} and fermionic~\cite{NakajimaNP2016} systems.
However, quantum-state-transfer protocols have so far been formulated predominantly within idealized lattice models. 
This leaves open the question of how they can be efficiently implemented in the continuum descriptions.
The exceptional tunability of ultracold atomic platforms makes them particularly well suited for addressing this question.
Their parameters can be tuned to investigate how the underlying spectral structure affects transfer fidelity, while the driving protocol itself can be dynamically adapted to the instantaneous properties of the system.

Motivated by these advantages, in this work we demonstrate efficient edge-to-edge transfer of localized winding states in a quasiperiodic ultracold atomic gas.
By simulating a realistic continuum model, we identify experimentally relevant regimes of high-fidelity operation and establish practical guidelines for their implementation. 
Our optimized protocols thus provide a bridge between idealized state-transfer schemes and their realization in realistic ultracold-atom experiments.

We consider a one-dimensional ultracold atomic gas confined in a bichromatic optical lattice that realizes a quasiperiodic Aubry--Andr\'e--Harper (AAH) system~\cite{Harper:1955,Aubry:1980,Modugno:2009,Kraus:2012, Fangli:2015,Liu:2017,Liu:2021}, as schematically illustrated in Fig.~\ref{fig:schematics}.
We first construct locally adiabatic transfer protocols from the instantaneous spectrum and extend this construction to higher protocol orders, which control how strongly the evolution slows down near small-gap regions.
We focus in particular on the lowest two orders, which yield tangent and Roland-Cerf protocols, respectively, and compare their performance across experimentally relevant parameters.
We then implement these protocols directly in the continuum and simulate the transfer dynamics under a realistic experimental Hamiltonian beyond the tight-binding approximation.
For this purpose, we employ the multi-configurational time-dependent Hartree method for indistinguishable particles (MCTDH)~\cite{Streltsov:2006, Streltsov:2007, Alon:2007, Alon:2008}, as implemented in the MCTDH-X software~\cite{Lode:2012, Lode:2016, Fasshauer:2016, Lode:2020,Lin:2020, Molignini:2025-SciPost, MCTDHX}.
To interpret the resulting dynamics, we additionally derive an effective two-level description based on the dominant avoided crossing.
For the Roland-Cerf protocol, this even leads to a closed-form expression for the fidelity and a simple analytical benchmark for the continuum results.

Our results show that edge-to-edge transfer fidelities can reach up to 99.6\% through an appropriate choice of protocol shape, transfer time, and lattice parameters.
This is achieved despite a highly nontrivial tradeoff between edge localization, spectral-gap size, transfer duration, and protocol order.
Furthermore, we show that even strongly localized states can be transferred efficiently by tailoring the protocol to the detailed structure of the quasiperiodic spectrum, with approximately invariant performance under lattice-depth rescaling.
Remarkably, the effective two-level description captures the main fidelity trends and coherent oscillatory features over a broad parameter regime, providing a simple predictive picture of the otherwise complex continuum dynamics.
Our results therefore provide both practical guidance for implementing quasiperiodic state-transfer protocols in ultracold atomic systems and simple effective models for interpreting their performance.

The rest of the paper is structured as follows. 
In Sec.~\ref{sec:model}, we present our quasiperiodic ultracold atomic gas model.
In Sec.~\ref{sec:methods:analytics}, we introduce our analytic optimization approach and the two transfer protocols we use for adiabatic quantum state transfer of edge-localized states.
Section~\ref{sec:methods:numerics} is devoted to the description of the continuum implementation of the protocols and the full numerics based on first-principles simulations using MCTDH-X.
We discuss the performance of different transfer protocols in Sec.~\ref {sec:results}.
Finally, we conclude with a summary and outlook in Sec.~\ref{sec:conclusions}.

\section{Model: Ultracold atoms in a quasiperiodic potential}
\label{sec:model}

We consider a one-dimensional ultracold atomic gas confined in a bichromatic optical lattice generated by two standing waves with incommensurate wavelengths~\cite{Bloch:2008,Roati:2008,Modugno:2009,Xiao:2017,Luschen:2018,Molignini:2025-1,Molignini:2025-2}.
The system is described by the second-quantized continuum Hamiltonian
\begin{align}
\hat{H} &= \int \mathrm{d}x \, \hat{\Psi}^{\dagger}(x) \left[ -\frac{\hbar^{2}}{2m}\frac{\mathrm{d}^2}{\mathrm{d}x^2} + V(x) \right] \hat{\Psi}(x) \nonumber\\
&\quad + \frac{g}{2} \int \mathrm{d}x \,
\hat{\Psi}^{\dagger}(x) \hat{\Psi}^{\dagger}(x) \hat{\Psi}(x) \hat{\Psi}(x), 
\label{eq:continuum_hamiltonian}
\end{align}
where $\hat{\Psi}(x)$ is a bosonic field operator at position $x$, $\hbar$ is the reduced Planck constant, $m$ is the mass of the ultracold atomic species, and $g$ is the contact interaction strength.
In the present work, we focus on the noninteracting regime $g=0$, which already captures the essential quasiperiodic pumping physics~\cite{ThoulessPRB1983, Kraus:2012}.
Such a noninteracting system can be realized by tuning the scattering length near a Feshbach resonance~\cite{Chin:2010}. 
The bichromatic optical-lattice potential $V(x)$  reads
\begin{align}
V(x) &= V_p(x) + V_d(x) \nonumber \\
&= \frac{V_p}{2} \cos(2k_p x) + \frac{V_d}{2} \cos(2k_d x+\varphi),
\label{eq:continuum_potential}
\end{align}
where $V_p$ ($V_d$) denotes the depth of the primary (detuning) lattice, and the corresponding wavevectors are $k_j = 2\pi/\lambda_j$ with $j=p,d$ and wavelengths $\lambda_j$.
The parameter $\varphi$ is a controllable phase offset, commonly referred to as the \emph{phason} angle~\cite{JagannathanRMP2021}.
We additionally define the ratio of the detuning-lattice to primary-lattice depths as $\rho=V_d/V_p$,
henceforth referred to as the \emph{detuning strength} for simplicity.
To study the controlled transport of edge modes, we consider a finite number of sites in the optical lattice throughout this work, i.e., we implement hard-wall boundary conditions.
Experimentally, such finite systems with sharp boundaries can be approximated using optical box traps, which have been realized for
ultracold Bose gases~\cite{MeyrathPRA2005, GauntPRL2013}.

Quasiperiodicity emerges from the irrational ratio between the two lattice wavevectors~\cite{Aubry:1980}, which we choose as
\begin{equation}
\beta = \frac{k_d}{k_p} = \frac{\sqrt{5}-1}{2},
\label{eq:beta}
\end{equation}
corresponding to the inverse golden ratio.
Because irrational ratios cannot be represented exactly in finite systems, the quasiperiodic potential is approximated through successive Fibonacci rational approximants,
\begin{equation}
\beta \approx \frac{F_{n-1}}{F_n},
\end{equation}
where $F_n$ denotes the $n$-th Fibonacci number.

For the $n$-th Fibonacci approximant, we consider an interval $x \in [0, L]$ containing $N_w=F_n$ minima of the primary lattice, and accordingly $k_p = N_w \pi/L$.
We label the center of each lattice well from left to right by $j=1, \cdots, N_w$, with coordinates $x_j=L\left( j - \frac{1}{2} \right)/N_w$.
This construction naturally generates finite quasiperiodic chains with lengths $N_w = \dots,13,21,34,55,\dots$, which converge toward the irrational quasiperiodic limit for increasing system size.

In the deep optical lattice regime $V_p \gg E_r$, where $E_r = \hbar^2 k_p^2/2m$ is the recoil energy of the primary lattice, the continuum system can be projected onto the lowest-band Wannier basis $\{w_j(x)\}$ of the primary lattice, where each Wannier function is centered around the minimum of each well~\cite{Modugno:2009} (see Appendix~\ref{app:wannier} for more details).
Therefore, the field operator is expanded as
\begin{equation}
\hat{\Psi}(x) = \sum_{j=1}^{N_w} w_j(x)\hat{c}_j,
\label{eq:wannier_expansion}
\end{equation}
where $\hat{c}_{j}^{\dagger}$ ($\hat{c}_{j}$) are the creation (annihilation) operators for site $j$.
Inserting this expansion into the continuum Hamiltonian \eqref{eq:continuum_hamiltonian}, we obtain an effective nearest-neighbor tight-binding Hamiltonian of AAH type,
\begin{align}
\hat{H}_{\mathrm{AAH}}(\varphi)
&= -\sum_{j=1}^{N_w-1} J_j(\varphi)
\left( \hat{c}_{j+1}^{\dagger}\hat{c}_{j} + \mathrm{H.c.} \right) \nonumber\\
&\quad +\sum_{j=1}^{N_w} \delta_j(\varphi)\, \hat{c}_{j}^{\dagger}\hat{c}_{j}.
\label{eq:gAAH}
\end{align}
Here, $J_j(\varphi)$ and $\delta_j(\varphi)$ denote the site-dependent nearest-neighbor hopping and onsite energy, respectively. 
They are obtained from the finite-size Wannier
overlap integrals
\begin{align}
\delta_j(\varphi) &= \int \mathrm{d}x\, w_j^{*}(x) \left[ \hat{H}_0+V_d(x;\varphi) \right] w_j(x), \\
J_j(\varphi) &= -\int \mathrm{d}x\, w_{j+1}^{*}(x) \left[ \hat{H}_0+V_d(x;\varphi) \right] w_j(x),
\end{align}
where the single-particle Hamiltonian associated with the primary lattice is
\begin{align}
\hat{H}_{0} &= -\frac{\hbar^{2}}{2m}\frac{\mathrm{d}^2}{\mathrm{d}x^2} + V_p(x). \label{eq:hamiltonian:single_particle}
\end{align}
It is useful to separate the contributions from the primary and detuning lattices as
\begin{align}
\delta_j(\varphi) &= \delta_j^{(0)}+\delta_j^{(d)}(\varphi), \\
J_j(\varphi) &= J_j^{(0)}+J_j^{(d)}(\varphi).
\end{align}
For the deep primary lattices and weak detuning considered here, the detuning-induced correction to the hopping is negligible, $|J_j^{(d)}|\ll J_j^{(0)}$, while $J_j^{(0)}$ approaches a site-independent bulk value $J$ away from the boundaries.
After subtracting the uniform bulk contribution to the onsite energy, the remaining onsite term can be written approximately as
\begin{equation}
\delta_j(\varphi) \simeq \Delta_j^{\mathrm{boundary}} + \Delta_j \cos\left[ 2\pi\beta\left(j-\frac{1}{2}\right)+\varphi \right],
\end{equation}
where $\Delta_j^{\mathrm{boundary}} = \delta_j^{(0)}-\delta_{\mathrm{bulk}}$ accounts for the finite-size boundary correction.
In the bulk and deep-lattice limit, $\Delta_j\rightarrow\Delta$ and $\Delta_j^{\mathrm{boundary}}\rightarrow0$,
whereas close to the hard-wall boundaries the corrections remain of the same order as the hopping amplitude (see Appendix~\ref{app:wannier} for details).

\begin{figure}[t!]
    \centering
    \includegraphics[width=\linewidth]{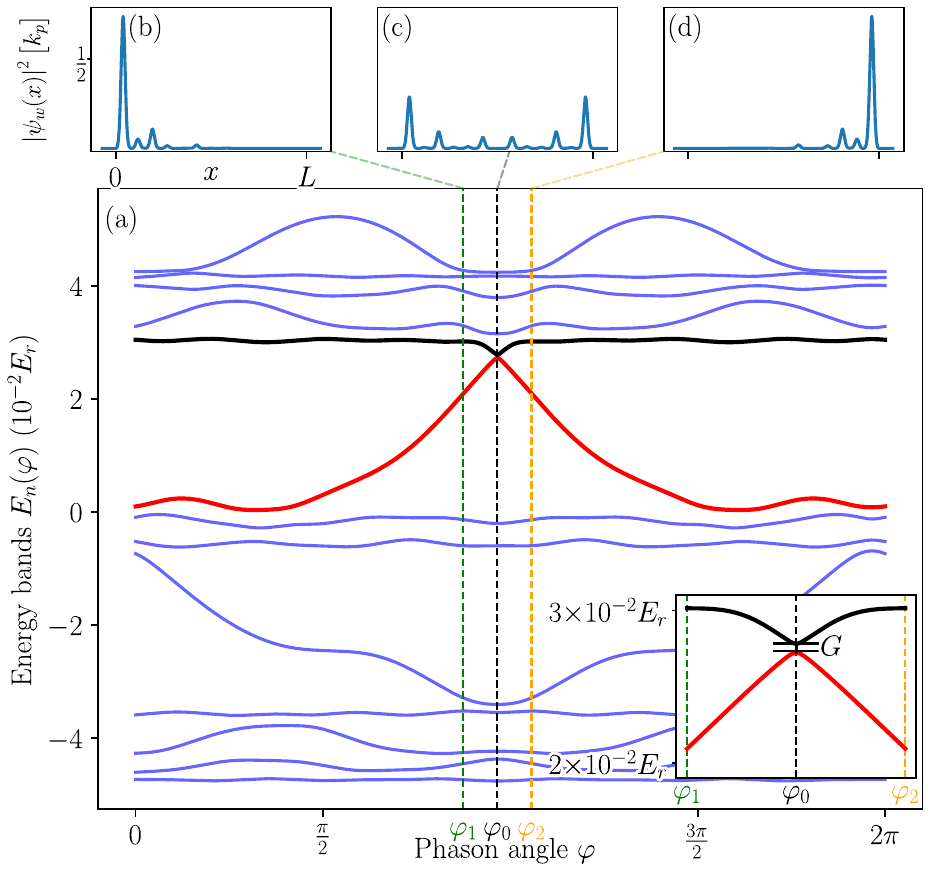}
    \caption{\textbf{Phason-dependent spectrum and avoided crossing.}
    (a) Energy spectrum as a function of the phason angle $\varphi$ for an AAH quasiperiodic chain with $N_w=13$ Wannier centers and lattice depth $V_p=10E_r$. 
    The detuning strength is set to $\rho\approx 0.007$, and in our choice of units ($k_p=N_w\pi/L$), the length is $\pi N_w$. 
    Black and red curves highlight edge-localized states in the largest gap of the quasiperiodic chain, where the latter is the edge-localized state of interest involved in the transfer protocol.
    All other states are shown in blue.
    As $\varphi$ is varied, the edge state traverses the bulk gap and undergoes an avoided crossing near $\varphi=\varphi_0$.
    The inset on the bottom right shows a magnified view of the avoided crossing, where the minimum energy gap $G$ is indicated.
    The vertical dashed lines mark the initial, crossing, and final phason angles $\varphi_1$, $\varphi_0$, and $\varphi_2$, respectively.
    (b)-(d) Winding state density at the three phason angles (with shared axes) indicated in panel (a).} 
    \label{fig:spec_example}
\end{figure}

Unless stated otherwise, energies are expressed in units of $E_r$, lengths in units of $k_p^{-1}$, and times in units of the inverse hopping of the bulk of the system $\hbar/J_{\mathrm{bulk}}$.
We further set the primary potential depth to $10 E_r$, which is deep enough to obtain good agreement between the continuum potential and its tight-binding approximation. 
We discuss the effects of varying the primary lattice depth in Appendix~\ref{app:ChangingS}, where we show that the transfer performance remains approximately invariant under lattice-depth rescaling provided that the detuning strength and protocol duration are adjusted accordingly.

The quasiperiodic spectrum exhibits topological~\cite{Kraus:2012,Kraus:2012-2,MadsenPRB2013,VarjasPRL2019,RaiPRB2021,ElsePRX2021,FanFP2022} edge-localized states inside the quasiperiodic gaps~\footnote{We remark that the finite approximants considered here retain the relevant bulk gaps and associated edge-localized subgap states, continuously connecting to the quasiperiodic limit as the system size increases.}.
These states can be identified in the eigenspectrum. 
In Fig.~\ref{fig:spec_example}(a), we show the eigenvalues $E_n(\varphi)$ of the Hamiltonian in Eq.~\eqref{eq:gAAH} as a function of the phason angle $\varphi$.
When the phason angle $\varphi$ is varied continuously from 0 to $2\pi$, the eigenspectrum exhibits in-gap states (denoted by red and black curves for the largest gap), which traverse the quasiperiodic gaps and connect neighboring quasibands.
Because their energies wind across the spectrum during the phason cycle, these modes are commonly referred to as \emph{winding states}~\cite{Kraus:2012, Kraus:2012-2, Verbin:2013}, where the integer winding can be connected to a Chern number~\cite{MadsenPRB2013,MoustajCondMat2025,JagannathanPRB20205,JagannathanArxiv2026}.
The localization of the winding states depends on the phason angle~\cite{MarsalArxiv2026}. 
We illustrate the eigenstate density as a function of position in Figs.~\ref{fig:spec_example}(b) and \ref{fig:spec_example}(d) for the phason angles $\varphi_1$ and $\varphi_2$, which lie symmetric on opposite sides of the gap.
These figures show that the localization of the winding state switches from one edge to the other as we move from $\varphi_1$ to $\varphi_2$, despite becoming highly delocalized around the gap minimum at $\varphi_0$, as shown in \ref{fig:spec_example}(c).
Importantly, the winding states can transfer their localization from one edge of the chain to the other continuously as $\varphi$ is tuned adiabatically~\cite{ghosh2025quantum}.
This edge-to-edge transport mechanism serves as the basis for the fully time-dependent quasiperiodic transfer protocol studied in this work.

\section{Analytical calculations: optimized transfer protocols}
\label{sec:methods:analytics}
We now discuss the analytical methods and derivations used in our work to develop and characterize the adiabatic state transfer protocols for the edge-localized winding states. 
Specifically, we first derive an optimized transfer protocol in terms of transfer time and fidelity, then an effective two-level approximation, and finally a useful protocol parametrization and effective model to facilitate experimental comparison.

\subsection{Transfer protocol and adiabaticity}
\label{subsec:protocols}
The goal of the protocol is to initialize the system in an edge-localized winding state at an initial phason angle $\varphi_1$  and then vary $\varphi(t)$ such that the state is transported to the opposite edge at a final phason angle $\varphi_2$ [see Fig.~\ref{fig:spec_example}(d)] with maximal fidelity.
This dynamical process is governed by the time-dependent Schr\"odinger equation
\begin{equation}
i\hbar \partial_t \ket{\Psi(t)}
=
\hat{H}\bigl[\varphi(t)\bigr]\ket{\Psi(t)} ,
\label{eq:tdse_phi}
\end{equation}
where $\hat{H}[\varphi(t)]$ denotes the instantaneous Hamiltonian at phason angle $\varphi(t)$.
For each fixed value of $\varphi$, we define the instantaneous eigenvalue problem
\begin{equation}
\hat{H}(\varphi)\ket{\psi_n(\varphi)} = E_n(\varphi)\ket{\psi_n(\varphi)} .
\label{eq:instantaneous_eigenproblem}
\end{equation}
The target winding state is denoted by $\ket{\psi_w(\varphi)}$, with instantaneous energy $E_w(\varphi)$.
The standard adiabatic condition~\cite{Born:1928cqs, Comparat:2009, Bradlyn:2022} requires that transitions from the winding state to all other instantaneous eigenstates remain suppressed during the evolution.
A useful criterion to avoid these non-adiabatic excitations can be found by defining the time-independent function $\lambda^{(0)}(\varphi)$~\cite{Jansen:2006,Liu:2022}
\begin{align}
\dot{\varphi}(t)\lambda^{(0)}(\varphi)
\equiv\dot{\varphi}(t)
\sum_{n\neq w}
\left|
\frac{
\mel{\psi_w(\varphi)}{\partial_\varphi \hat{H}}{\psi_n(\varphi)}
}{
E_w(\varphi)-E_n(\varphi)
}
\right|.
\label{eq:lambda0}
\end{align}
The quantity \(\dot\varphi\,\lambda^{(0)}\) sets a characteristic nonadiabatic transition rate and must remain small compared with the relevant spectral frequency scale.
This means that the phason velocity $\dot{\varphi}(t)$ needs to be small when $\lambda^{(0)}(\varphi)$ is large.
As the time scale of the system will be related to the relative eigenenergies in the spectrum, we further consider the family of adiabatic weight functions
\begin{equation}
\lambda^{(N)}(\varphi) = \sum_{n\neq w} \left| \frac{ \mel{\psi_w(\varphi)}{\partial_\varphi \hat{H}}{\psi_n(\varphi)}
}{\left[E_w(\varphi)-E_n(\varphi)\right]^{N+1}} \right| ,
\label{eq:lambdaN}
\end{equation}
where the parameter $N$ amplifies the effect of relative band energies on the metric and controls how strongly the protocol slows down close to small gaps.
The phason trajectory $\varphi^{(N)}(t)$ is then chosen from the differential condition
\begin{equation}
\dot{\varphi}^{(N)}(t)\lambda^{(N)}\!\left(\varphi^{(N)}(t)\right) = \frac{I^{(N)}}{T},
\label{eq:protocol_condition}
\end{equation}
where $T$ is the total transfer time and
\begin{equation}
I^{(N)} \equiv \int_{\varphi_1}^{\varphi_2} \lambda^{(N)}(\varphi)\,\mathrm{d}\varphi.
\label{eq:IN}
\end{equation}
This construction distributes the chosen adiabatic weight along the state-transfer path.
By defining the relevant minimum gap between the winding state and the closest adjacent band
\begin{equation}
G =  E_{w+1}(\varphi_0)-E_w(\varphi_0),
\label{eq:gap}
\end{equation}
we can introduce the dimensionless adiabatic parameter
\begin{equation}
\tau_{\mathrm{A}}^{(N)} = G^{N} I^{(N)},
\label{eq:JN}
\end{equation}
so that the protocol is expected to become adiabatic when
\begin{equation}
T \gg T^{(N)}_{\mathrm{A}} \equiv (N+1) \hbar \tau_{\mathrm{A}}^{(N)}/G
\label{eq:adiabatic_condition_final}
\end{equation}
The quantity $T^{(N)}_{\mathrm{A}}$, which has units of time, is thus also referred to as the \emph{adiabatic boundary} for $T$. 
The factor of $N+1$ was found heuristically, as this definition of $T_{A}^{(N)}$ gives a better lower limit. 

To quantify the quality of state transfer, we compute the final fidelity
\begin{equation}
\mathcal{F} = \left| \braket{\psi_w(\varphi_2)}{\Psi(T)} \right|^2 ,
\label{eq:fidelity}
\end{equation}
where $\ket{\Psi(T)}$ is the obtained time-evolved state at the end of the protocol and $\ket{\psi_w(\varphi_2)}$ is the desired target winding state at the final phason angle.
Perfect adiabatic transfer corresponds to $\mathcal{F}=1$, while deviations from unity quantify nonadiabatic leakage into other instantaneous eigenstates.

\begin{figure*}
    \begin{minipage}{0.49\textwidth}
        \centering
        \includegraphics[width=\linewidth]{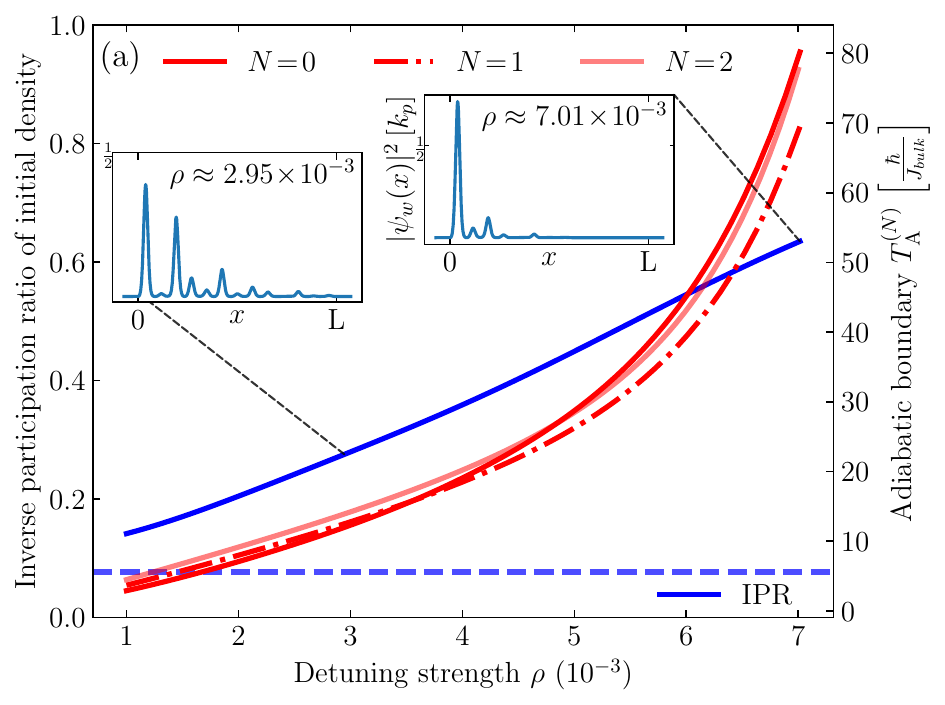}
    \end{minipage}
    \begin{minipage}{0.49\textwidth}
        \centering
        \includegraphics[width=\linewidth]{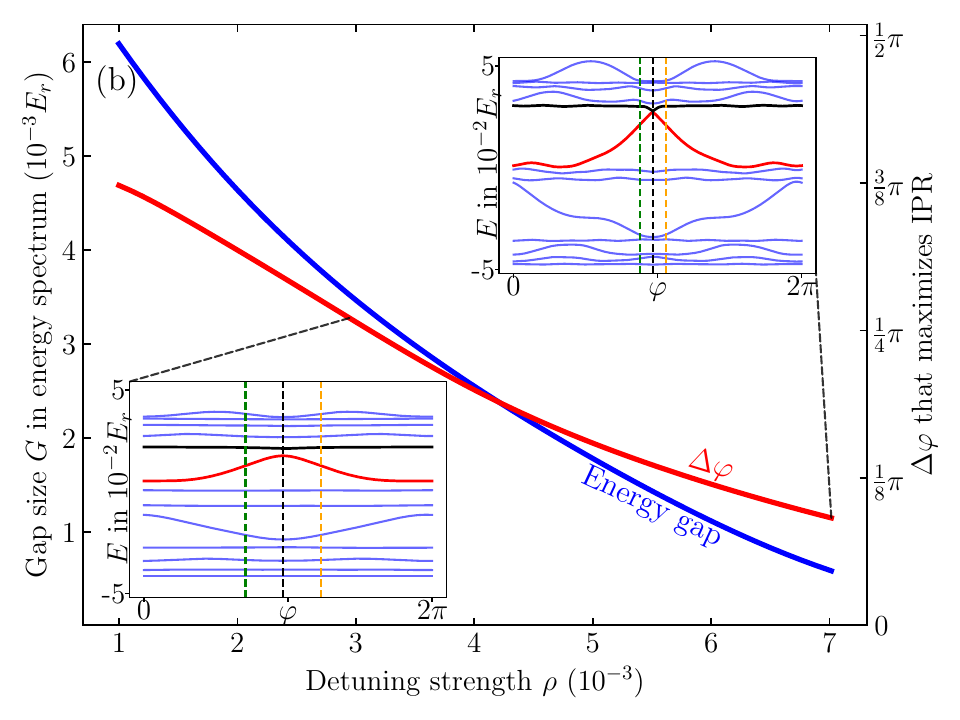}
    \end{minipage}
        \caption{ 
        \textbf{Localization-adiabaticity tradeoff in phason-mediated state transfer.}
        (a) Inverse participation ratio (left axis) and adiabatic boundary $T_A^{(N)}$ (right axis) as functions of the detuning strength $\rho$ for a system of $N_w=13$ sites.
        The different curves for $T_A^{(N)}$ correspond to different locally adiabatic protocols labeled by the order $N$.
        The increasing IPR indicates stronger localization of the winding states for larger quasiperiodic modulation, shown in the insets for two examples (low detuning $\rho \approx2.95 \times 10^{-3}$ and larger detuning $\rho \approx 7.01\times 10^{-3}$). 
        Smaller detuning strengths further weaken the edge localization.
        The horizontal dashed blue line indicates the smallest possible value for the IPR ($1/13$). 
        (b) Minimum gap $G$ (left axis) and phason interval $\Delta\varphi$ (right axis) as functions of the detuning strength $\rho$. The insets show the energy spectrum as a function of the phason angle corresponding to the chosen $\rho$ in the insets of panel (a).
        } 
        \label{fig:IPRvsJ}
\end{figure*}

\subsection{Optimizing localization, gap size, and protocol geometry}
\label{subsec:localization}

The efficiency of the transfer protocol is controlled not only by its shape and the total transfer time $T$, but also by several properties of the selected winding state.

First, we quantify the spatial localization of the initial winding state through its inverse participation ratio (IPR)~\cite{EdwardsIPR_1972},
\begin{equation}
\mathrm{IPR}\left[\ket{\psi_w(\varphi_1)} \right]
= \sum_{j=1}^{N_w} \left| \left<w_j \middle| \psi_w(\varphi_1) \right> \right|^4,
\label{eq:ipr}
\end{equation}
where $\{\ket{w_j}\}$ denotes the Wannier basis.
For a state localized on a single site, $\mathrm{IPR}=1$, while for a state uniformly spread over $N_w$ sites, $\mathrm{IPR}=1/N_w$.
The IPR therefore provides a direct measure of how strongly the winding state is localized (at the edge) at the beginning of the protocol.

The initial phason angle $\varphi_1$ is then chosen such that the IPR of the winding state is maximal, corresponding to the most strongly edge-localized configuration.
The central angle $\varphi_0$ denotes the angle around which the energy spectrum is symmetric. From the condition
\begin{equation}
    V(x,\varphi_0+\delta\varphi)=V(L-x,\varphi_0-\delta \varphi), \quad \forall \delta\varphi
    \label{eq:symmetry_condition}
\end{equation}
this central angle can be determined to be $\varphi_0=-\pi\beta N_w\pmod{\pi}$.
The final phason angle is then chosen symmetrically as
\begin{equation}
\varphi_2 = \varphi_0 + \Delta\varphi,
\qquad
\Delta\varphi=\varphi_0-\varphi_1.
\label{eq:delta_phi}
\end{equation}
Thus, the protocol transports the winding state across the avoided crossing region over the interval $\varphi\in[\varphi_1,\varphi_2]$.

The quantities $\mathrm{IPR}$, $G$, and $\Delta\varphi$ all play distinct and partially competing roles in the state-transfer dynamics.
Over the parameter range considered here, the $\mathrm{IPR}$ increases monotonically with the detuning strength $\rho$.
A large IPR (and thus a large $\rho$) is generally desirable because it corresponds to a strongly localized edge state with minimal hybridization
with the bulk.
Within the same parameter range, though, increasing $\rho$ also reduces the minimum gap $G$, making the dynamics more susceptible to nonadiabatic transitions and therefore requiring slower protocols to maintain high fidelity.
At the same time, increasing $\rho$ decreases the required phason interval $2\Delta\varphi$, meaning that the winding state needs to travel over a shorter distance in phason space before reaching the avoided crossing region, which is generally preferable.

These competing trends are summarized in Fig.~\ref{fig:IPRvsJ}.
Optimizing the state-transfer protocol requires balancing these opposite tendencies rather than simply maximizing a single quantity.
As a result, the adiabatic boundary $T_A^{(N)}$ (also shown in Fig.~\ref{fig:IPRvsJ}) exhibits a nontrivial dependence on both $\rho$ and the protocol order $N$.
We remark that, while higher protocol orders ($N \ge 2$)  can reduce the adiabatic parameter, they may also enhance coherent fidelity oscillations, making the optimal choice of protocol nontrivial.
We therefore focus on the $N=0$ and $N=1$ cases in the rest of this work, briefly considering higher protocol orders in appendix~\ref{app:two-level}, and leaving a more in-depth analysis as an interesting direction for future work.

\subsection{Two-level approximation}
\label{subsec:two-level-approx}

The dominant nonadiabatic losses typically occur near the minimum gap between the winding state and the closest adjacent band.
This motivates an effective two-level description in the subspace spanned by the states $\left\{ \ket{\psi_w(\varphi)},\, \ket{\psi_{w+1}(\varphi)} \right\}$.
Within this approximation, the time-evolved state can be written as (see Appendix~\ref{app:two-level} for details)
\begin{align}
&\ket{\Psi(t)} \approx c_w(t) e^{-\frac{i}{\hbar} \int_0^t E_w[\varphi(t')]\,\mathrm{d}t'} \ket{\psi_w[\varphi(t)]} \nonumber\\
& + c_{w+1}(t) e^{-\frac{i}{\hbar} \int_0^t E_{w+1}[\varphi(t')]\,\mathrm{d}t'} \ket{\psi_{w+1}[\varphi(t)]},
\label{eq:two_level_ansatz}
\end{align}
where $c_w$ and $c_{w+1}$ are the coefficients corresponding to the winding state and its adjacent state, respectively.

A main advantage of the two-level picture is that, for the $N=1$ protocol, it yields a closed expression for the fidelity,
\begin{align}
\mathcal{F}_{\mathrm{2L}}^{(1)} &= 1 - \frac{4\left[\tau_{\mathrm{A}}^{(1)}\right]^2}{\left( \frac{GT}{\hbar} \right)^2+4\left[\tau_{\mathrm{A}}^{(1)}\right]^2} \nonumber \\
&\qquad \times \sin^2 \left[ \frac{f}{2} \sqrt{ \left( \frac{GT}{\hbar} \right)^2+4\left[\tau_{\mathrm{A}}^{(1)}\right]^2} \right],
\label{eq:two_level_fidelity}
\end{align}
where
\begin{equation}
f = \frac{1}{GT} \int_0^T \left[ E_{w+1}[\varphi(t)] - E_w[\varphi(t)] \right] \mathrm{d}t .
\label{eq:f1}
\end{equation}
We use this expression as an analytical benchmark for the full numerical simulations.
Good agreement with Eq.~\eqref{eq:two_level_fidelity} indicates that the state transfer dynamics are dominated by the winding state and the nearest adjacent band, while deviations signal leakage into additional bulk states or limitations of the two-level approximation. 
Interestingly, Eq.~\eqref{eq:two_level_fidelity} predicts unit fidelity for finite transfer time when the sine term of $\mathcal{F}_{\mathrm{2L}}^{(1)}$ vanishes, which occurs for
\begin{equation}
    T_n=\frac{\hbar}{G}\sqrt{\left(\frac{2\pi n}{f}\right)^2-4\left[\tau_{\mathrm{A}}^{(1)}\right]^2}, \quad n \in \mathbb{N}_{>0}.
    \label{eq:Unit_fidelity_times}
\end{equation}
For the $N=0$ protocol, the two-level approximation does not yield a simple closed analytical expression for the fidelity.
Instead, the coefficients $c_w(t)$ and $c_{w+1}(t)$ are obtained by numerically integrating the effective two-level evolution equations derived in Appendix~\ref{app:two-level}.
The corresponding fidelity is then given by $\mathcal{F}_{\mathrm{2L}}^{(0)}=|c_w(T)|^2$, assuming the system is initialized in the winding state with $c_w(0)=1$ and $c_{w+1}(0)=0$.

\begin{figure*}[t!]
   \includegraphics[width=\linewidth]{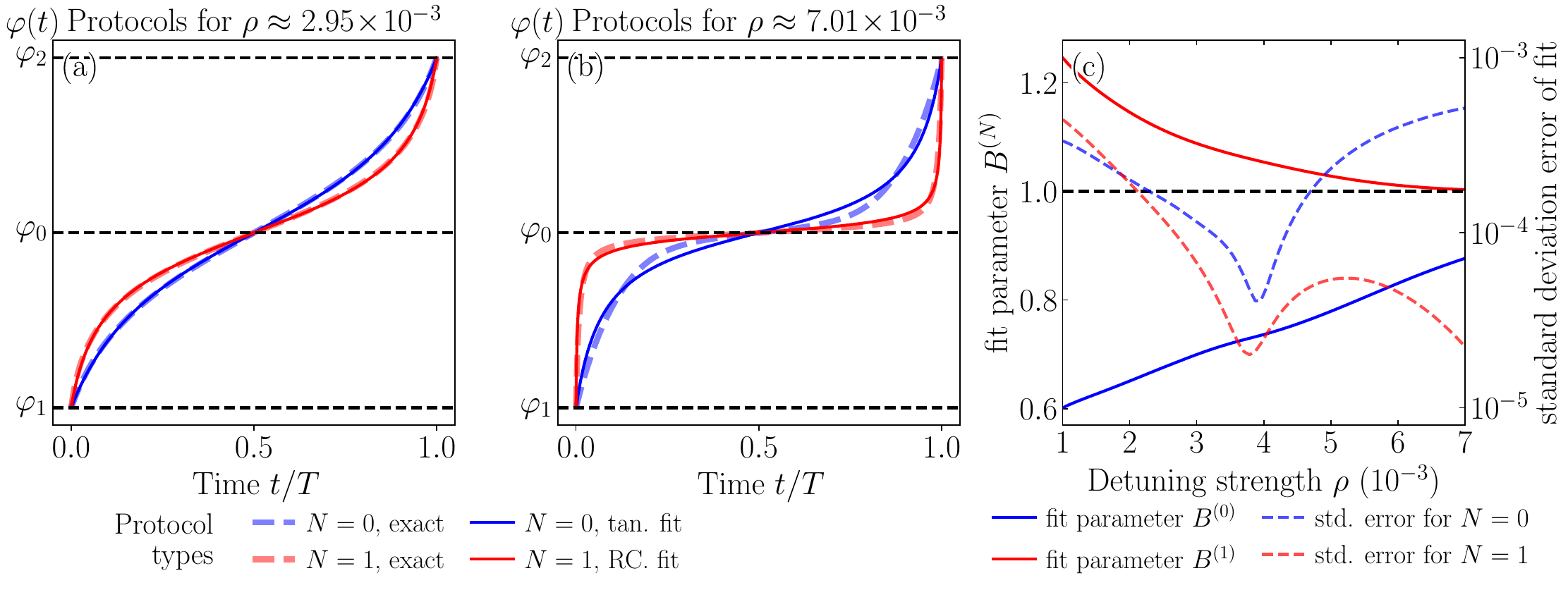}
   \vspace{-0.9cm}
    \caption{
    (a)-(b) Representative phason protocols $\varphi(t)$ used for quasiperiodic state transfer for two different values of detuning strength $\rho$. 
    The different colors correspond to distinct choices (different values of $N$) of the local adiabatic condition in Eq.~\eqref{eq:protocol_condition}.
    The dashed lines are the exact numerical solutions for Eq.~\eqref{eq:protocol_condition}, while the solid lines are fits of the tangent-protocol (blue) in Eq.~\eqref{eq:Tangentprot} and the Roland-Cerf protocol (red) in Eq.~\eqref{RolandCerfProt}. 
    (c) Fit parameters $B^{(0)}$ and $B^{(1)}$ as functions of $\rho$ for the $N=0$ tangent protocol (left axis, solid blue) and the $N=1$ Roland-Cerf protocol (left axis, solid red), respectively. The black line at $B^{(N)}=1$ indicates the respective upper/lower limit for $B^{(0)}$/$B^{(1)}$. 
    The corresponding standard deviation errors are depicted as dashed lines using the right axis for scale.}
    \label{fig:protocols}
\end{figure*}

\subsection{Analytical protocol parametrizations}
\label{subsec:protocol_parametrizations}

The optimal locally adiabatic protocols obtained from
Eq.~\eqref{eq:protocol_condition} generally require a numerical solution of the instantaneous eigenvalue problem.
However, for experimental implementations, it is more desirable to construct simple closed-form parametrizations that  reproduce the exact numerical protocols.
Building on the previous assumption that the state transfer dynamics are dominated by an avoided crossing between the winding state and the nearest adjacent band, we can compare our model locally with the Landau--Zener Hamiltonian~\cite{Landau:1932, Zener:1932,Glasbrenner:2023}
\begin{equation}
H_{\mathrm{LZ}}(t) = \Phi(t)\sigma_z + \alpha \sigma_x,
\end{equation}
where $\sigma_x$ and $\sigma_z$ are the respective Pauli matrices, $\Phi\in [-\infty,\infty]$ is a sweep parameter similar to $\varphi$, and $\alpha$ is the coupling between the two levels.
The instantaneous eigenenergies are $E_{\pm}(\Phi)=\pm\sqrt{\Phi^2+\alpha^2}$.
Hence, the instantaneous level separation is $E_{+}-E_{-}=2\sqrt{\Phi^2+\alpha^2}$, and the minimum gap at the avoided crossing $\Phi=0$ is $G_{\mathrm{LZ}}=2|\alpha|$.

Within the Landau-Zener approximation, the generalized adiabaticity measure introduced in Eq.~\eqref{eq:lambdaN} becomes
\begin{equation}
\lambda_{\mathrm{LZ}}^{(N)}(\Phi) = \frac{|\alpha|}{2^{N+1} \left(\Phi^2+\alpha^2\right)^{(N+2)/2}}.
\label{eq:lambda_LZ}
\end{equation}
The corresponding locally adiabatic protocols can be obtained analytically.
For $N=0$, the solution takes the tangent form
\begin{equation}
\Phi^{(0)}(t)
=
A^{(0)}
\tan\left[
\pi B^{(0)}
\left(
\frac{t}{T}-\frac12
\right)
\right],
\label{eq:Tangentprot}
\end{equation}
with the parameters $A^{(0)}=\alpha$ and $B^{(0)}=1$.
This functional form is closely related to the tangent and hyperbolic sweep protocols appearing in Demkov--Kunike models
~\cite{Robinson:1985,Suominen:1992,Vitanov:1999},
which provide analytically tractable descriptions of nonadiabatic population transfer in driven two-level systems.
For $N=1$, we recover the Roland-Cerf protocol~\cite{Roland:2002,Malossi:2013,Stefanatos:2020},
\begin{equation}
\Phi^{(1)}(t) = A^{(1)} \frac{
2\left(\frac{t}{T}-\frac12\right)
}{\sqrt{ \left(B^{(1)}\right)^2
- 4\left(\frac{t}{T}-\frac12\right)^2}},
\label{RolandCerfProt}
\end{equation}
again with $A^{(1)}=\alpha$ and $B^{(1)}=1$.
Remarkably, we find that the optimal numerical phason protocols $\varphi(t)-\varphi_0$ can be fitted by these analytical forms with $B^{(N)}$ as a fit parameter while using the parameter $A^{(N)}$ to impose the boundary conditions $\varphi^{(N)}(0)=\varphi_1$ and $\varphi^{(N)}(T)=\varphi_2$.
Thus, we will refer to the $N=0$ and $N=1$ protocols as the tangent and Roland-Cerf protocols, respectively.

In Figs.~\ref{fig:protocols}(a)-(b) we compare the exact numerical protocol (dashed curves) for different protocol orders $N$ with the analytical approximations (solid curves) for two different detuning strengths $\rho$. 
In Fig.~\ref{fig:protocols}(c), we show the dependence of the fit parameters and the corresponding fitting error as a function of the detuning strength $\rho$.

\section{Continuum simulations and numerical implementation}
\label{sec:methods:numerics}
To illustrate the performance of a realistic continuum implementation of our protocols, we solve the continuum time-dependent Schrödinger equation with the multi-configurational time-dependent Hartree method for indistinguishable particles (MCTDH)~\cite{Streltsov:2006,Streltsov:2007, Alon:2007,Alon:2008}.
In particular, we employ the MCTDH-X software~\cite{Lode:2012,Lode:2016,Fasshauer:2016,Lode:2020,Lin:2020,Molignini:2025-SciPost,MCTDHX}, which has been successfully applied to a wide range of ultracold atomic systems, such as setups with optical lattices, harmonic traps, and more complex geometries~\cite{Xiang:2023, Beinke:2018, Roy:2018, Dutta:2019, Schaefer:2020, Lode:2021, Lode:2021-10, Debnath:2024, Roy:2023, Dutta:2023, Aloqali:2024, Dutta:2024, Haldar:2024, Chakrabarti:2024, Bhowmik:2025, Chakrabarti:2025-2, Roy:2025, Roy:2025-7, Dutta:2025, Roy:2025-4, Roy:2026, Fischer:2015, Chatterjee:2018, Chatterjee:2019, Bera:2019, Bera:2019-symm, Chatterjee:2020, Roy:2022, Hughes:2023, Bilinskaya:2024, Molignini:2024-2, Roy:2024-annals, Roy:2024-epjp, Chakrabarti:2025, Molignini:2025-JPCM, Lode:2017, Lode:2018, Molignini:2018, Lin:2019, Lin:2020-PRA, Lin:2021, Molignini:2022, Rosa-Medina:2022, Ortuno-Gonzalez:2025}.
A more detailed overview of the method and implementation details is provided in Appendix~\ref{app:MCTDHX}.
Our numerical strategy consists of embedding the lattice-model winding states into a continuum representation and subsequently evolving them under the full continuum Hamiltonian that more faithfully describes experimental realizations in ultracold atomic platforms. 

For the effectively single-particle system considered here, we can set $N_p=1$, $g=0$.
In this case, the MCTDH-X ansatz is exact with a single
time-dependent orbital ($M=1$). 
The propagation therefore reduces to solving
\begin{equation}
i\hbar\partial_t\psi(x,t)
=
\left[
-\frac{\hbar^2}{2m}\frac{\mathrm{d}^2}{\mathrm{d}x^2}
+
V\bigl(x;\varphi(t)\bigr)
\right]\psi(x,t),
\label{eq:continuum_tdse}
\end{equation}
where $\psi(x,t)$ is the continuum time-dependent wave-function of the problem.
While MCTDH-X is typically employed for correlated problems, it nevertheless provides (i) a convenient continuum
implementation of the time-dependent protocol with an adaptive solver and (ii) a direct route toward future extensions involving interactions and genuinely correlated many-body dynamics.

\begin{figure*}[t!]
 \begin{minipage}{0.49\textwidth}
        \centering
        \includegraphics[width=\linewidth]{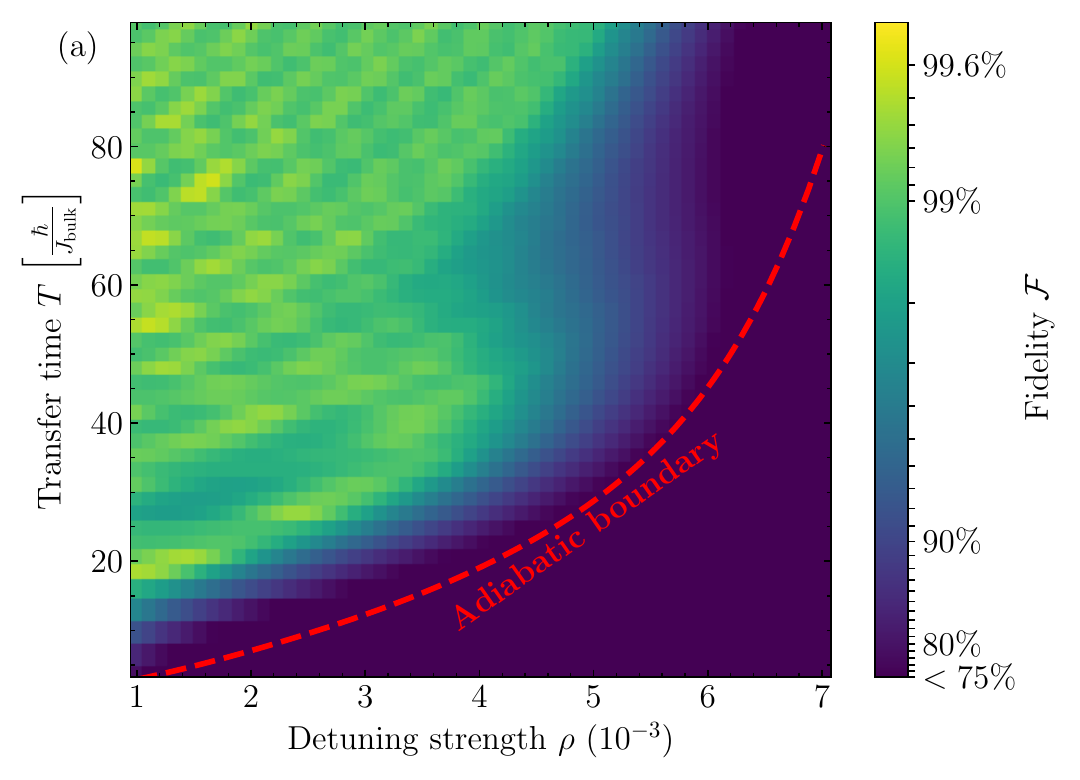}
    \end{minipage}
	\centering
     \begin{minipage}{0.49\textwidth}
        \centering
        \includegraphics[width=\linewidth]{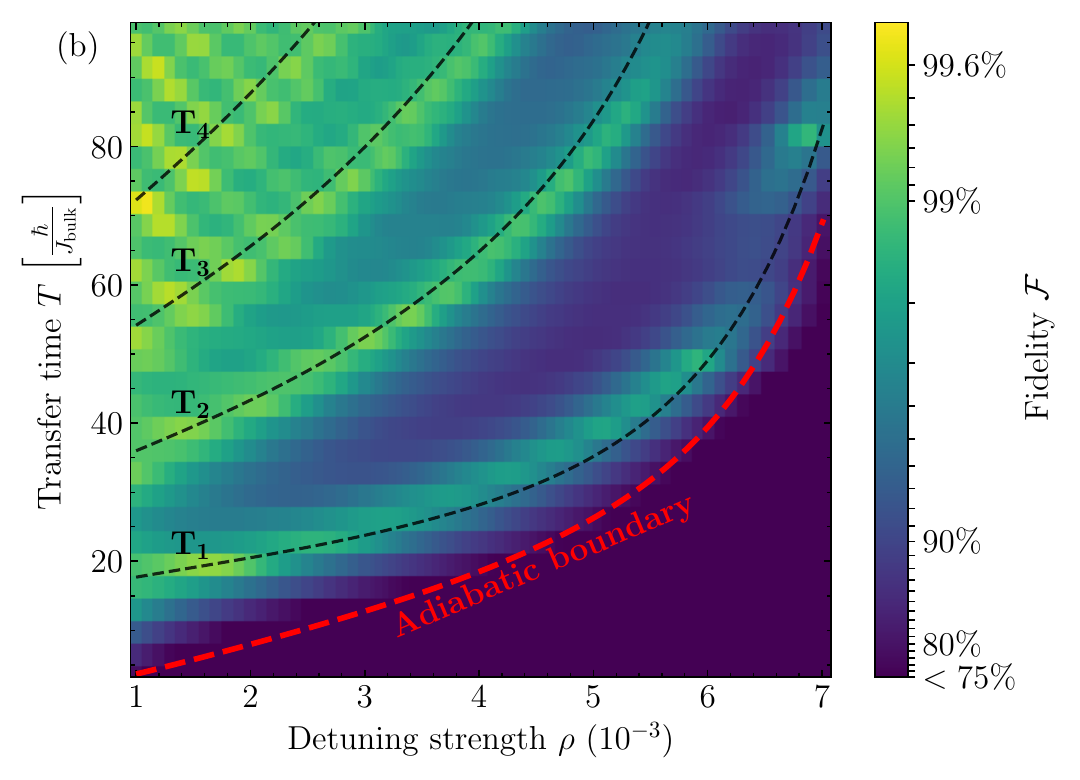}
    \end{minipage}
    \caption{\textbf{Performance diagram for the state-transfer fidelity.} 
    Final-state fidelity $\mathcal{F}$ as a function of the dimensionless transfer time $TJ_{\mathrm{bulk}}/\hbar$ and detuning strength $\rho$ for (a) the tangent protocol ($N=0$) and (b) the Roland-Cerf protocol ($N=1$). 
    Each panel was obtained with 2160 MCTDH-X simulations over a $72 \times 30$ grid in $(\rho, T)$ space.
    The color scale is logarithmic, and fidelities below $\mathcal{F}=0.75$ are omitted to emphasize the high-fidelity region. 
    The dashed red line indicates the adiabatic boundary from Eq.~\eqref{eq:adiabatic_condition_final}. 
    In the right panel, we additionally show the times $T_n$ from Eq.~\eqref{eq:Unit_fidelity_times} for which the two-level assumptions predict unit fidelity.
    }
     \label{fig:PhaseDiagram}
\end{figure*}

\subsection{Finite-size Wannier basis and state preparation}
\label{subsec:Wannier}

The first step in our simulations is to construct a state representation for the continuum system.
This is obtained from the Wannier functions associated with the finite primary optical lattice.
Unlike standard infinite-system Wannier functions~\cite{Marzari:2012}, the orbitals employed here are generated directly from the finite-size continuum setup with hard-wall boundary conditions~\cite{Dutta:2022, Hughes:2023}.
As a result, they naturally incorporate finite-size corrections associated with the boundaries (see Appendix~\ref{app:wannier} for more details).
Starting from the continuum single-particle Hamiltonian $\hat{H}_0$ in Eq.~\eqref{eq:hamiltonian:single_particle},  we compute the lowest-band single-particle eigenstates and construct a set of localized Wannier functions $\left\{ w_j(x) \right\}$, centered around the minima of the primary optical lattice.
For a given initial phason angle $\varphi_1$, we then identify the desired winding state of the tight-binding Hamiltonian and express it in the continuum Wannier basis by replacing the discrete site occupations with the corresponding finite-size Wannier functions,
\begin{align}
    \psi_w(x, 0) &\equiv \left<x|\psi_w(\varphi_1)\right> \nonumber \\
    &=\sum_{j=1}^{N_w} C_{j} \left<x|w_j\right> = \sum_{j=1}^{N_w} C_{j} w_j(x),
    \label{eq:lattice_state_expansion}
\end{align}
where $C_j=\braket{w_j}{\psi_w(\varphi_1)}$ are the expansion coefficients of the lattice winding state in the Wannier basis.
The resulting full continuum wave function $\psi_w(x,0)$ is then used as the (single) initial orbital needed for initializing the noninteracting MCTDH-X simulation.

\subsection{Time evolution and fidelity evaluation}
\label{subsec:time-evol}
Once the initial continuum state has been constructed, we load it into MCTDH-X and propagate it under the full continuum Hamiltonian of Eq.~\eqref{eq:continuum_hamiltonian}.
The time dependence enters through the phason parameter $\varphi(t)$ according to the locally adiabatic protocols with different adiabatic orders $N$, described in Sec.~\ref{subsec:protocols}.

During the time evolution, this single orbital is propagated according to the MCTDH-X equations of motion, which for $N_p=M=1$ reproduce the exact single-particle continuum dynamics generated by the time-dependent Hamiltonian.
The simulations are performed for different values of the detuning strength $\rho$ and total transfer time $T$. 
In the main text, we focus on the case where the system size consists of $N_w=13$ lattice sites while in Appendix~\ref{app:convexity} we also discuss the next Fibonacci approximant $N_w=21$ and mention briefly what might differ for even larger system sizes ($N_w=34,55,\dots$).

At the end of the evolution, the obtained final continuum state $\ket{\Psi(T)}$ is compared with the desired target winding state at the final phason angle $\varphi_2$ by calculating the fidelity in Eq.~\eqref{eq:fidelity} numerically.
The target state is also obtained by diagonalizing the lattice Hamiltonian at $\varphi_2$ and mapping the corresponding lattice winding state into the continuum Wannier representation according to Eq.~\eqref{eq:lattice_state_expansion}.
Thus, both the initial and target states are represented as full continuum wave functions constructed within the same finite-size Wannier basis.

\section{Transfer protocol performance}
\label{sec:results}
We now present the results of the time evolution of the winding states under the different adiabatic protocols and compare their performance.
We compute the fidelity $\mathcal{F}$ as a function of transfer time $T$ and detuning strength $\rho$ using the tangent protocol ($N=0$) and the Roland-Cerf protocol ($N=1$).
We consider times up to $T=100\hbar/J_{\mathrm{bulk}}$, where $J_{\mathrm{bulk}}$ represents the nearest-neighbor hopping amplitude in the central region of the lattice. The detuning strength is chosen between very small detuning strenghts ($\rho=0.001$), where the wavefunction is weakly edge localized, up to strong detuning strengths ($\rho=0.007$) with large $\mathrm{IPR}$ (see Fig.~\ref{fig:IPRvsJ}).

We begin by examining the fidelity landscape as a function of the detuning strength $\rho$ and transfer time $T$ for the tangent protocol in Fig.~\ref{fig:PhaseDiagram}(a) and the Roland-Cerf protocol in Fig.~\ref{fig:PhaseDiagram}(b).
Overall, both protocols exhibit similar qualitative trends. First, the fidelity generally increases with the transfer time, since slower protocols suppress nonadiabatic transitions. 
Second, the dependence on the detuning strength is nonmonotonic. 
At small and intermediate $\rho$, increasing the detuning strength can improve the transfer by producing more strongly localized initial and target winding states. 
At larger $\rho$, however, the minimum gap $G$ decreases~\cite{ghosh2025quantum}, and the resulting enhancement of nonadiabatic transitions causes the fidelity to deteriorate unless the transfer time is increased accordingly. 
These competing effects are consistent with the localization--gap tradeoff discussed in Sec.~\ref{subsec:localization}. 

Both protocols reach an overall maximum fidelity of around $99.6\%$. 
The two-level treatment of the Roland-Cerf protocol predicts unit fidelity for the times $T_n$ [Eq.~\eqref{eq:Unit_fidelity_times}], and we in fact observe that the fidelity reaches a peak around the black dashed lines in Fig.~\ref{fig:PhaseDiagram}(b).
We attribute the remaining discrepancy primarily to the finite but small hybridization between the winding state and additional states, as well as the inexactness of the tight-binding approximation.
A smaller contribution may also arise from the numerical discretization of the continuum propagation.

\begin{figure}
    \centering
    \includegraphics[width=1.0\linewidth]{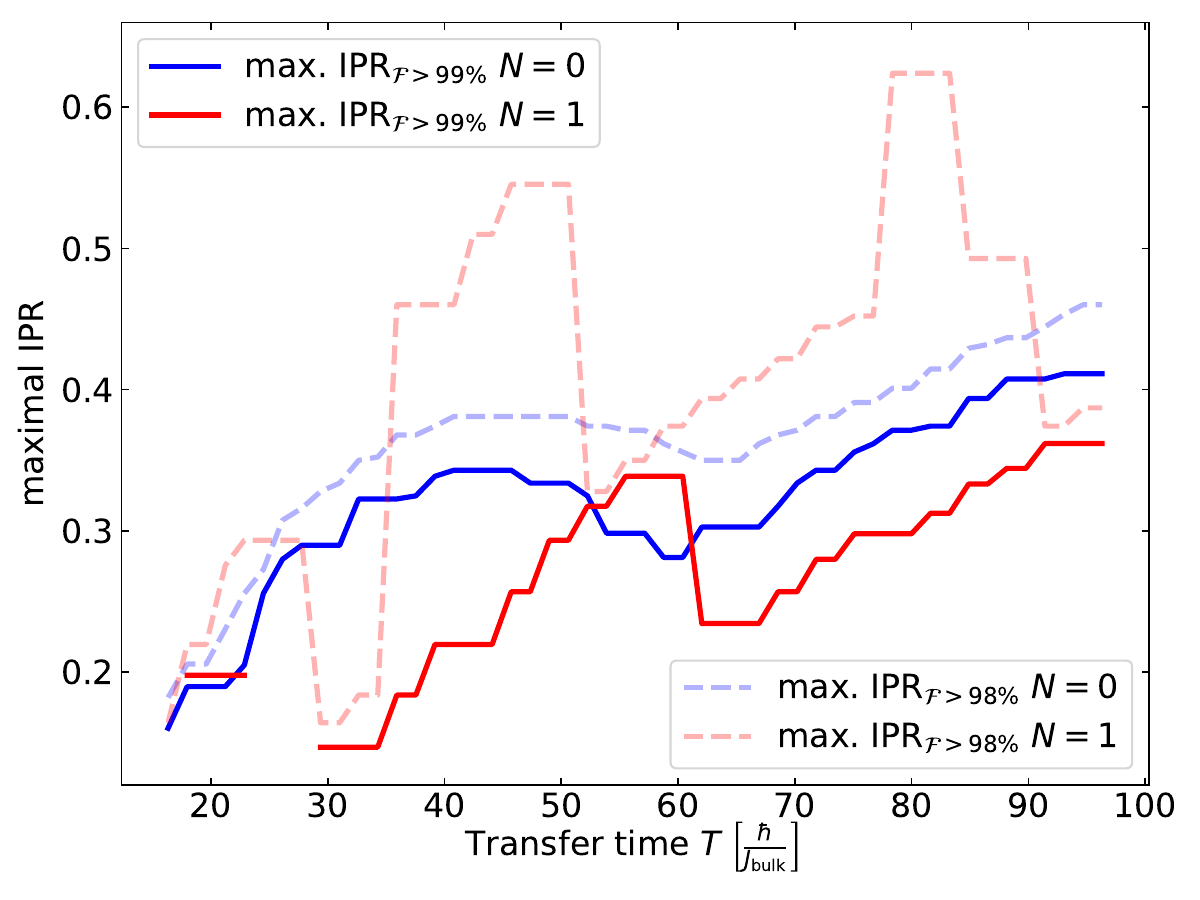}
    \caption{\textbf{Maximum transferable localization at fixed fidelity}. 
    For each value of the transfer time $T$, we plot the IPR of the most localized state that can be transported with a transport fidelity above $99\%$ (solid lines) and $98\%$ (dashed lines) for the two different protocols (colors).
    Due to the oscillatory behavior of the $N=1$ protocol, the fidelity is more sensitive to small deviations arising from the continuum--tight-binding mismatch and from numerical discretization.
    }
    \label{fig:fid_threshold}
\end{figure}

In regions of short transfer time and large detuning strength, both protocols fail to transfer the winding state with high fidelity because the evolution becomes strongly nonadiabatic and population is transferred to other instantaneous eigenstates.
These parameter regions should therefore be avoided in experimental implementations.
We omit the fidelity heatmaps below $\mathcal{F}=0.75$ to focus on the successful state transfer regions.

\begin{figure*}
    \begin{minipage}{0.49\textwidth}
        \centering
        \includegraphics[width=\linewidth]{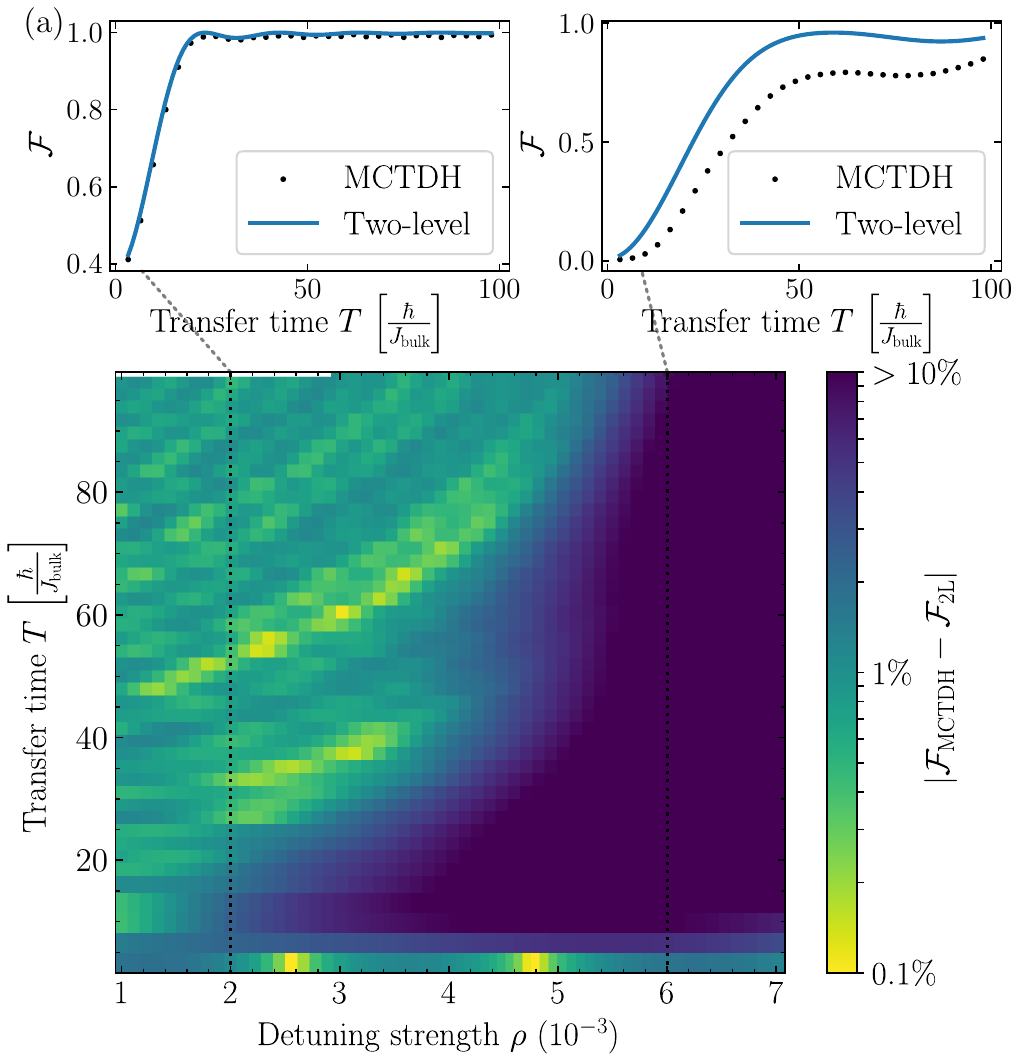}
    \end{minipage}
      \begin{minipage}{0.49\textwidth}
        \centering
        \includegraphics[width=\linewidth]{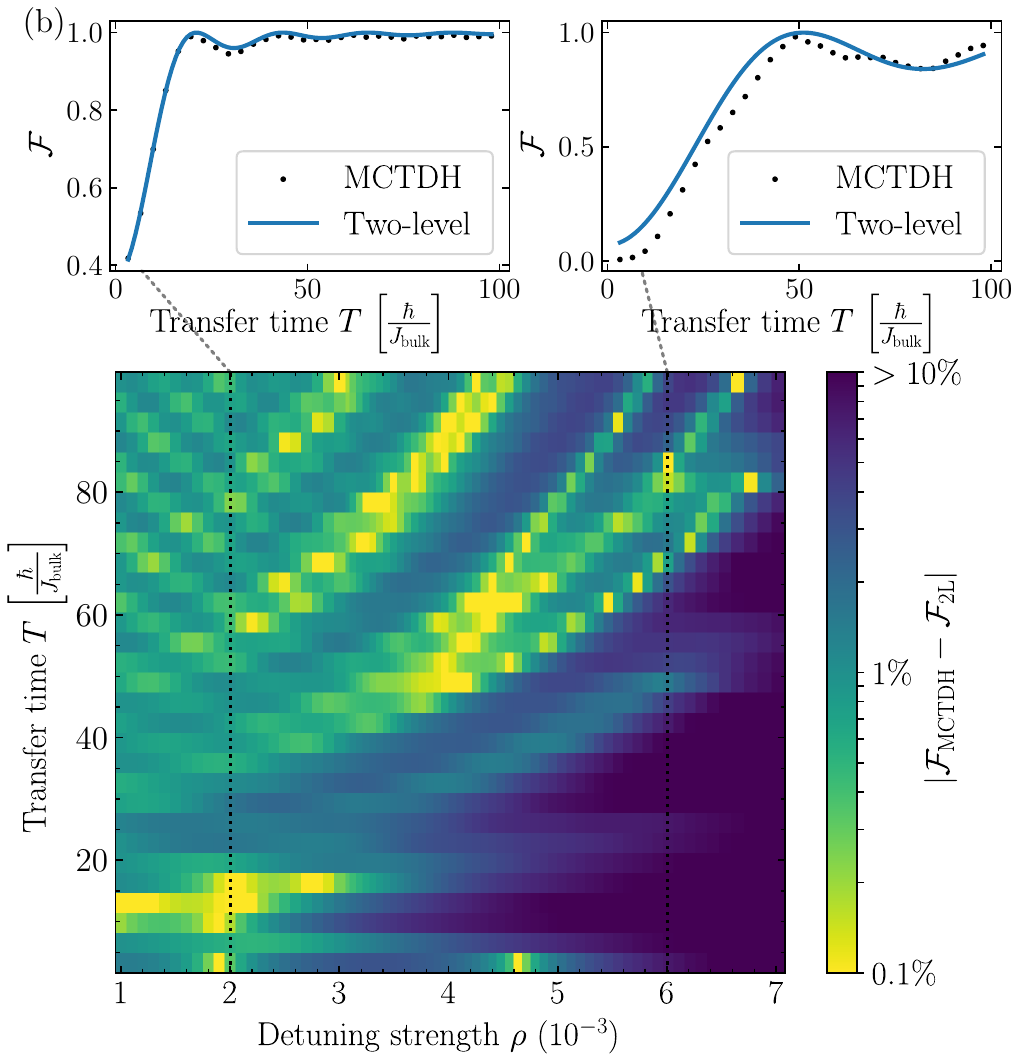}
    \end{minipage}
    \caption{\textbf{Comparison between numerical results and analytical two-level picture.} 
    The top panels show the transfer fidelity $\mathcal{F}$ as a function of the transfer time $T$ for the (a) tangent protocol ($N=0$, left) and (b) the Roland-Cerf protocol ($N=1$, right) for two fixed detuning strengths $\rho$.
    Continuum MCTDH-X calculations are shown with black dots, while the two-level approximation is depicted by a solid blue line.
    The bottom panels show the absolute fidelity error of the two-level approximation $|\mathcal{F}_{\mathrm{MCTDH}} - \mathcal{F}_{2L}|$ over the $(T, \rho)$ plane.
    Dark blue colors indicate a poor performance of the protocol, which typically occurs for small energy gaps $G$ (larger detuning strength $\rho$) and short transfer times $T$.
    Each panel was obtained with 2160 MCTDH-X simulations over a $72 \times 30$ grid in $(\rho, T)$ space.
    }
     \label{fig:PhaseDiagram-comparison}
\end{figure*}

The dashed red line shows the characteristic adiabatic timescale $T=T_{\mathrm{A}}^{(N)}$ obtained from
Eq.~\eqref{eq:adiabatic_condition_final}.
We note that, while the adiabatic boundary  broadly tracks the crossover from predominantly low- to
high-fidelity transfer, it should not be interpreted as a sharp boundary, since adiabaticity formally requires
$T\gg T_{\mathrm{A}}^{(N)}$.
Moreover, coherent oscillations and transitions involving additional states produce visible deviations from this simple estimate.

The high-fidelity regimes are quite extensive in parameter space.
Remarkably, high fidelities can already be achieved with a relatively fast transfer time of around $20$--$40$ time units, provided $\rho$ is low enough~\footnote{
For a more concrete time measure, we can consider ${}^{87}\mathrm{Rb}$ atoms in a primary optical lattice with wavelength $\lambda_p=1064\,\mathrm{nm}$. 
At the lattice depth $V_p=10E_r$ used here, the bulk hopping is approximately $J_{\mathrm{bulk}}\simeq0.0227E_r$, corresponding to $J_{\mathrm{bulk}}/h\simeq47\,\mathrm{Hz}$. A dimensionless transfer time $TJ_{\mathrm{bulk}}/\hbar=40$ therefore corresponds to a physical duration of approximately $T\simeq136\,\mathrm{ms}$, which lies well within experimentally accessible timescales. Actual experimental operating times will additionally be constrained by heating and coherence timescales, the available phason-control bandwidth, and imperfections in state preparation.}.
These fast-transfer regimes correspond to winding states that are hardly localized at the boundaries. 
Such states may still be useful for edge-to-edge transport, particularly when short protocol durations and robustness against nonadiabatic losses are more important than single-site localization.
Conversely, transferring states localized almost entirely at the boundary requires larger detuning strengths and therefore longer protocol durations. 

Both protocols can achieve high fidelities in this regime, but, interestingly, the Roland-Cerf protocol remains effective over a broader range of $\rho$ and can reach fidelities close to unity for the most strongly localized states at very large values of $\rho$.
In general, the Roland-Cerf protocol outperforms the tangent protocol in high-localization regimes, in terms of achieving higher fidelities with faster protocols (lower values of $T$).

A disadvantage of the Roland-Cerf protocol seems to be that it is less stable against numerical distortions like inaccuracies coming from the tight-binding description. 
We show this in Fig.~\ref{fig:fid_threshold} where we plot the IPR of the most edge-localized state that can be transported with a transport fidelity above a certain threshold for the two protocols as a function of the transfer time. 
The fact that, for most transfer times, the tangent protocol can  transport more localized states with a transport fidelity above the threshold $99\%$ can be attributed to the mentioned distortions. 
When we lower the threshold to $98\%$, we again see how the Roland-Cerf protocol outperforms the tangent protocol.

We assume that the susceptibility of the $N=1$ protocol emerges from  the pronounced oscillatory dependence of the fidelity on the transfer time $T$. 
In contrast to the tangent protocol, where the fidelity stays close to unity once the transfer time is well above the adiabatic boundary, the fidelity of the Roland-Cerf protocol exhibits contour lines of roughly equal fidelity that follow the hyperbolic shape of the adiabatic boundary.
These oscillations arise from coherent interference between nonadiabatic transition amplitudes accumulated during the evolution.
Consequently, a naive increase in the total duration of the protocol does not necessarily lead to improved state-transfer fidelity.
To preserve a good state transfer while increasing the localization of a winding state (i.e., increasing $\rho$), a useful strategy would be to increase the transfer time along the hyperbolic curves $T_n$ instead of simply increasing it proportionally to $\rho$.

Next, we demonstrate the comparison between the MCTDH-X simulations and the two-level approximation in detail.
For experimental implementations, it is important to have such simple tools to estimate the fidelity. 
In Fig.~\ref{fig:PhaseDiagram-comparison}, we thus show the discrepancy between the numerical results and the analytical two-level description introduced in Sec.~\ref{subsec:two-level-approx} and Appendix~\ref{app:two-level}.

We quantify the absolute fidelity error as
\begin{equation}
\delta \mathcal{F}(T,\rho) =
\left| \mathcal{F}_{\mathrm{MCTDH-X}}(T,\rho) - \mathcal{F}_{\mathrm{2L}}(T,\rho) \right|.
\label{eq:two_level_error}
\end{equation}
Fig.~\ref{fig:PhaseDiagram-comparison} shows $\delta \mathcal{F}$ over the full parameter space for the tangent protocol in panel~(a) and the Roland-Cerf protocol in panel~(b).
The insets additionally compare the two fidelities along cuts at small and large detuning strengths, respectively.
We emphasize that $\delta \mathcal{F}$ measures the total discrepancy between the analytical two-level description and the full continuum simulation. 
It therefore contains contributions both from the truncation of the dynamics to two instantaneous states and from residual differences between the effective tight-binding model and the continuum Hamiltonian.

Overall, we find that the two-level approximation agrees well with the numerical results for sufficiently long transfer times $T$ and small detuning strengths $\rho$.
This is particularly true for the Roland-Cerf protocol.
Furthermore, the two-level approximation is also able to qualitatively reproduce the fidelity oscillations observed in the full continuum simulations.
The approximation becomes less accurate for short transfer times and, especially, for larger values of $\rho$, where transitions involving additional states become increasingly relevant.
This behavior is consistent with what is shown in Fig.~\ref{fig:IPRvsJ}, where small values of $\rho$ lead to larger spectral gaps $G$, which improves the accuracy of the two-level approximation due to weaker interband coupling. 
Slow transfer protocols further suppress dynamical excitations to other bands due to adiabaticity. 

While the qualitative localization--gap tradeoff is general, changes to the setup geometry in terms of different primary potential depth $V_p$ and system size $N_w$ can further affect the quantitative behavior of the fidelity.
We discuss the impact of these changes in Appendices~\ref{app:ChangingS} and \ref{app:convexity}, respectively.
In particular, the findings of Appendix~\ref{app:convexity} reveal that the transfer protocol performance can be impacted by the convexity of neighboring bands in the energy spectrum, which can depend nontrivially on the system size $N_w$.
For example, for $N_w=21$, the winding state and the neighboring band have the same curvature near the avoided crossing (as opposed to the opposite curvature exhibited by the $N_w=13$ system).
Consequently, the numerically obtained $N=0$ and $N=1$ locally adiabatic protocols no longer exhibit the simple tangent and Roland-Cerf shapes found for $N_w=13$.
This change in spectral geometry also reverses their relative performance: whereas the $N=1$ protocol is generally more effective for $N_w=13$, the $N=0$ protocol performs more robustly for $N_w=21$, including at large detuning strengths.
The optimal protocol order is therefore not universal, but must be determined from the spectrum of the particular finite-size approximant.

\section{Conclusions and Outlook}
\label{sec:conclusions}
To summarize, in this work we have investigated quantum state transfer mediated by edge-localized winding states in a one-dimensional quasiperiodic ultracold atomic system.
Motivated by the high degree of control available in ultracold-atom experiments, we constructed locally adiabatic protocols from the instantaneous spectral properties of the corresponding tight-binding model and assessed their performance through continuum MCTDH-X simulations.
To obtain analytical insights into the transfer dynamics, we additionally derived an expression for the transfer fidelity within an effective two-level framework.

Our results reveal a fundamental tradeoff between the localization of the initial and target states and the minimum spectral gap encountered during the transfer: stronger edge localization is generally accompanied by a smaller gap and therefore requires longer protocol durations to suppress nonadiabatic transitions.
We further extended the locally adiabatic construction to higher protocol orders, which modify how strongly the evolution is slowed down near small-gap regions.
For the main system size ($N_w=13$) considered here, the stronger slowing down of the $N=1$ Roland-Cerf protocol near the minimum gap makes it particularly effective at large detuning strengths, where the winding states are strongly localized but nonadiabatic transitions are otherwise more difficult to suppress.
In this regime, it maintains high fidelity over a broader parameter range and can achieve transfer fidelities close to unity.

The continuum results are well explained by the analytical predictions over a broad parameter regime.
In particular, the two-level approximation captures the main fidelity trends and the coherent oscillatory structure.
Transitions involving additional states and residual differences between the tight-binding and continuum descriptions become relevant outside this regime.
We also find that the relative performance of different protocol orders depends nontrivially on the finite-size approximant.
The most effective protocol must therefore be selected according to the spectral geometry of the specific system rather than assumed to be universal.

Our work reinforces the view of quasiperiodic ultracold atomic gases as a promising platform for investigating controlled quantum state transfer.
We provide practical guidelines for choosing the protocol order and experimental parameters, together with simple analytical models that can support the interpretation and benchmarking of future experiments.

While the present work focuses on single-particle state transfer in a noninteracting ultracold atomic gas, the MCTDH-X framework provides a natural route toward interacting many-body extensions.
Quasiperiodic systems with various interactions constitute an interesting and active field of research~\cite{MaceSciPost2019, ChiaracanePRB2021, SandbergPRB2024, KobialkaPRB2024, WangPRB2024, MariusPRL2025, Mondal:2026, BonselQuantum2026}.
For instance, interactions are known to produce qualitatively new forms of topological transport, including fractionally quantized pumping~\cite{MariusPRL2025}.
Particularly relevant to the present setting, recent work on interacting bosons in a quasiperiodic Aubry--Andr\'e lattice has shown that the quantization of the pumped charge can already break down for weak interactions but can re-emerge in the hard-core limit~\cite{Mondal:2026}. Additionally, interactions may influence the localization properties of the quasiperiodic bound states~\cite{BonselQuantum2026}.
These results motivate investigating how interactions reshape the winding states, avoided crossings, and locally adiabatic protocols considered here.
Strong interactions may also enable the controlled transfer of bound, correlated, or entangled many-body states, substantially extending the single-particle mechanism studied in this work.

A second important direction is to determine the stability of the transfer protocols against realistic perturbations.
End-to-end transfer protocols in Fibonacci-type chains have already been shown to exhibit remarkable resilience to disorder~\cite{ghosh2025quantum, ghosh2026FCwaveguide}.
Moreover, quantized transport in a driven quasiperiodic chain was found to survive bounded local disorder even after the relevant instantaneous spectral gaps close~\cite{GottlobPRXQ2025}.
These results motivate a systematic investigation of the tangent, Roland-Cerf, and higher-order locally adiabatic protocols in the presence of static spatial disorder, temporal fluctuations of the phason angle and lattice depths, imperfect state preparation, finite temperature, and dissipation channels.

More broadly, quasiperiodicity itself can qualitatively reshape topological transport.
It has been shown experimentally to compete with, and under suitable driving cycles even induce, Thouless pumping in ultracold atomic systems~\cite{NakajimaNP2021}.
Quasiperiodic modulations have likewise been predicted to generate quantized pumping in parameter regimes that are otherwise topologically trivial~\cite{PadhanMishraPRB2024}.
Complementary photonic experiments have demonstrated topological transport in continuum incommensurate potentials and clarified its connection to successive rational approximants~\cite{YangPNAS2024, PengELight2025}.
These developments suggest that the interplay between quasiperiodicity, finite-size spectral structure, and protocol geometry may provide additional routes for controlling and optimizing state transfer beyond the specific setup considered here.

\begin{acknowledgments}
P.M. acknowledges support by the Swedish Research Council (Grant 2024-05213) and by the Olle Engkvist Foundation.
We thank Adonis Haxhijaj for useful discussions.
Computation time on the Sunrise Compute Cluster of Stockholm University is gratefully acknowledged.
The simulation data are available upon reasonable request to the corresponding author.
\end{acknowledgments}

\appendix
\section{Finite-size Wannier function construction and tight-binding limit}
\label{app:wannier}

In order to identify edge-localized states, as discussed in Sec.~\ref{sec:model} and Sec.~\ref{subsec:Wannier}, we employ the tight-binding approximation, which maps the continuum quasiperiodic potential onto an effective lattice model using localized Wannier functions $w_i$. 
For the finite-size systems considered here, this procedure yields an Aubry--André--Harper (AAH) model~\cite{Harper:1955,Aubry:1980,Modugno:2009,Kraus:2012,Fangli:2015,Liu:2017,Liu:2021}. 
In this Appendix, we briefly outline the construction of the Wannier basis and the resulting lattice Hamiltonian following the procedure presented in the Supplemental Material of Ref.~\cite{Dutta:2022}. 

As our potential assumes hard-wall boundary conditions, it is natural to expand the Hilbert space in the basis of particle-in-a-box eigenfunctions, i.e., a sine basis,
\begin{equation}
    \alpha_m(x)=\sqrt{\frac{2}{L}}\sin\left(m\pi \frac{x}{L}\right), \quad m=1,\dots, M_{\mathrm{max}}
\end{equation}
where $M_{\mathrm{max}}$ indicates a sufficiently large but finite truncation.
We have verified that $M_{\mathrm{max}}=800$ is sufficient to obtain converged results.
In this basis, the kinetic energy operator becomes diagonal:
\begin{equation}
    \bra{\alpha_m}\hat{K}\ket{\alpha_n}=\delta_{mn}E_r\left(\frac{m}{N_w}\right)^2,
\end{equation}
while the potential energy from the primary lattice becomes:
\begin{equation}
    V_{mn}=\bra{\alpha_m} V_p(\hat{x})\ket{\alpha_n}.
\end{equation}

The eigenvectors of the $M_{\mathrm{max}}\times M_{\mathrm{max}}$
Hamiltonian matrix correspond to the discrete eigenstates of the
finite primary-lattice potential. 
Although the hard-wall boundaries break translational invariance, in the deep-lattice limit these eigenstates can be grouped into manifolds that continuously connect to the Bloch bands of the corresponding infinite periodic lattice.
We select the $N_w$ lowest-energy eigenstates, which form the finite-system counterpart of the lowest Bloch band, and define the projector onto this subspace as $\hat{\Pi}$. 
This subspace provides the basis from which the localized Wannier functions are constructed.

\begin{figure}[t!]
    \centering
\includegraphics[width=1.0\linewidth]{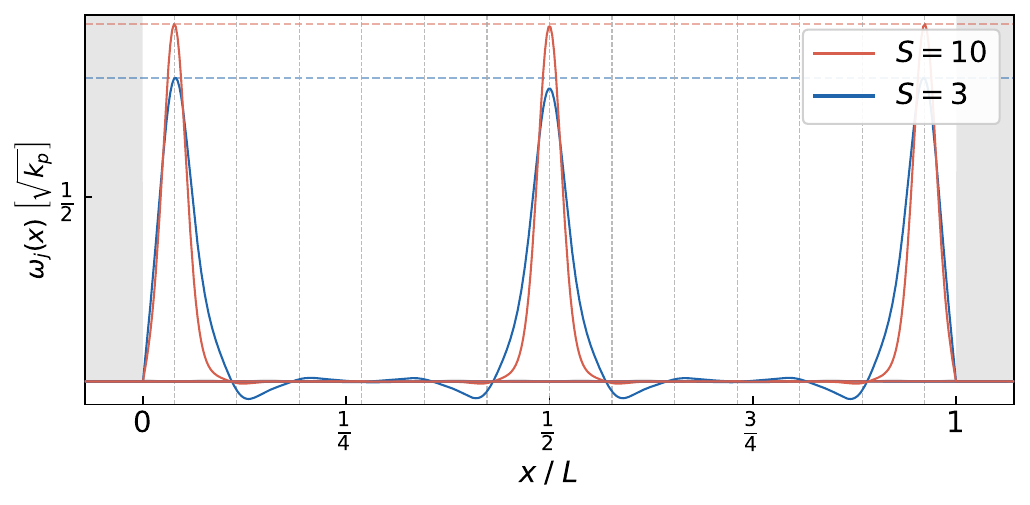}
    \caption{\textbf{Wannier function localization in a finite-size system}. We show examples of Wannier functions in the bulk and at the boundaries of the system, for a potential depth of $S=3$ (blue) and $S=10$ (red) recoil energies, respectively. }
    \label{fig:Wannierfunctions}
\end{figure}

\begin{figure}[t!]
    \centering
    \includegraphics[width=1.0\linewidth]{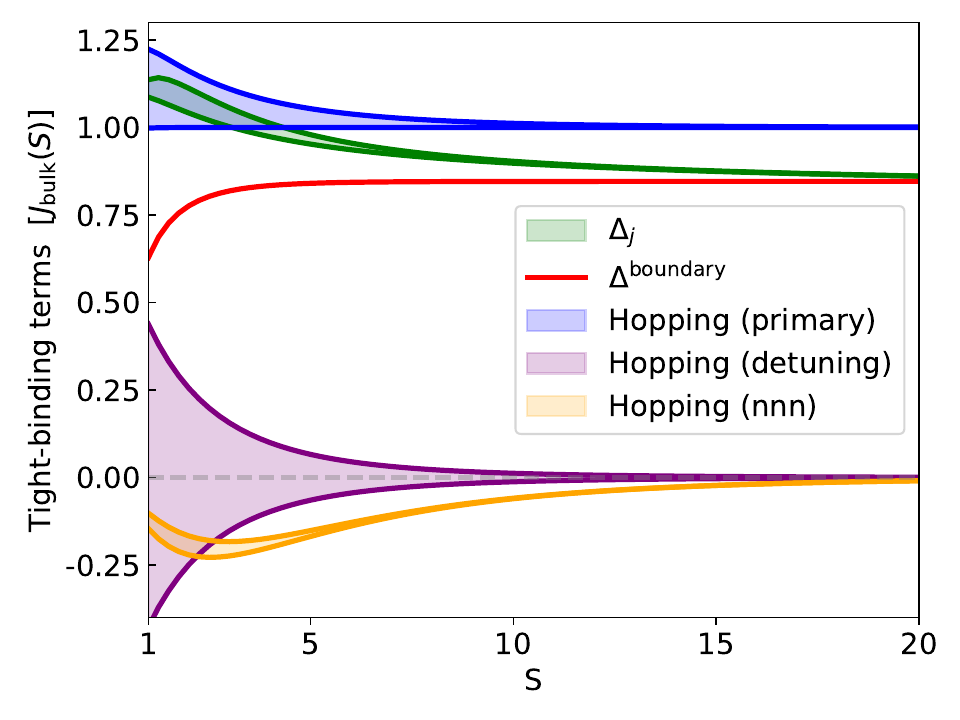}
    \caption{\textbf{Tight-binding parameters versus primary lattice depth.}
    Tight-binding parameters entering Eq.~\eqref{eq:tight_binding_lattice_general} as functions of the primary lattice depth $S$ expressed in units of the recoil energy. 
    The detuning strength $\rho$ is chosen such that the shape parameter $s(S,\rho)$ [Eq.~\eqref{eq:shape_factor}] remains constant.
    }
    \label{fig:Lattice_model_terms}
\end{figure}

The next step follows from the Marzari--Vanderbilt construction~\cite{Marzari:1997,Marzari:2012}, according to which Wannier functions are obtained by minimizing their spatial spread. 
In one dimension, this is equivalent to diagonalizing the projected position operator $\hat{X}_P$ 
\begin{equation}
    \hat{X}_P=\hat{\Pi}^\dagger \hat{X}\hat{\Pi}.
\end{equation}
Let $U\in\mathbb{C}^{N_w\times N_w}$ be the unitary matrix that diagonalizes this operator. 
The resulting Wannier functions are then given by
\begin{equation}
w_i(x)=\sum_{j=1}^{N_w}\sum_{m=1}^{M_{\mathrm{max}}}
U_{ij}\,\Pi_{jm}\,
\left<x\middle|\alpha_m\right>,
\end{equation}
where $\Pi_{jm}$ denotes the expansion coefficient of the $j$th projected eigenstate in the sine basis.
Figure~\ref{fig:Wannierfunctions} visualizes the profile of a few Wannier functions obtained with our procedure, with the vertical dashed line indicating the position of the minima of the potential. 
The Wannier functions have peaks around the potential minima. 
For a deeper lattice ($S=10$, where $S=V_p/E_r$ denotes the number of recoil energies), the Wannier functions (red) are more localized around the potential minima, while for a shallower lattice ($S=3$), the Wannier functions (blue) are more spread out in space, such that next-nearest-neighbor overlaps might become important.
Furthermore, the horizontal dashed lines demonstrate that the Wannier functions at the boundaries are higher than in the bulk, although this height difference reduces for larger system sizes but remains of the order of the hopping terms (see Fig.~\ref{fig:Lattice_model_terms}).

In the tight-binding approximation, the Wannier functions form an
orthonormal basis for the low-energy single-particle Hilbert space.
The field operator can therefore be expanded as
\begin{equation}
\hat{\Psi}(x)
=
\sum_{j=1}^{N_w}w_j(x)\hat{c}_j,
\end{equation}
where $\hat{c}_j^\dagger$ ($\hat{c}_j$) creates
(annihilates) a particle in the Wannier state localized around the
$j$th lattice site.

We denote the full single-particle continuum Hamiltonian by
\begin{align}
\hat{H}(\varphi) &= \hat{H}_0+V_d(\hat{x};\varphi), \nonumber \\
\hat{H}_0 &= -\frac{\hbar^2}{2m}\frac{\mathrm{d}^2}{\mathrm{d}x^2}
+V_p(\hat{x}).
\end{align}
Projecting $\hat{H}(\varphi)$ onto the lowest-band Wannier basis and
neglecting next-nearest-neighbor and longer-range hopping terms gives
\begin{align}
\hat{H}_{\mathrm{tb}}(\varphi) &= \sum_{j=1}^{N_w}
\delta_j(\varphi)\hat{c}_j^\dagger\hat{c}_j \nonumber \\
&- \sum_{j=1}^{N_w-1}
J_j(\varphi) \left(
\hat{c}_{j+1}^\dagger\hat{c}_j+\mathrm{H.c.}
\right).
\label{eq:tight_binding_lattice_general}
\end{align}
The tight-binding coefficients are therefore defined as
\begin{align}
\delta_j(\varphi) &= \int \mathrm{d}x\,
w_j^*(x) \left[ \hat{H}_0+V_d(x;\varphi) \right] w_j(x), \\
J_j(\varphi) &= -\int \mathrm{d}x\, w_{j+1}^*(x) \left[ \hat{H}_0+V_d(x;\varphi) \right] w_j(x).
\end{align}
It is useful to separate the contributions originating from the
primary and detuning lattices as
\begin{align}
\delta_j(\varphi) &= \delta_j^{(0)}+\delta_j^{(d)}(\varphi),
\\
J_j(\varphi) &= J_j^{(0)}+J_j^{(d)}(\varphi),
\end{align}
where
\begin{align}
\delta_j^{(0)} &=
\int \mathrm{d}x\, w_j^*(x)\hat{H}_0w_j(x), \\
\delta_j^{(d)}(\varphi) &= \int \mathrm{d}x\, w_j^*(x)V_d(x;\varphi)w_j(x), \\
J_j^{(0)} &= -\int \mathrm{d}x\, w_{j+1}^*(x)\hat{H}_0w_j(x), \\
J_j^{(d)}(\varphi) &= -\int \mathrm{d}x\, w_{j+1}^*(x)V_d(x;\varphi)w_j(x).
\label{eq:tb_contributions}
\end{align}

The conventional AAH model follows after making additional bulk and deep-lattice approximations. 
It assumes that, sufficiently far from the boundaries, the Wannier functions are related by translations, $w_j(x)\simeq w(x-x_j)$, where the coordinates are defined as in the main text as $x_j=a\left(j-\frac12\right)$ with $a=\frac{L}{N_w}=\frac{\pi}{k_p}$.
Furthermore, it is assumed that $w(x)$ is real and reflection symmetric. 
Under these assumptions, the primary-lattice contribution to the onsite energy is site independent and can be removed by an overall energy shift, while the primary-lattice hopping becomes approximately uniform
\begin{equation}
\delta_j^{(0)}\simeq0, \qquad J_j^{(0)}\simeq J.
\end{equation}
The detuning-lattice contribution to the onsite energy then becomes
\begin{equation}
\delta_j^{(d)}(\varphi) \simeq \Delta \cos\left[ 2\pi\beta\left(j-\frac12\right)+\varphi \right],
\label{eq:onsite_bulk_modulation}
\end{equation}
with
\begin{equation}
\Delta = \frac{V_d}{2} \int_{-\infty}^{\infty}\mathrm{d}y\, |w(y)|^2\cos(2k_dy).
\label{eq:Delta_overlap}
\end{equation}
The detuning lattice also produces an off-diagonal quasiperiodic correction~\cite{Boers:2007, Dominguez-Castro:2009, Modugno:2009, Biddle:2009},
\begin{equation}
J_j^{(d)}(\varphi) \simeq J' \cos\left(2\pi\beta j+\varphi\right),
\label{eq:detuning_hopping}
\end{equation}
where
\begin{align}
J'=
-\frac{V_d}{2} \int_{-\infty}^{\infty}\mathrm{d}y\,
w^*\!\!\left(y-\frac{a}{2}\right)\! w\!\left(y+\frac{a}{2}\right) \cos(2k_dy).
\label{eq:Jprime}
\end{align}
The different phase offsets in Eqs.~\eqref{eq:onsite_bulk_modulation} and
\eqref{eq:detuning_hopping} arise because the onsite term is centered
at the well position $x_j=a(j-1/2)$, whereas the hopping term is
centered on the bond between sites $j$ and $j+1$, at position $ja$.

For a deep primary lattice and weak detuning lattice, the overlap entering $J'$ is small and one may neglect the off-diagonal quasiperiodicity, $J'\simeq0$. 
Neglecting boundary corrections as well then gives the conventional AAH Hamiltonian reported in the main text,
\begin{align}
\hat{H}_{\mathrm{AAH}} &= -J(S) \sum_{j=1}^{N_w-1}
\left( \hat{c}_{j+1}^\dagger\hat{c}_j+\mathrm{H.c.} \right) \nonumber\\
&\quad+ \Delta(S,\rho)\! \sum_{j=1}^{N_w} \cos\left[ 2\pi\beta\!\left(j\!-\!\frac12\right)\!+\!\varphi \right]\! \hat{c}_j^\dagger\hat{c}_j.
\label{eq:AAHgeneric}
\end{align}

Within this conventional bulk AAH approximation, closed-form expressions for $J(S)$ and $\Delta(S,\rho)$ can be obtained
in the deep-lattice limit~\cite{Xiao:2017,SinghPRA2015,Bloch:2008}
\begin{align}
    J(S)&=\frac{4}{\sqrt{\pi}}E_r S^{(3/4)}e^{-2\sqrt{S}}, \\
    \Delta(S,\rho)&=\frac{\rho S E_r}{2}e^{-\frac{\beta^2}{\sqrt{S}}}.
    \label{eq:AAHterms}
\end{align}
In brief, the expression for $\Delta$ follows from a harmonic approximation of the primary lattice around its minima, which yields Gaussian Wannier functions and an analytical result in the deep-lattice limit. 
The derivation of $J$ is more involved and exploits properties of the Mathieu equation~\cite{Mathieu}.

After measuring energies in units of $J(S)$, the dimensionless lattice Hamiltonian $h_{\mathrm{AAH}} = H_{\mathrm{AAH}} / J(S)$ becomes
\begin{align}
h_{\mathrm{AAH}} &= -\sum_{j=1}^{N_w-1}c_{j+1}^\dagger c_j+\mathrm{H.c.} \nonumber \\
&+ s\sum_{j=1}^{N_w}\cos\left(2\pi\beta j - \pi \beta +\varphi\right)c_j^\dagger c_j,
\label{eq:DimensionlessHam}
\end{align}
where the only remaining independent parameter is the dimensionless \emph{shape factor}
\begin{equation}
    s(S,\rho)\equiv\frac{\Delta(S,\rho)}{J(S)}
    =\frac{\sqrt{\pi}\rho S^{1/4}}{8}
    e^{2\sqrt{S}-\frac{\beta^2}{\sqrt{S}}}.
    \label{eq:shape_factor}
\end{equation}
This quantity determines the structure of the eigenstates and the phason spectrum and is used in the study of the generic AAH model to detect phase transitions~\cite{Dominguez-Castro:2009,Modugno:2009}. 
By increasing (decreasing) $S$ and simultaneously adjusting $\rho$ preserving the fraction $\frac{\Delta(S,\rho)}{J(S)}$, we only expect the energy scale to decrease (increase) while the shape of the spectrum and the eigenstates remain constant. 
We make use of this and explain it in more detail in Appendix~\ref{app:ChangingS}.

Equations~\eqref{eq:AAHgeneric}--\eqref{eq:shape_factor} describe the ideal bulk AAH limit. 
For the finite hard-wall system, the exact tight-binding coefficients are instead obtained from the overlap integrals in Eq.~\eqref{eq:tb_contributions}.
The onsite energies may be written approximately as
\begin{equation}
\delta_j(\varphi) \simeq \Delta_j^{\mathrm{boundary}}
+\Delta_j\cos\left[2\pi\beta\left(j-\frac12\right)+\varphi\right],
\label{eq:finite_onsite}
\end{equation}
where
\begin{equation}
\Delta_j^{\mathrm{boundary}} = \int \mathrm{d}x\,w_j^*(x)\hat{H}_0w_j(x) - \delta_{\mathrm{bulk}},
\end{equation}
and $\delta_{\mathrm{bulk}}$ denotes the corresponding onsite energy in the bulk. 
Assuming that the Wannier density is approximately reflection symmetric around the well center $x_j$, the modulation amplitude is
\begin{equation}
\Delta_j = \frac{V_d}{2}\int \mathrm{d}x\,|w_j(x)|^2
\cos\left[2k_d(x-x_j)\right].
\end{equation}
The site dependence of $\Delta_j$ and $\Delta_j^{\mathrm{boundary}}$ accounts for finite-size effects.
In the bulk and deep-lattice limit,
\begin{equation}
\Delta_j\longrightarrow\Delta, \qquad \Delta_j^{\mathrm{boundary}}\longrightarrow0,
\end{equation}
although the boundary correction remains of order \(J(S)\), i.e. it vanishes asymptotically on the same scale as \(J(S)\).
Similarly, the primary-lattice hopping $J_j^{(0)}$ may vary close to the hard-wall boundaries but approaches $J$ in the bulk.
Finite-size distortions of the Wannier functions also cause $J_j^{(d)}(\varphi)$ to deviate from the uniform-amplitude cosine in Eq.~\eqref{eq:detuning_hopping}, particularly for the outermost bonds.
The bulk expression is recovered away from the boundaries.
In our numerics, however, the entire detuning-induced hopping correction becomes negligible since $|J'|\ll J$. 
Figure~\ref{fig:Lattice_model_terms} illustrates how the exact finite-size coefficients approach the conventional AAH limit as the primary lattice is made deeper.

\begin{figure*}[t!]
    \centering
    \includegraphics[width=1.0\linewidth]{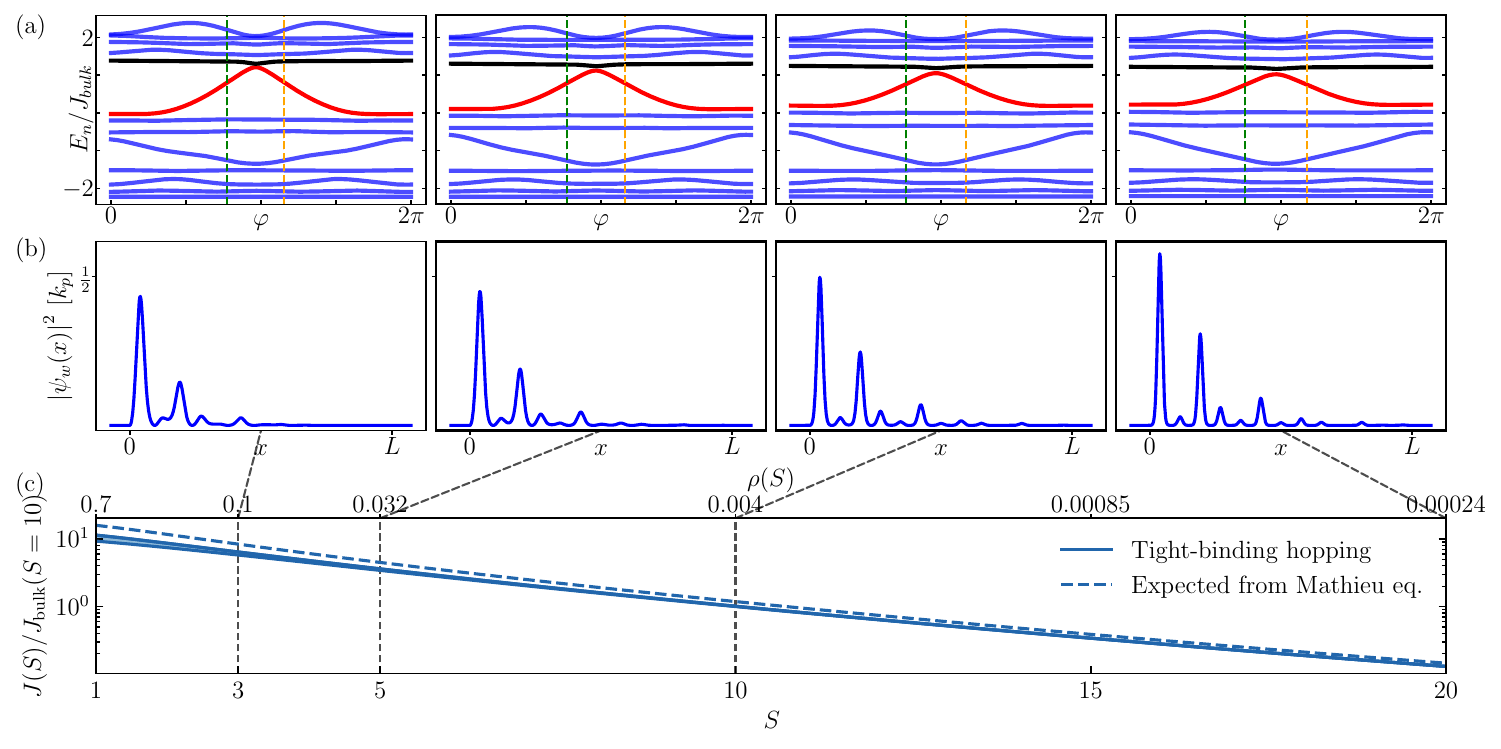}
    \vspace{-0.7cm}
    \caption{
    \textbf{Spectral invariance of the bulk AAH model at fixed shape factor.}
    The dimensionless energy spectrum $E_n(\varphi)/J_{\mathrm{bulk}}$ (a) and the localized tight-binding eigenstate (b) are plotted for different values of the potential depth $S$ expressed in units of the recoil energy.
    The selected values of $S$ are indicated by dashed lines from panel (c), where we additionally plot the hopping term (which gives the energy scale) in the range $S \in [1,20]$.
    For every value of $S$, we have chosen $\rho(S)$ such that the shape factor $s(S,\rho)$ stays constant. 
    The dashed line corresponds to the result from Eq.~\eqref{eq:AAHterms} that is derived from the Mathieu equation.}
    \label{fig:invarianceunderS}
\end{figure*}

\section{Invariance of the phason spectrum}
\label{app:ChangingS}
In Appendix~\ref{app:wannier}, we derived how the conventional bulk AAH Hamiltonian depends on a single shape factor $s(S,\rho)=\Delta(S,\rho)/J(S)$  when expressed in units of the hopping amplitude $J(S)$.
In this Appendix, we demonstrate that the overall shape of the phason spectrum remains unchanged when the depth of the primary lattice $S$ is varied, provided that the detuning strength $\rho$ is adjusted accordingly, thereby supplementing the discussions in Sec.~\ref{sec:model}.
This finding thus generalizes the results we presented in Sec.~\ref{sec:results} for $S=10$ to a much broader family of system parameters.

We remark that for the finite continuum-derived model, this invariance is only approximate.
Finite-size boundary corrections, spatial variations of the hopping and onsite terms, and corrections beyond the nearest-neighbor tight-binding approximation may retain an additional dependence on $S$. 
These corrections become progressively smaller as the primary lattice is made deeper.

Along a contour in $S$--$\rho$ space where $s(S,\rho)$ remains constant, the dimensionless bulk Hamiltonian $h_{\mathrm{AAH}}(\varphi)$ in Eq.~\eqref{eq:DimensionlessHam} remains unchanged. 
Consequently, its eigenvectors and dimensionless eigenvalues are invariant, while the physical energy scale changes proportionally to $J(S)$.
The contours that leave $s$ invariant are given by
\begin{equation}
       \rho(S')=\rho(S)\cdot\left(\frac{S}{S'}\right)^{1/4}e^{-2\left(\sqrt{S'}-\sqrt{S}\right)+\beta^2\left(\frac{1}{\sqrt{S'}}-\frac{1}{\sqrt{S}}\right)},
\end{equation}
meaning that the \textit{shape} of the energy spectrum and eigenstates remain unchanged on these contours as long as $S$ and $S'$ lie within the tight-binding regime.
This theoretical result is validated in Fig.~\ref{fig:invarianceunderS}, where we vary $S$ between 1 and 20 while keeping $s$ constant, and plot the corresponding phason spectrum and winding state.
We find that the dimensionless phason spectrum and the selected winding state are already nearly invariant for $S=5$, with the agreement improving further as the primary lattice is made deeper, while the physical energy scale decreases according to $J(S)$.
For smaller values of $S$, the Wannier functions become less localized around the potential minima, so next-nearest-neighbor and longer-range hopping terms become increasingly important and the nearest-neighbor tight-binding approximation becomes less accurate.

\begin{figure*}
 \begin{minipage}{0.49\textwidth}
        \centering
        \includegraphics[width=\linewidth]{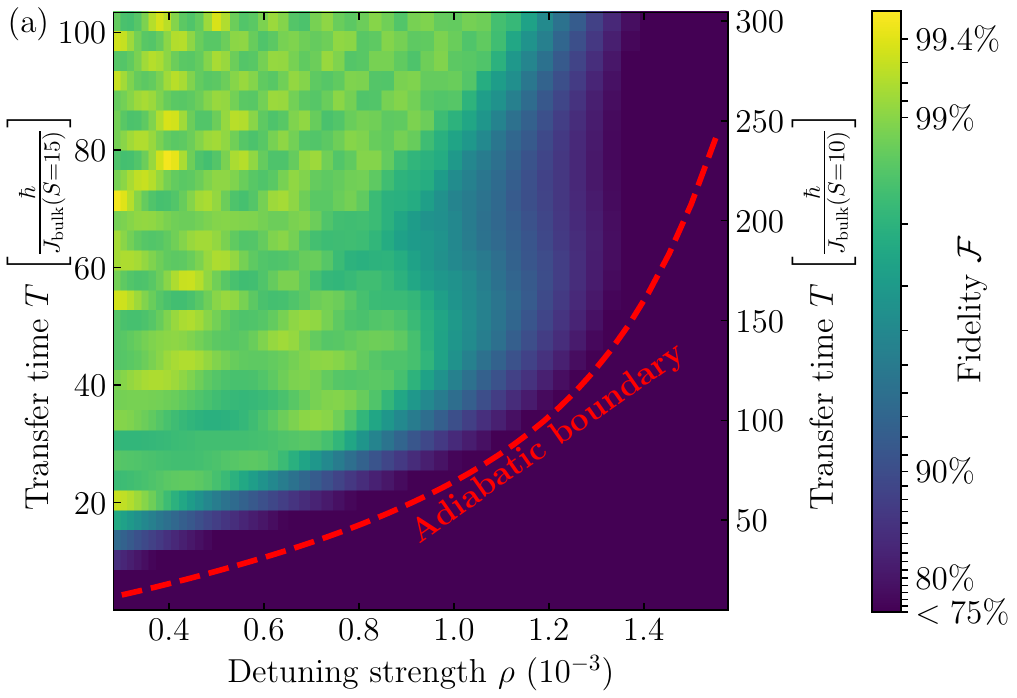}
    \end{minipage}
	\centering
     \begin{minipage}{0.49\textwidth}
        \centering
        \includegraphics[width=\linewidth]{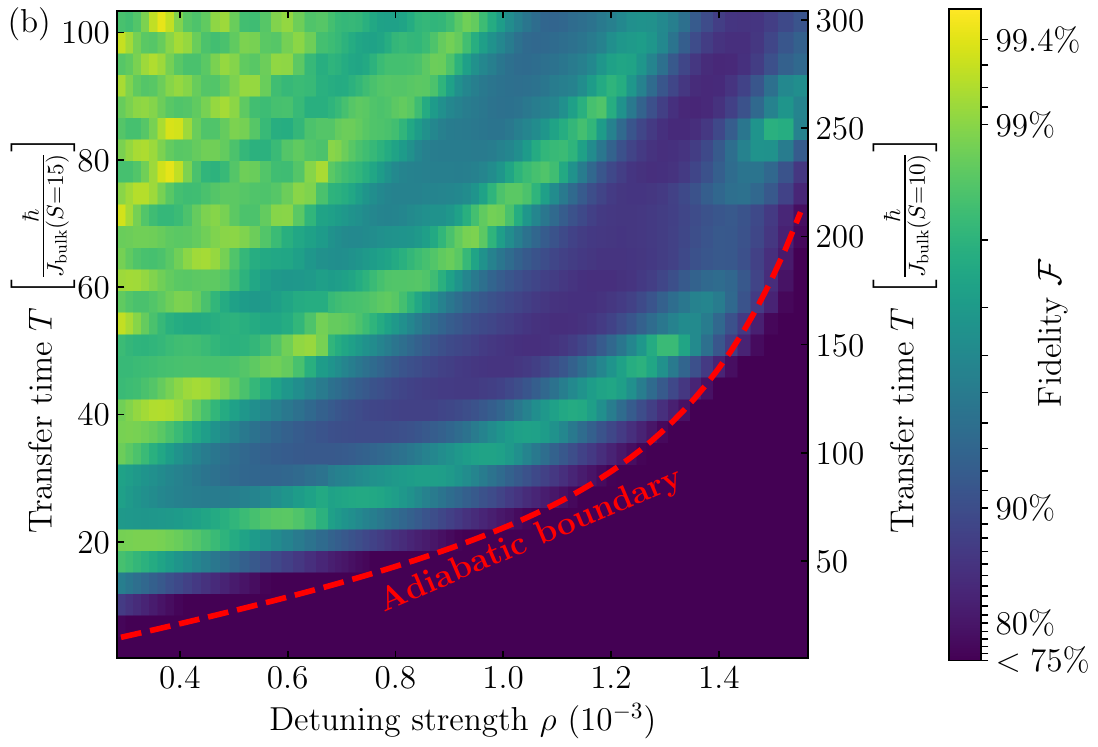}
    \end{minipage}
     \caption{
     \textbf{State-transfer fidelity for a deeper lattice.} 
     The fidelity as a function of the detuning strength $\rho$ and transfer time $T$ for (a) the tangent protocol and (b) the Roland-Cerf protocol, obtained for a primary lattice depth of $S=15$ recoil energies. 
     By scaling the transfer time up by a factor of $J(10)/J(15)\approx 3.05$ (right $y$-axis) and the detuning strength down by the factor $\rho(15)/\rho(10)\simeq 0.21$, see Eq.~\eqref{eq:scaleinvariance}, the fidelity is expected to stay unchanged as compared with the main-text Fig.~\ref{fig:PhaseDiagram} obtained with $S=10$. 
     Discrepancies arise from imperfect application of the tight-binding approximation and numerical truncation in the MCTDH-X simulation (for instance, finite size grids). 
     For example, the color bar reaches up to $99.4\%$ instead of $99.6\%$ as in Fig.~\ref{fig:PhaseDiagram}. 
     We attribute this to the fact that for $S=15$ the Wannier functions become \textit{narrower} and thus there are fewer grid points in MCTDH-X covering the Wannier peak.
     Notice also that the unit of time we used throughout the paper ($\hbar$ divided by the hopping term) depends on the lattice depth $S$ and thus we show the transfer time in two different but properly rescaled units.}
     \label{fig:PS-S=15}
\end{figure*}

This result has important implications for the state-transfer fidelity $\mathcal{F}$. 
Throughout this work, we set $S=10$, a value for which the system is already well within the tight-binding regime and which is commonly used in experimental realizations~\cite{LohseNP2016, Deissler2011}. 
Consider now a change of lattice depth $S\to S'$, where both $S$ and $S'$ lie within the tight-binding regime. 
As shown in Appendix~\ref{app:two-level}, in the high-fidelity regime $\mathcal{F}$ depends only on the product $TG$, where $T$ is the transfer time and $G$ is the relevant energy gap. 
Since $G$ has dimensions of energy, it scales proportionally to the hopping amplitude $J(S)$.
Consequently, the fidelity at the end of the protocol can be preserved under a change of lattice depth by rescaling the protocol duration according to $T\propto 1/J(S)$.
As a result, in the tight-binding limit, the state-transfer fidelity is invariant under the transformation
\begin{align}
S& \longrightarrow S' \nonumber \\
\rho&\longrightarrow \rho'=\rho\times\left(\frac{S}{S'}\right)^{1/4}e^{-2\left(\sqrt{S'} - \sqrt{S}\right) + \beta^2\left(\frac{1}{\sqrt{S'}}-\frac{1}{\sqrt{S}}\right)} \nonumber \\
T& \longrightarrow T'=T\frac{J(S)}{J(S')}
\label{eq:scaleinvariance}
\end{align}
This scaling is demonstrated in Fig.~\ref{fig:PS-S=15}, where we increase the primary-lattice depth from $S=10$ to $S=15$ and adjust $\rho$ and $T$ according to Eq.~\eqref{eq:scaleinvariance}.

\section{Adiabatic condition and two-level approximation}
\label{app:two-level}
In this Appendix, we provide additional explanations for two important concepts used extensively throughout the main text, in particular the development of the analytic approach in Sec.~\ref{subsec:two-level-approx} and the discussion in Appendix~\ref{app:ChangingS}.
The first concept is how to measure non-adiabatic transitions to derive optimized pump protocols.
To this end, we use the following quantity
\begin{equation}
    \lambda^{(N)}(\varphi)
\equiv
\sum_{n\neq w}
\left|
\frac{
\mel{\psi_n(\varphi)}{\partial_\varphi \hat{H}}{\psi_w(\varphi)}
}{
\left(E_n(\varphi)-E_w(\varphi)\right)^{N+1}
}
\right|.
\label{lambdaN}
\end{equation}
The second concept is the two-level approximation
\begin{align}
\ket{\Psi(t)} &\approx c_w(t) e^{-\frac{i}{\hbar} \int_0^t E_w[\varphi(t')]\,\mathrm{d}t'} \ket{\psi_w[\varphi(t)]} \nonumber\\
& + c_{w+1}(t) e^{-\frac{i}{\hbar} \int_0^t E_{w+1}[\varphi(t')]\,\mathrm{d}t'} \ket{\psi_{w+1}[\varphi(t)]}
\label{eq:two_level_ansatz2}
\end{align}
which allowed us to approximate the fidelity over a certain range of $(\rho,T)$. 

During our transfer protocol, a state $\ket{\Psi(t)}$ evolves according to the Schrödinger equation
\begin{equation}
    i\hbar \partial_t \ket{\Psi(t)} = H[\varphi(t)]\ket{\Psi(t)},
    \label{eq:Schroedinger-full-appD}
\end{equation}
with the time-dependent Hamiltonian $H[\varphi(t)]$.
In general, a state initialized in a given eigenstate does not remain on the same energy band during the evolution due to non-adiabatic transitions. 
It is therefore convenient to expand the time-evolved state in the instantaneous eigenbasis defined by
\begin{equation}
    E_n[\varphi(t)]\ket{\psi_n[\varphi(t)]}
    =
    H[\varphi(t)]\ket{\psi_n[\varphi(t)]}.
    \label{eq:Schroedinger-appD}
\end{equation}
This leads to the decomposition
\begin{equation}
    \ket{\Psi(t)}=\sum_nc_n(t) e^{i\xi_n(t)}\ket{\psi_n[\varphi(t)]}.
    \label{eq:instexpansion}
\end{equation}
Here, the expansion coefficients $c_n$ fulfill the initial condition $c_n(t=0)=\delta_{n,w}$ when the system is prepared in a winding state $\ket{\psi_w}$.

By inserting this ansatz into the time-dependent Schrödinger equation~\eqref{eq:Schroedinger-full-appD}, one can obtain the equation of motion for $c_n$. 
The phase $\xi_n$ can be chosen to simplify the resulting differential equations. 
By setting:
\begin{equation}
    \xi_n(t)=-\frac{1}{\hbar}\int_{0}^tE_n[\varphi(t')]\mathrm{d}t'+\int_{\varphi(0)}^{\varphi(t)}\gamma_n(\varphi)\mathrm{d}\varphi
\end{equation}
with the last term being an integral over the Berry connection $\gamma_n=i\bra{\psi_n}\partial_{\varphi}\ket{\psi_n}$~\cite{WilczekBerryPhase1989}, the diagonal terms in the derivation of this differential equation will vanish, and eventually, one obtains:
\begin{equation}
    i\frac{\partial c_n}{\partial t}=i\dot{\varphi}\sum_{m\neq n}c_me^{i(\xi_m-\xi_n)}\frac{\bra{\psi_n}\partial_\varphi H\ket{\psi_m}}{E_n-E_m},
    \label{eq:Diffull}
\end{equation}
where we have now dropped the arguments for ease of notation. 
Non-adiabatic excitations between the bands are avoided by demanding that the derivative of the initial amplitude $c_w$ stays small during time evolution. 
This can be achieved by imposing: 
\begin{equation}
    \left|\frac{\partial c_w}{\partial t}\right|\leq \dot{\varphi} \sum_{m\neq w}\left|\frac{\bra{\psi_w}\partial_\varphi H\ket{\psi_m}}{E_w-E_m}\right|\overset{!}{=}\frac{G}{\hbar C}
    \label{eq:protocol_dif}
\end{equation}
for some dimensionless number $C\gg1$. 
The inequality was obtained using the normalization of the coefficients and the monotonicity of the protocol $\dot{\varphi} \ge 0$.
The factor $G$, which was already defined in Eq.~\eqref{eq:gap} and has units of energy, was chosen such that the dimensions on both sides of the equation match. 
Using the definition $\lambda^{(0)} \equiv \sum_{m\neq w}\left|\frac{\bra{\psi_w}\partial_\varphi H\ket{\psi_m}}{E_w-E_m}\right|$ [already shown in the main text in Eq.~\eqref{eq:lambda0}], Eq.~\eqref{eq:protocol_dif} can be solved by separation of variables.
In particular, for a transfer time $T$, we have
\begin{equation}
    \hbar C\int_{\varphi(0)}^{\varphi(T)}\lambda^{(0)}(\varphi)\mathrm{d}\varphi=TG ,
\end{equation}
and because adiabaticity requires $C\gg 1$ it follows that:
\begin{equation}
    TG\gg \hbar \int_{\varphi(0)}^{\varphi(T)}\lambda^{(0)}(\varphi)\mathrm{d}\varphi\equiv \hbar I^{(0)},
\end{equation}
where we have also defined the integrated adiabatic weight $I^{(0)}$.
Higher protocol orders ($N>0$) arise from the observation that the oscillation of the phase in Eq.~\eqref{eq:Diffull}, which depends on the relative energies and becomes large for large $T$, governs the dynamics of the system and consequently the instantaneous unit of time. 
We can thus amplify its influence on the protocol by considering the more generalized condition
\begin{equation}
    \dot{\varphi}
    \sum_{m\neq w}
    \left|
    \frac{\bra{\psi_w}\partial_\varphi H\ket{\psi_m}}
    {(E_w-E_m)^{N+1}}
    \right|
    =
    \frac{G^{1-N}}{\hbar C}.
    \label{eq:protocol_dif2}
\end{equation}
Increasing $N$ in the denominator progressively concentrates the evolution around the avoided crossing.

The same observation also motivates a two-level approximation. 
Since the couplings in Eq.~\eqref{eq:Diffull} are weighted by inverse energy differences, the dominant contribution comes from the band $E_{w+1}(\varphi)$ closest to the winding state $E_w(\varphi)$. 
Neglecting all other bands in the interval $[\varphi_1,\varphi_2]$ yields the effective two-level description:
\begin{align} 
        i\frac{\partial}{\partial t} &\begin{pmatrix} c_w \\ c_{w+1} \end{pmatrix} 
        = {}  \dot{\varphi} \frac{\bra{\psi_{w+1}}\partial_\varphi H\ket{\psi_{w}}}{E_{w+1}-E_{w}} \nonumber\\
        & \times \begin{pmatrix} 0 & ie^{i(\xi_w-\xi_{w+1})} \\ -ie^{-i(\xi_w-\xi_{w+1})} & 0 \end{pmatrix} 
        \begin{pmatrix} c_w \\ c_{w+1} \end{pmatrix}
        \label{eq:2levelmatrixequation} 
\end{align}
We can simplify this two-level equation by fixing a gauge for the instantaneous eigenstates. 
This is justified by the observation that, due to the appearance of $\gamma_n(\varphi)$ in the phase $\xi_n$, the ansatz in Eq.~\eqref{eq:instexpansion} is invariant under a gauge transformation $\ket{\psi_n(\varphi)}\mapsto e^{i\chi_n(\varphi)}\ket{\psi_n(\varphi)}$. 
In particular, because the Hamiltonian is real and symmetric, it is possible to fix a gauge such that the components of the tight-binding eigenvectors are real.
In this gauge $\gamma_n(\varphi)$ vanishes, which can be derived by a simple calculation.
The remaining sign freedom can then be used to ensure that the off-diagonal matrix element does not change sign along the protocol.
Within the two-level truncation, we define the corresponding adiabatic weight
\begin{equation}
\lambda_{\mathrm{2L}}^{(N)}(\varphi) = \frac{\left|
\bra{\psi_{w+1}(\varphi)} \partial_\varphi H
\ket{\psi_w(\varphi)} \right|}{\left[E_{w+1}(\varphi)-E_w(\varphi) \right]^{N+1} }.
\label{eq:lambdaN_2L}
\end{equation}
This quantity corresponds to retaining only the $w+1$ contribution in the full adiabatic weight $\lambda^{(N)}$.
In the parameter regime where the dynamics are dominated by these two states, $\lambda_{\mathrm{2L}}^{(N)}\simeq\lambda^{(N)}$.
The two-level equation of motion then becomes
\begin{align}
i\frac{\partial}{\partial t} 
\begin{pmatrix} c_w \\ c_{w+1} \end{pmatrix}
&= \dot{\varphi}\, \lambda_{\mathrm{2L}}^{(N)} \Delta E^N \nonumber\\ &\quad\times
\begin{pmatrix}
0 & i e^{\frac{i}{\hbar}\int_0^t\Delta E\,\mathrm{d}t'} \\
-i e^{-\frac{i}{\hbar}\int_0^t\Delta E\,\mathrm{d}t'} & 0
\end{pmatrix}
\begin{pmatrix}
c_w \\ c_{w+1}
\end{pmatrix},
\label{eq:2levelmatrixequation2}
\end{align}
where $\Delta E=E_{w+1}-E_w$.
Since the protocol itself is constructed from the full weight $\lambda^{(N)}$ through Eq.~\eqref{eq:protocol_condition}, in the two-level regime we may use
\begin{equation}
\dot{\varphi}\lambda_{\mathrm{2L}}^{(N)} \simeq \dot{\varphi}\lambda^{(N)} = \frac{I^{(N)}}{T}.
\end{equation}

\begin{figure}
    \centering
    \includegraphics[width=1.0\linewidth]{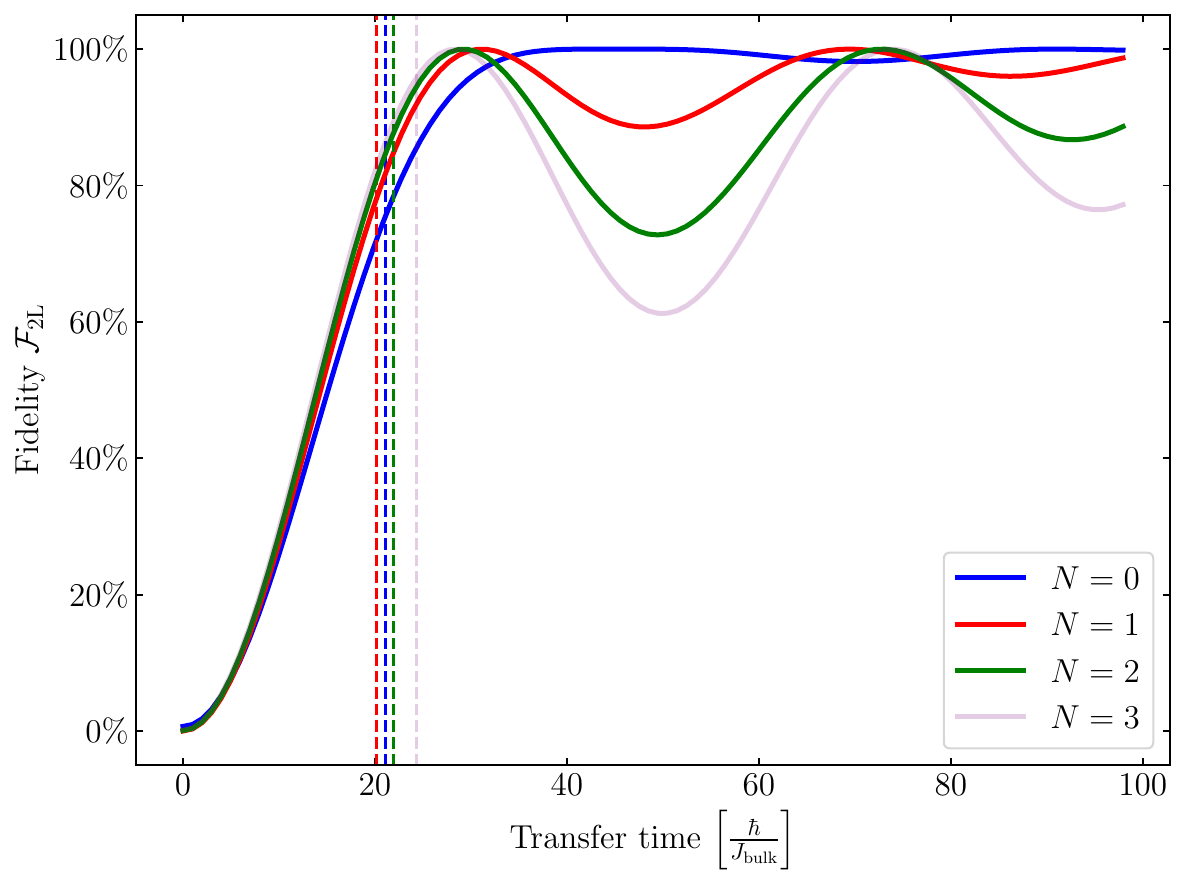}
    \caption{\textbf{Fidelity for higher protocol orders $\mathbf{N}$}: We plot the fidelity that results from the two-level assumption for the representative values $S=10$ and $\rho=0.0042$. 
    It can clearly be seen that the amplitude of the fidelity oscillations increases for $N=2$ and $N=3$. 
    The vertical dashed lines show the adiabatic boundaries for the corresponding values of $N$. }
    \label{fig:higher_N_oscillations}
\end{figure}

Introducing the dimensionless quantities $\Delta \epsilon=\Delta E/G$, $\tau_{\mathrm{A}}^{(N)}=G^N I^{(N)}$, and $\tau=t/T$, Eq.~\eqref{eq:2levelmatrixequation2} can be further rewritten as
\begin{align}
        i \frac{\partial}{\partial \tau } &\begin{pmatrix}
            c_w \\ c_{w+1}
        \end{pmatrix} = \, \tau_{\mathrm{A}}^{(N)}\Delta \epsilon^N\nonumber\\
        &\times \begin{pmatrix}
            0 & i e^{\frac{iTG}{\hbar} \int \Delta \epsilon \mathrm{d}\tau}\\
            -i e^{-\frac{iTG}{\hbar} \int \Delta \epsilon \mathrm{d}\tau} &0
        \end{pmatrix}
        \begin{pmatrix}
            c_w \\ c_{w+1}
        \end{pmatrix}.
    \label{eq:2levelequation3}
\end{align}
Within the two-level approximation, the dynamics therefore depend on the protocol duration $T$ and minimum gap $G$ only through the combination $TG$. We show examples of the numerical solution of this equation in Fig.~\ref{fig:higher_N_oscillations}.
Consequently, the fidelity is expected to be a function of this product alone.
For the special case $N=1$, Eq.~\eqref{eq:2levelequation3} can be simplified further by introducing
\begin{equation}
    f(\tau)=\int_0^{\tau}\Delta \epsilon(\tau')\,\mathrm{d}\tau'.
\end{equation}
This transforms the evolution equation into
\begin{align}
    i\frac{\partial}{\partial f}&\begin{pmatrix}
        c_w \\ c_{w+1}
    \end{pmatrix}= \tau_{\mathrm{A}}^{(1)}
    \nonumber\\
    &\times\begin{pmatrix}
        0 & ie^{\frac{iTGf}{\hbar}}\\
        -ie^{-\frac{iTGf}{\hbar}}& 0
    \end{pmatrix}
    \begin{pmatrix}
        c_w \\ c_{w+1}
    \end{pmatrix},
\end{align}
which can be solved analytically, yielding Eq.~\eqref{eq:two_level_fidelity}.

\section{MCTDH-X}
\label{app:MCTDHX}
In this Appendix, we provide additional details on the multiconfigurational time-dependent Hartree method for indistinguishable particles (MCTDH)~\cite{Streltsov:2006,Streltsov:2007, Alon:2007,Alon:2008} and on its implementation in MCTDH-X~\cite{Lode:2012, Lode:2016, Fasshauer:2016, Lode:2020, Lin:2020, Molignini:2025-SciPost, MCTDHX}, as discussed in Sec.~\ref{sec:methods:numerics}.
The method uses an adaptive ansatz and the time-dependent variational principle~\cite{TDVM81} to solve the many-body Schr\"odinger equation directly in the continuum for time-dependent Hamiltonians containing kinetic energy, external potentials, and, when present, interparticle interactions.
MCTDH-X has been very successful in studying a wide variety of ultracold atomic systems, such as noninteracting but driven systems~\cite{Xiang:2023}, short-range interacting gases~\cite{Beinke:2018,Roy:2018,Dutta:2019, Schaefer:2020,Lode:2021,Lode:2021-10,Debnath:2024,Roy:2023,Dutta:2023,Aloqali:2024,Dutta:2024,Haldar:2024,Chakrabarti:2024,Bhowmik:2025,Chakrabarti:2025-2,Roy:2025,Roy:2025-7,Dutta:2025,Roy:2025-4,Roy:2026}, long-range interacting atoms and molecules~\cite{Fischer:2015,Chatterjee:2018, Chatterjee:2019,Bera:2019,Bera:2019-symm,Chatterjee:2020,Roy:2022,Hughes:2023,Bilinskaya:2024,Molignini:2024-2,Roy:2024-annals,Roy:2024-epjp,Molignini:2025-1,Molignini:2025-2,Molignini:2025-3,Chakrabarti:2025,Molignini:2025-JPCM}, and even ultracold gases coupled to optical cavities~\cite{Lode:2017,Lode:2018,Molignini:2018,Lin:2019,Lin:2020-PRA,Lin:2021,Molignini:2022,Rosa-Medina:2022,Ortuno-Gonzalez:2025}.

In the present work, we use MCTDH-X as a continuum propagation framework for the quasiperiodic transfer protocol.
Although the simulations reported here are performed in the noninteracting single-particle limit, the same numerical infrastructure can be directly extended to interacting many-body pumping problems.

The MCTDH ansatz expands the many-body wavefunction in a time-dependent many-body basis,
\begin{equation}
\ket{\Psi(t)} = \sum_{\vec{n}} C_{\vec{n}}(t) \ket{\vec{n};t},
\label{eq:app_mctdh_ansatz}
\end{equation}
where the configurations are labeled by occupation vectors
\begin{equation}
\vec{n}=(n_1,\dots,n_M),
\qquad
\sum_{j=1}^{M} n_j=N_p.
\end{equation}
Here $N_p$ is the particle number and $M$ is the number of time-dependent single-particle orbitals.
For bosons, the configurations are permanents, while for fermions they are Slater determinants. 
Both can be written in second-quantized form as 
\begin{equation}
\ket{\vec{n};t} = \mathcal{N} \prod_{j=1}^{M}  \left[\hat{b}_j^\dagger(t)\right]^{n_j}
\ket{\mathrm{vac}},
\label{eq:app_permanents}
\end{equation}
where $\hat{b}_j^\dagger(t)$ creates a particle in the orbital $\varphi_j(x,t)$, and the normalization factor is $\mathcal{N} = \prod_{j=1}^M \frac{1}{\sqrt{n_j!}}$ for bosons and $\mathcal{N}=1$ for fermions.
The orbitals are kept orthonormal at all times, $\braket{\varphi_i(t)}{\varphi_j(t)}=\delta_{ij}$.

The equations of motion are obtained from the time-dependent Dirac-Frenkel variational principle,
\begin{equation}
\mel{\delta \Psi}{i\hbar\partial_t-\hat{H}}{\Psi}=0,
\label{eq:app_dirac_frenkel}
\end{equation}
where variations are taken with respect to both the coefficients $C_{\vec{n}}(t)$ and the orbitals $\varphi_j(x,t)$.
This yields a coupled set of first-order differential equations for the coefficients and nonlinear integro-differential equations for the orbitals.
In the noninteracting case, the coefficient equation has the form
\begin{equation}
i\hbar \dot{C}_{\vec{n}}(t) = \sum_{\vec{m}} \mel{\vec{n};t}{\hat{H}}{\vec{m};t} C_{\vec{m}}(t),
\label{eq:app_coeff_eom}
\end{equation}
while the orbital equations can be written schematically as
\begin{equation}
i\hbar \partial_t \ket{\varphi_j} = \hat{P} \left[ \hat{h}\ket{\varphi_j} \right],
\label{eq:app_orb_eom}
\end{equation}
where $\hat{h}$ is the one-body Hamiltonian and $\hat{P} = 1-\sum_{k=1}^{M}\ket{\varphi_k}\bra{\varphi_k}$ projects the evolution onto the subspace orthogonal to the occupied orbitals, ensuring orthonormality of the time-dependent basis.

Generally speaking, the MCTDH ansatz becomes formally exact for $M\to\infty$. 
For finite $M$, convergence can be assessed by increasing the number of orbitals or by monitoring natural-orbital occupations.
However, in the noninteracting single-particle limit, $N_p=1, g=0$, the MCTDH ansatz is already exact for a single orbital, $M=1$, since there are no many-body correlations to resolve.
The MCTDH-X propagation therefore reduces to the continuum time evolution of a single optimized orbital under the time-dependent one-body Hamiltonian,
\begin{equation}
\hat{h}(t) = -\frac{\hbar^2}{2m}\frac{\mathrm{d}^2}{\mathrm{d}x^2} + V[x;\varphi(t)].
\label{eq:app_one_body_hamiltonian}
\end{equation}
Using MCTDH-X in this limit is nevertheless useful because it allows us to propagate the same continuum Hamiltonian that would be used for future interacting many-body extensions.

In the results presented in the main text, the continuum wave function is represented on a spatial grid of 512 points over the interval $x\in[0,L]$. 
Time propagation is performed using a Modified Craig–Sneyd (MCS) integrator with an error tolerance of $10^{-9}$. 
These numerical parameters are sufficient to resolve the qualitative structure of the phase diagrams reported here.
The absolute fidelity values, particularly very close to unity, may retain a small sensitivity to the spatial discretization, and sub-percent differences should therefore be interpreted with some caution.

\section{System-size dependence and spectral curvature}
\label{app:convexity}
\begin{figure}
    \centering
    \includegraphics[width=0.7\linewidth]{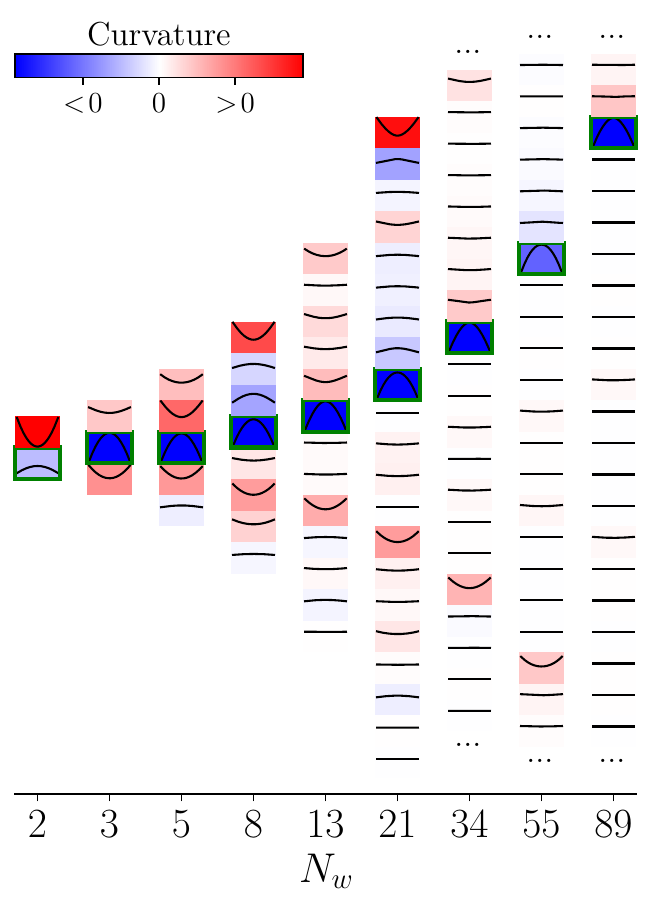}
    \caption{\textbf{Schematic plot of the curvature behavior:} We show the curvature of the energy bands around the phason angle $\varphi_0$ for different system sizes. The winding state is highlighted in green. One can see that for $N_w=5,13,34,89$ the curvature of the adjacent topmost band is opposite to the winding state, whereas for $N_w=8,21,55$ it is the same.}
    \label{fig:pyramid}
\end{figure}
In this Appendix, we investigate how the finite-size Fibonacci
approximant affects the spectral structure and the resulting
state-transfer protocols, providing additional details on the discussions in Sec.~\ref{subsec:time-evol} and in Sec.~\ref{sec:results}.
The main text focuses on $N_w=13$, which already supports well-localized winding states and a simple avoided-crossing geometry.
Nonetheless, larger system sizes may be experimentally relevant. 
We therefore examine the next Fibonacci approximant $N_w=21$, and briefly discuss the broader system-size dependence.

In general, the winding state and the adjacent band can exhibit different relative curvatures around the symmetry point $\varphi=\varphi_0$ depending on the finite-size Fibonacci approximant. 
Empirically, we observe an alternating pattern: for $N_w=5,13,34,89,\ldots$, the two bands have opposite curvature near $\varphi_0$ (see Fig.~\ref{fig:pyramid}), whereas for $N_w=8,21,55,\ldots$, they have the same curvature. 
This observation is based on the system sizes examined here and is not intended as a general analytical classification.
\begin{figure}
    \centering
    \includegraphics[width=1.0\linewidth]{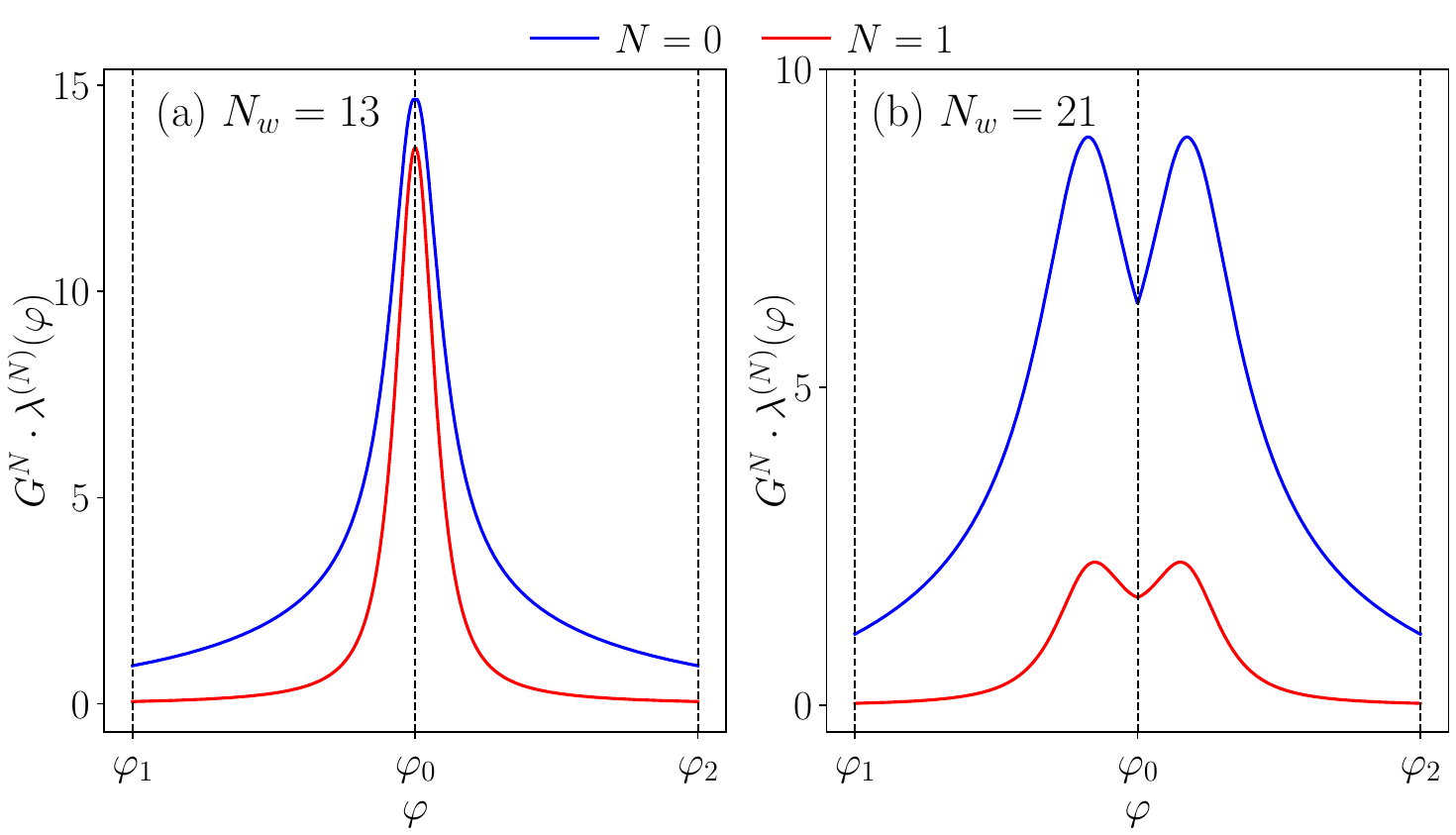}
    \caption{\textbf{The function} $\mathbf{G^N\lambda^{(N)}(\varphi)}$ \textbf{for different system sizes:} (a) For $N_w=13$ the winding state and the adjacent band have opposite curvatures. 
    As a consequence the adiabatic metrics (both $N=0$ and $N=1$) are monotonic from $\varphi_1$ to $\varphi_0$. 
    This is not the case in (b) for $N_w=21$ lattice sites. 
    Because the winding state and the adjacent band have the same curvature around $\varphi_0$ $\lambda^{(N)}(\varphi)$ has a more complicated structure leading to more complicated protocols. (In both cases: $S=10$ and $\rho=0.006$)}
    \label{fig:Lambda_comparision}
\end{figure}

\begin{figure}[h!]
    \centering
    \includegraphics[width=1.0\linewidth]{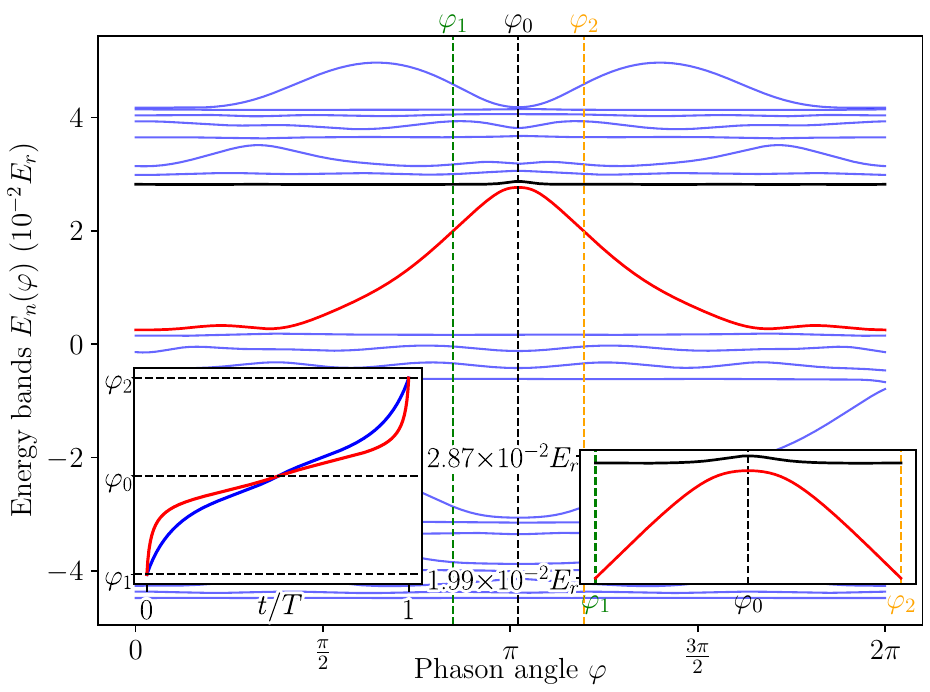}
    \caption{\textbf{Energy spectrum for} $\mathbf{N_w=21}$. 
    The tight-binding energy bands are plotted as a function of the phason angle for $N_w=21$ lattice sites ($S=10$, $\rho=0.006$).
    The red curve highlights the winding state, while the black curve indicates the adjacent band.
    All other bands are colored in blue.
    Right inset: Magnified view of the winding state and adjacent band around the energy gap. 
    In contrast to the case $N_w=13$ lattice sites, the two bands have the same-sign curvature around $\varphi_0$ (compare with Fig.~\ref{fig:spec_example}). 
    Left inset: state-transfer protocols $\varphi(t/T)$ for $N=0$ (blue curve) and $N=1$ (green curve). 
    Due to the different curvature of the energy spectrum around the gap, the protocols no longer have the simple shape obtained in Fig.~\ref{fig:protocols}.}
    \label{fig:Phason_spectrumNw21}
\end{figure}

\begin{figure*}
\begin{minipage}{0.49\textwidth}
        \centering
        \includegraphics[width=\linewidth]{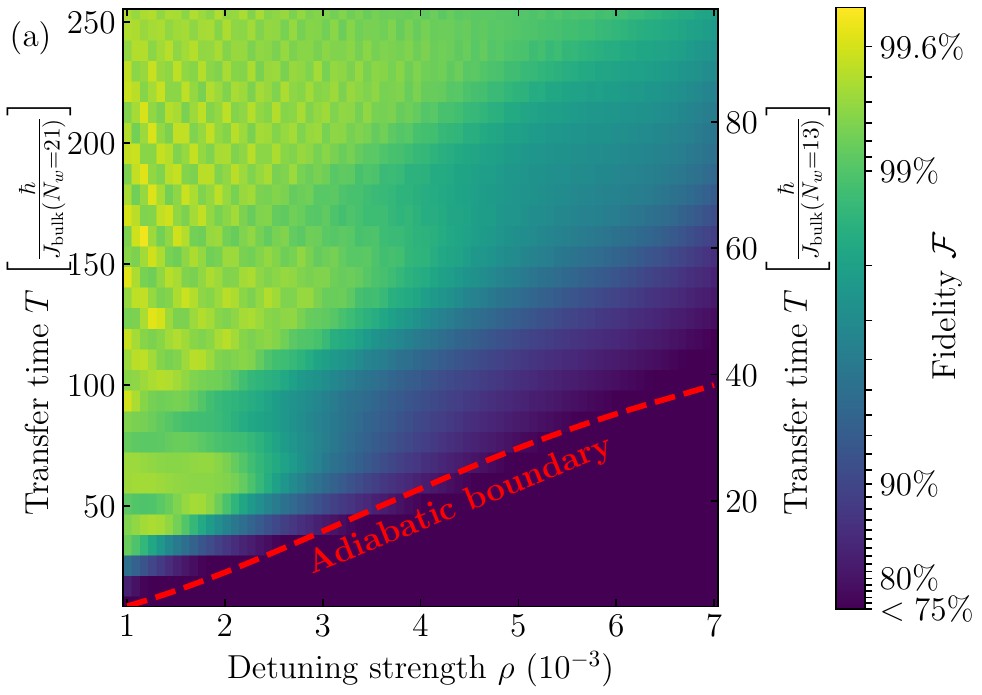}
    \end{minipage}
    \begin{minipage}{0.49\textwidth}
        \centering
        \includegraphics[width=\linewidth]{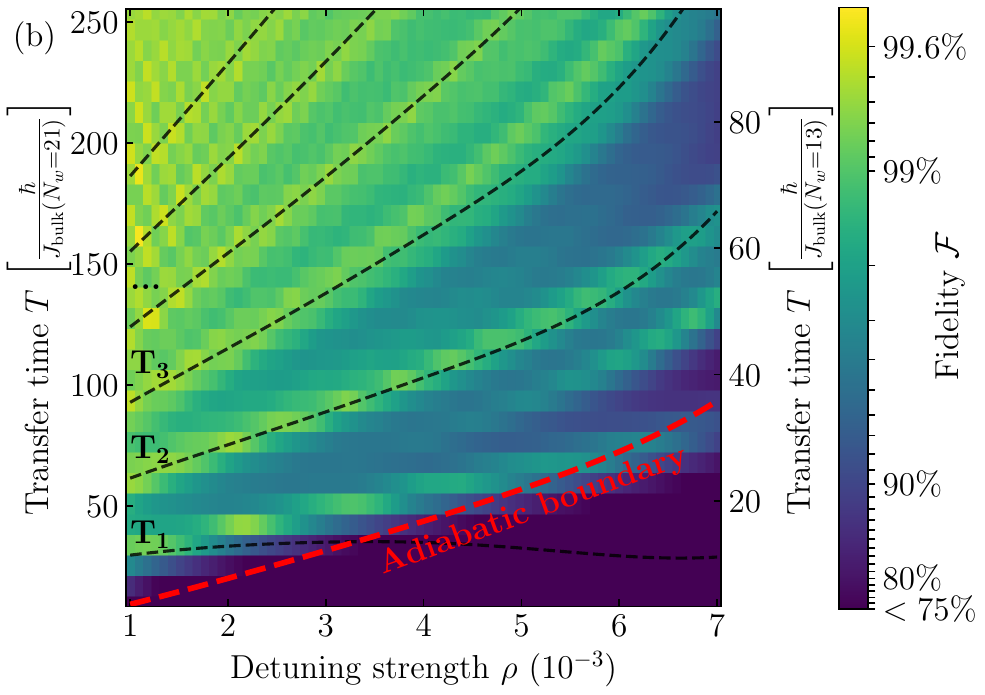}
    \end{minipage}
    \caption{
    \textbf{Fidelity diagram for} $\mathbf{N_w=21}$ \textbf{lattice sites}.
    Plot of the state-transfer fidelity as a function of detuning strength $\rho$ and transfer time $T$ for (a) $N=0$ and (b) $N=1$.
    The red dashed lines indicate the adiabatic boundary and the black lines are again the times from Eq.~\eqref{eq:Unit_fidelity_times}. 
    When the time $T_1$ crosses the adiabatic boundary it no longer describes the high-fidelity branch as the two-level approximation fails. 
    }
    \label{PS21}
\end{figure*}

The relative curvature directly affects the instantaneous level separation $\Delta E(\varphi) = E_{w+1}(\varphi)-E_w(\varphi)$, and therefore the shape of the adiabatic weight $\lambda^{(N)}(\varphi)$ (see Fig.~\ref{fig:Lambda_comparision}). 
When the two bands have opposite curvature, as for $N_w=13$, $\Delta E(\varphi)$ has a single dominant minimum near $\varphi_0$. 
Consequently, $\lambda^{(N)}(\varphi)$ develops a single dominant peak, and the protocol slows down primarily in one avoided-crossing region. 
This produces the simple tangent- and Roland-Cerf-like trajectories discussed in the main text.

When the two bands have the same curvature, the situation is more subtle. 
In this case, the local behavior of the level separation is controlled by the difference between the magnitudes of the two curvatures rather than by their signs alone. 
Consequently, $\varphi_0$ need not correspond to a sharply defined isolated minimum of $\Delta E(\varphi)$: the gap may become comparatively flat around $\varphi_0$ or develop additional off-center minima through cancellations of the quadratic terms and higher-order variations of the two bands. 
The corresponding adiabatic weight $\lambda^{(N)}(\varphi)$ can therefore acquire a broader or multi-peaked structure, causing the locally adiabatic protocol to slow down over several distinct portions of the transfer interval. 

For concreteness, we focus on the $N_w=21$ case, which is the next Fibonacci approximant after $N_w=13$ and exhibits the same-curvature spectral structure discussed above.
Fig.~\ref{fig:Phason_spectrumNw21} shows the relevant winding state and adjacent band, together with the corresponding numerical state-transfer protocols in the left inset.
In contrast to the $N_w=13$ case, the adiabatic weight contains several relevant features across the transfer interval, causing the protocol to slow down in multiple phason regions and acquire a multistage shape.

Fig.~\ref{PS21} shows the corresponding fidelity landscapes for the $N_w=21$ system.
Because the adiabatic weight has a multipeaked structure, the numerical solutions of Eq.~\eqref{eq:protocol_condition} can no longer be accurately fitted by the tangent and Roland-Cerf parametrizations used for $N_w=13$.
We therefore employ the numerically obtained $N=0$ and $N=1$ protocols directly in the continuum propagation.
The fidelity landscapes show behavior similar to the $N_w=13$ case. 
In regions of large detuning strengths, the $N=1$ protocol turns out to reach a larger transport fidelity for small transfer times, although it exhibits stronger fidelity oscillations. 
Thus, for $N_w=21$, the $N=1$ protocol can still yield larger peak fidelities at selected short transfer times, whereas the $N=0$ protocol provides a broader and less oscillatory high-fidelity region.
It is in this latter sense that the $N=0$ protocol is more robust.

In both cases, the adiabatic boundary from the definition in Eq.~\eqref{eq:adiabatic_condition_final} appears to describe a useful crossover between low- and high-fidelity regions.
The transfer times $T_n$ for which Eq.~\eqref{eq:two_level_fidelity} predicts unit fidelity capture the oscillation peaks for the $N=1$ fidelity plot. 
A difference to Fig.~\ref{fig:PhaseDiagram} is that the transfer time $T_1(\rho)$ crosses the adiabatic boundary at around $\rho \simeq 0.003$ and fails to describe the first fidelity peak for larger values of $\rho$ as this peak is apparently no longer in a region where the two-level approximation is valid.


\bibliographystyle{apsrev4-2mod}
\bibliography{biblio.bib}

@article{Mondal:2026,
      title={Breakdown of bosonic Thouless pump due to interaction in a quasiperiodic lattice}, 
      author={Suman Mondal and Emmanuel Gottlob and Fabian Heidrich-Meisner and Ulrich Schneider},
      journal={Phys. Rev. B},
      volume={113},
      pages={235131},
      url={https://doi.org/10.1103/6gdw-nmjk2601.18229},
      doi={10.1103/6gdw-nmjk2601.18229},
      year={2026}
}

@article{MariusPRL2025,
  title = {Multiband Fractional Thouless Pumps},
  author = {J\"urgensen, Marius and Steiner, Jacob and Refael, Gil and Rechtsman, Mikael C.},
  journal = {Phys. Rev. Lett.},
  volume = {135},
  issue = {16},
  pages = {166601},
  numpages = {6},
  year = {2025},
  month = {Oct},
  publisher = {American Physical Society},
  doi = {10.1103/4d5s-n4gn},
  url = {https://link.aps.org/doi/10.1103/4d5s-n4gn}
}

@book{WilczekBerryPhase1989,
    author = {Wilczek, F and Shapere, A},
    title = {Geometric Phases in Physics},
    publisher = {WORLD SCIENTIFIC},
    year = {1989},
    doi = {10.1142/0613},
    address = {},
    edition   = {},
    URL = {https://www.worldscientific.com/doi/abs/10.1142/0613}
}

@article{EdwardsIPR_1972,
    doi = {10.1088/0022-3719/5/8/007},
    url = {https://doi.org/10.1088/0022-3719/5/8/007},
    year = {1972},
    month = {apr},
    publisher = {},
    volume = {5},
    number = {8},
    pages = {807},
    author = {J T Edwards and D J Thouless},
    title = {Numerical studies of localization in disordered systems},
    journal = {J. Phys. C: Solid State Phys.}
}

@article{Modugno:2009,
  author = {Modugno, M.},
  title = {Exponential localization in one-dimensional quasiperiodic optical lattices},
  journal = {New J. Phys.},
  volume = {11},
  pages = {033023},
  year = {2009},
  doi = {10.1088/1367-2630/11/3/033023}
}

@article{Roati:2008,
  author = {Roati, G. and D'Errico, C. and Fallani, L. and Fattori, M. and Fort, C. and Zaccanti, M. and Modugno, G. and Modugno, M. and Inguscio, M.},
  title = {Anderson localization of a non-interacting Bose--Einstein condensate},
  journal = {Nature},
  volume = {453},
  pages = {895},
  year = {2008},
  doi = {10.1038/nature07071}
}

@article{Bloch:2008,
  author = {Bloch, I. and Dalibard, J. and Zwerger, W.},
  title = {Many-body physics with ultracold gases},
  journal = {Rev. Mod. Phys.},
  volume = {80},
  pages = {885},
  year = {2008},
  doi = {10.1103/RevModPhys.80.885}
}

@article{Luschen:2018,
  author = {L{\"u}schen, H. P. and Scherg, S. and Kohlert, T. and Schreiber, M. and Bordia, P. and Li, X. and Das Sarma, S. and Bloch, I.},
  title = {Single-particle mobility edge in a one-dimensional quasiperiodic optical lattice},
  journal = {Phys. Rev. Lett.},
  volume = {120},
  pages = {160404},
  year = {2018},
  doi = {10.1103/PhysRevLett.120.160404}
}

@article{Xiao:2017,
  title = {Mobility edges in one-dimensional bichromatic incommensurate potentials},
  author = {Li, Xiao and Li, Xiaopeng and Das Sarma, S.},
  journal = {Phys. Rev. B},
  volume = {96},
  issue = {8},
  pages = {085119},
  numpages = {18},
  year = {2017},
  month = {Aug},
  publisher = {American Physical Society},
  doi = {10.1103/PhysRevB.96.085119},
  url = {https://link.aps.org/doi/10.1103/PhysRevB.96.085119}
}

@article{Molignini:2025-1,
  author = {Molignini, Paolo},
  title = {Stability of quasicrystalline ultracold fermions to dipolar interactions},
  journal = {Phys. Rev. Research},
  volume = {7},
  pages = {L032026},
  year = {2025},
  doi = {10.1103/szdc-61nl},
  url = {https://doi.org/10.1103/szdc-61nl}
}

@article{Molignini:2025-2,
  author = {Molignini, Paolo and Chakrabarti, Barnali},
  title = {Stability of dipolar bosons in a quasiperiodic potential},
  journal = {Phys. Rev. Research},
  volume = {7},
  pages = {023237},
  year = {2025},
  doi = {10.1103/PhysRevResearch.7.023237},
  url = {https://link.aps.org/doi/10.1103/PhysRevResearch.7.023237}
}

@article{Chin:2010,
  author = {Chin, C. and Grimm, R. and Julienne, P. and Tiesinga, E.},
  title = {Feshbach resonances in ultracold gases},
  journal = {Rev. Mod. Phys.},
  volume = {82},
  pages = {1225},
  year = {2010},
  doi = {10.1103/RevModPhys.82.1225}
}

@article{Aubry:1980,
  author  = {Aubry, Serge and Andr{\'e}, Gilles},
  title   = {Analyticity breaking and Anderson localization in incommensurate lattices},
  journal = {Ann. Israel Phys. Soc.},
  volume  = {3},
  pages   = {133},
  year    = {1980},
  url = {https://chaos.if.uj.edu.pl/~delande/Lectures/files/An.Is.Phys.Soc.pdf}
}

@article{Kraus:2012,
  author = {Kraus, Y. E. and Zilberberg, O.},
  title = {Topological equivalence between the Fibonacci quasicrystal and the Harper model},
  journal = {Phys. Rev. Lett.},
  volume = {109},
  pages = {116404},
  year = {2012},
  doi = {10.1103/PhysRevLett.109.116404}
}

@article{Harper:1955,
  author = {Harper, P. G.},
  title = {Single Band Motion of Conduction Electrons in a Uniform Magnetic Field},
  journal = {Proc. Phys. Soc. A},
  volume = {68},
  pages = {874},
  year = {1955},
  doi = {10.1088/0370-1298/68/10/304}
}

@article{Liu:2017,
doi = {10.1088/1361-6455/aa98d6},
url = {https://doi.org/10.1088/1361-6455/aa98d6},
year = {2017},
month = {dec},
publisher = {IOP Publishing},
volume = {51},
number = {2},
pages = {025301},
author = {Liu, Tong and Wang, Pei and Chen, Shu and Xianlong, Gao},
title = {Phase diagram of a generalized off-diagonal Aubry–André model with p-wave pairing},
journal = {J. Phys. B},
}

@article{Liu:2021,
  title = {Dual mapping and quantum criticality in quasiperiodic Su-Schrieffer-Heeger chains},
  author = {Liu, Tong and Xia, Xu},
  journal = {Phys. Rev. B},
  volume = {104},
  issue = {13},
  pages = {134202},
  numpages = {6},
  year = {2021},
  month = {Oct},
  publisher = {American Physical Society},
  doi = {10.1103/PhysRevB.104.134202},
  url = {https://link.aps.org/doi/10.1103/PhysRevB.104.134202}
}

@article{Fangli:2015,
  title = {Localization and adiabatic pumping in a generalized Aubry-Andr\'e-Harper model},
  author = {Liu, Fangli and Ghosh, Somnath and Chong, Y. D.},
  journal = {Phys. Rev. B},
  volume = {91},
  issue = {1},
  pages = {014108},
  numpages = {9},
  year = {2015},
  month = {Jan},
  publisher = {American Physical Society},
  doi = {10.1103/PhysRevB.91.014108},
  url = {https://link.aps.org/doi/10.1103/PhysRevB.91.014108}
}

@article{Biddle:2009,
  author = {Biddle, Justin and Wang, B. and Priour, D. J. Jr. and Das Sarma, S.},
  title = {Predicted mobility edges in one-dimensional incommensurate optical lattices: An exactly solvable model of Anderson localization},
  journal = {Phys. Rev. A},
  volume = {80},
  pages = {021603(R)},
  year = {2009},
  doi = {10.1103/PhysRevA.80.021603}
}

@article{Boers:2007,
  author = {Boers, D. J. and Goedeke, S. and Hinrichs, D. and Holthaus, M.},
  title = {Mobility edges in bichromatic incommensurate optical lattices},
  journal = {Phys. Rev. A},
  volume = {75},
  pages = {063404},
  year = {2007},
  doi = {10.1103/PhysRevA.75.063404}
}

@article{Dominguez-Castro:2009,
doi = {10.1088/1361-6404/ab1670},
url = {https://doi.org/10.1088/1361-6404/ab1670},
year = {2019},
publisher = {IOP Publishing},
volume = {40},
number = {4},
pages = {045403},
author = {Domínguez-Castro, G A and Paredes, R},
title = {The Aubry–André model as a hobbyhorse for understanding the localization phenomenon},
journal = {Eur. J. Phys.},
}

@article{Kraus:2012-2,
  author = {Kraus, Yaacov E. and Lahini, Yoav and Ringel, Zohar and Verbin, Mor and Zilberberg, Oded},
  title = {Topological States and Adiabatic Pumping in Quasicrystals},
  journal = {Phys. Rev. Lett.},
  volume = {109},
  pages = {106402},
  year = {2012},
  doi = {10.1103/PhysRevLett.109.106402}
}

@article{Verbin:2013,
  author = {Verbin, Mor and Zilberberg, Oded and Kraus, Yaacov E. and Lahini, Yoav and Silberberg, Yaron},
  title = {Observation of topological phase transitions in photonic quasicrystals},
  journal = {Phys. Rev. Lett.},
  volume = {110},
  pages = {076403},
  year = {2013},
  doi = {10.1103/PhysRevLett.110.076403}
}

@article{Bradlyn:2022,
	title={Lecture notes on Berry phases and topology},
	author={Barry Bradlyn and Mikel Iraola},
	journal={SciPost Phys. Lect. Notes},
	pages={51},
	year={2022},
	doi={10.21468/SciPostPhysLectNotes.51},
}

@article{Comparat:2009,
  title = {General conditions for quantum adiabatic evolution},
  author = {Comparat, Daniel},
  journal = {Phys. Rev. A},
  volume = {80},
  pages = {012106},
  year = {2009},
  doi = {10.1103/PhysRevA.80.012106},
}

@article{Born:1928cqs,
    author = "Born, M. and Fock, V.",
    title = "{Beweis des Adiabatensatzes}",
    doi = "10.1007/BF01343193",
    journal = "Z. Phys.",
    volume = "51",
    number = "3",
    pages = "165--180",
    year = "1928"
}

@article{Liu:2022,
  title = {Observation of edge-to-edge topological transport in a photonic lattice},
  author = {Liu, Weijie and Wu, Chaohua and Jia, Yuechen and Jia, Suotang and Chen, Gang and Chen, Feng},
  journal = {Phys. Rev. A},
  volume = {105},
  pages = {L061502},
  year = {2022},
  doi = {10.1103/PhysRevA.105.L061502},
  url = {https://link.aps.org/doi/10.1103/PhysRevA.105.L061502}
}

@article{Jansen:2006,
author = {Jansen, Sabine and Ruskai, Mary-Beth and Seiler, Ruedi},
year = {2007},
pages = {102111},
title = {Bounds for the adiabatic approximation with applications to quantum computation},
volume = {48},
journal = {Journal of Mathematical Physics},
doi = {10.1063/1.2798382}
}

@article{Landau:1932,
  author = {Landau, L. D.},
  title = {On the theory of transfer of energy at collisions II},
  journal = {Physics of the Soviet Union},
  volume = {2},
  pages = {46},
  year = {1932}
}

@article{Zener:1932,
  author = {Zener, C.},
  title = {Non-adiabatic crossing of energy levels},
  journal = {Proceedings of the Royal Society A},
  volume = {137},
  pages = {696},
  year = {1932},
  doi = {10.1098/rspa.1932.0165}
}

@article{Glasbrenner:2023,
doi = {10.1088/1361-6455/acc774},
url = {https://doi.org/10.1088/1361-6455/acc774},
year = {2023},
volume = {56},
pages = {104001},
author = {Glasbrenner, Eric P and Schleich, Wolfgang P},
title = {The Landau–Zener formula made simple},
journal = {Journal of Physics B: Atomic, Molecular and Optical Physics},
}

@article{Suominen:1992,
  title = {Population transfer in a level-crossing model with two time scales},
  author = {Suominen, K.-A. and Garraway, B. M.},
  journal = {Phys. Rev. A},
  volume = {45},
  pages = {374},
  year = {1992},
  publisher = {American Physical Society},
  doi = {10.1103/PhysRevA.45.374},
  url = {https://link.aps.org/doi/10.1103/PhysRevA.45.374}
}

@article{Vitanov:1999,
  title = {Nonlinear level-crossing models},
  author = {Vitanov, N. V. and Suominen, K.-A.},
  journal = {Phys. Rev. A},
  volume = {59},
  pages = {4580},
  year = {1999},
  publisher = {American Physical Society},
  doi = {10.1103/PhysRevA.59.4580},
  url = {https://link.aps.org/doi/10.1103/PhysRevA.59.4580}
}

@article{Robinson:1985,
  title = {Two-level systems driven by modulated pulses},
  author = {Robinson, E. J.},
  journal = {Phys. Rev. A},
  volume = {31},
  pages = {3986},
  year = {1985},
  publisher = {American Physical Society},
  doi = {10.1103/PhysRevA.31.3986},
  url = {https://link.aps.org/doi/10.1103/PhysRevA.31.3986}
}

@article{Roland:2002,
  title = {Quantum search by local adiabatic evolution},
  author = {Roland, J\'er\'emie and Cerf, Nicolas J.},
  journal = {Phys. Rev. A},
  volume = {65},
  issue = {4},
  pages = {042308},
  year = {2002},
  doi = {10.1103/PhysRevA.65.042308},
  url = {https://link.aps.org/doi/10.1103/PhysRevA.65.042308}
}

@article{Malossi:2013,
	title = {Quantum driving protocols for a two-level system: From generalized Landau-Zener sweeps to transitionless control},
	author = {Malossi, N. and Bason, M. G. and Viteau, M. and Arimondo, E. and Mannella, R. and Morsch, O. and Ciampini, D.},
	journal = {Phys. Rev. A},
	volume = {87},
	pages = {012116},
	year = {2013},
	doi = {10.1103/PhysRevA.87.012116},
	url = {https://link.aps.org/doi/10.1103/PhysRevA.87.012116}
}

@article{Stefanatos:2020,
doi = {10.1088/1751-8121/ab7423},
url = {https://doi.org/10.1088/1751-8121/ab7423},
year = {2020},
publisher = {IOP Publishing},
volume = {53},
pages = {115304},
author = {Stefanatos, Dionisis and Paspalakis, Emmanuel},
title = {Speeding up adiabatic passage with an optimal modified Roland–Cerf protocol},
journal = {Journal of Physics A: Mathematical and Theoretical},
}

@article{Deissler2011,
  author  = {Deissler, B. and Lucioni, E. and Modugno, M. and Roati, G.
             and Tanzi, L. and Zaccanti, M. and Inguscio, M. and Modugno, G.},
  title   = {Correlation function of weakly interacting bosons in a disordered lattice},
  journal = {New J. Phys.},
  volume  = {13},
  pages   = {023020},
  year    = {2011},
  doi     = {10.1088/1367-2630/13/2/023020}
}

@book{TDVM81,
	author = {Kramer, P. and Saraceno, M.},
	publisher = {Springer},
	series = {Lecture Notes in Physics},
	title = {Geometry of the Time-Dependent Variational Principle in Quantum Mechanics},
	volume = {140},
	year = {1981}}

@article{Streltsov:2006,
	author = {Alexej I. Streltsov and Ofir E. Alon and Lorenz S. Cederbaum},
	doi = {10.1103/PhysRevA.73.063626},
	journal = {Phys. Rev. A},
	pages = {063626},
	title = {General variational many-body theory with complete self-consistency for trapped bosonic systems},
	url = {https://doi.org/10.1103/PhysRevA.73.063626},
	volume = {73},
	year = {2006}}

@article{Streltsov:2007,
	author = {Alexej I. Streltsov and Ofir E. Alon and Lorenz S. Cederbaum},
	doi = {10.1103/PhysRevLett.99.030402},
	journal = {Phys. Rev. Lett.},
	pages = {030402},
	title = {Role of Excited States in the Splitting of a Trapped Interacting Bose-Einstein Condensate by a Time-Dependent Barrier},
	url = {https://doi.org/10.1103/PhysRevLett.99.030402},
	volume = {99},
	year = {2007}}

@article{Alon:2007,
	author = {Ofir E. Alon and Alexej I. Streltsov and Lorenz S. Cederbaum},
	doi = {10.1063/1.2771159},
	journal = {J. Chem. Phys.},
	pages = {154103},
	title = {Unified view on multiconfigurational time propagation for systems consisting of identical particles},
	url = {https://doi.org/10.1063/1.2771159},
	volume = {127},
	year = {2007},
}

@article{Alon:2008,
	author = {O. E. Alon and A. I. Streltsov and L. S. Cederbaum},
	doi = {10.1103/PhysRevA.77.033613},
	journal = {Phys. Rev. A},
	pages = {033613},
	title = {Multiconfigurational time-dependent Hartree method for bosons: Many-body dynamics of bosonic systems},
	url = {https://link.aps.org/doi/10.1103/PhysRevA.77.033613},
	volume = {77},
	year = {2008},
}

@article{Lode:2012,
    author = {Axel U. J. Lode and Kaspar Sakmann and Ofir E. Alon and Lorenz S. Cederbaum and Alexej I. Streltsov},
    title = {Numerically exact quantum dynamics of bosons with time-dependent interactions of harmonic type},
    journal = {Phys. Rev. A},
    volume = {86},
    pages = {063606},
    year = {2012},
    doi = {10.1103/PhysRevA.86.063606},
    url = {https://doi.org/10.1103/PhysRevA.86.063606}
}

@article{Lode:2016,
	author = {A. U. J. Lode},
	doi = {10.1103/PhysRevA.93.063601},
	journal = {Phys. Rev. A},
	pages = {063601},
	title = {Multiconfigurational time-dependent Hartree method for bosons with internal degrees of freedom: Theory and composite fragmentation of multicomponent Bose-Einstein condensates},
	url = {https://link.aps.org/doi/10.1103/PhysRevA.93.063601},
	volume = {93},
	year = {2016},
}

@article{Fasshauer:2016,
	author = {Elke Fasshauer and A. U. J. Lode},
	doi = {10.1103/PhysRevA.93.033635},
	journal = {Phys. Rev. A},
	pages = {033635},
	title = {Multiconfigurational time-dependent Hartree method for fermions: Implementation, exactness, and few-fermion tunneling to open space},
	url = {https://link.aps.org/doi/10.1103/PhysRevA.93.033635},
	volume = {93},
	year = {2016},
}

@article{Lin:2020,
	author = {R. Lin and P. Molignini and L. Papariello and M. C. Tsatsos and C. L{\'e}v{\^e}que and S. E. Weiner and E. Fasshauer and R. Chitra},
	doi = {10.1088/2058-9565/ab788b},
	journal = {Quantum Sci. Technol.},
	pages = {024004},
	title = {MCTDH-X: The multiconfigurational time-dependent Hartree method for indistinguishable particles software},
	url = {https://iopscience.iop.org/article/10.1088/2058-9565/ab788b},
	volume = {5},
	year = {2020},
}

@article{Lode:2020,
	author = {A. U. J. Lode and C. L{\'e}v{\^e}que and L. B. Madsen and A. I. Streltsov and O. E. Alon},
	doi = {10.1103/RevModPhys.92.011001},
	journal = {Rev. Mod. Phys},
	pages = {011001},
	title = {Colloquium: Multiconfigurational time-dependent Hartree approaches for indistinguishable particles},
	url = {https://journals.aps.org/rmp/abstract/10.1103/RevModPhys.92.011001},
	volume = {92},
	year = {2020},
}

@article{Molignini:2025-SciPost,
	title = {Many-body quantum dynamics with MCTDH-X},
	pages = {94},
	author = {Molignini, Paolo and Dutta, Sunayana and Fasshauer, Elke},
	journal = {SciPost Phys. Lect. Notes},
	year = {2025},
	publisher = {SciPost},
	doi = {10.21468/SciPostPhysLectNotes.94},
	url = {https://scipost.org/10.21468/SciPostPhysLectNotes.94}
}

@online{MCTDHX,
	author = {A. U. J. Lode and M. C. Tsatsos and E. Fasshauer and S. E. Weiner and R. Lin and L. Papariello and P. Molignini and C. L{\'e}v{\^e}que and M. B{\"u}ttner and J. Xiang and S. Dutta and Y. Bilinskaya},
	title = {MCTDH-X: The MultiConfigurational Time-Dependent Hartree Method for Indistinguishable Particles Software},
	url = {http://ultracold.org},
	year = {2012-2026}}

@article{Xiang:2023,
    author = {Jiabing Xiang and Paolo Molignini and Miriam Büttner and Axel U. J. Lode},
    title = {Pauli crystal melting in shaken optical traps},
    journal = {SciPost Phys.},
    volume = {14},
    pages = {003},
    year = {2023},
    url = {https://scipost.org/10.21468/SciPostPhys.14.1.003},
    doi = {10.21468/SciPostPhys.14.1.003}
}

@article{Beinke:2018,
    author = {Raphael Beinke and Lorenz S. Cederbaum and Ofir E. Alon},
    title = {Enhanced many-body effects in the excitation spectrum of a weakly interacting rotating Bose-Einstein condensate},
    journal = {Phys. Rev. A},
    volume = {98},
    pages = {053634},
    year = {2018},
    doi = {10.1103/PhysRevA.98.053634},
    url = {https://doi.org/10.1103/PhysRevA.98.053634}
}

@article{Roy:2018,
    author = {R. Roy and A. Gammal and M. C. Tsatsos and B. Chatterjee and B. Chakrabarti and A. U. J. Lode},
    title = {Phases, many-body entropy measures, and coherence of interacting bosons in optical lattices},
    journal = {Phys. Rev. A},
    volume = {97},
    pages = {043625},
    year = {2018},
    doi = {10.1103/PhysRevA.97.043625},
    url = {https://doi.org/10.1103/PhysRevA.97.043625}
}

@article{Dutta:2019,
    author = {S. Dutta and Marios C Tsatsos and Saurabh Basu and Axel U. J. Lode},
    title = {Management of the correlations of UltracoldBosons in triple wells},
    journal = {New J. Phys.},
    volume = {21},
    pages = {053044},
    year = {2019},
    doi = {10.1088/1367-2630/ab117d},
    url = {https://doi.org/10.1088/1367-2630/ab117d}
}

@article{Schaefer:2020,
    author = {Frank Schäfer and Miguel A. Bastarrachea-Magnani and Axel U. J. Lode and Laurent de Forges de Parny and Andreas Buchleitner},
    title = {Spectral Structure and Many-Body Dynamics of Ultracold Bosons in a Double-Well},
    journal = {Entropy},
    volume = {22},
    pages = {382},
    year = {2020},
    doi = {10.3390/e22040382},
    url = {https://doi.org/10.3390/e22040382}
}

@article{Lode:2021,
    author = {Axel U. J. Lode and Sunayana Dutta and Camille Lévêque},
    title = {Dynamics of Ultracold Bosons in Artificial Gauge Fields—Angular Momentum, Fragmentation, and the Variance of Entropy},
    journal = {Entropy},
    volume = {23},
    pages = {392},
    year = {2021},
    doi = {10.3390/e23040392},
    url = {https://doi.org/10.3390/e23040392}
}

@article{Lode:2021-10,
    author = {Axel U. J. Lode and Rui Lin and Miriam Büttner and Luca Papariello and Camille Lévêque and R. Chitra and Marios C. Tsatsos and Dieter Jaksch and Paolo Molignini},
    title = {Optimized observable readout from single-shot images of ultracold atoms via machine learning},
    journal = {Phys. Rev. A},
    volume = {104},
    pages = {L041301},
    year = {2021},
    doi = {10.1103/PhysRevA.104.L041301},
    url = {https://doi.org/10.1103/PhysRevA.104.L041301}
}

@article{Debnath:2024,
doi = {10.1088/1612-202X/ad21eb},
url = {https://doi.org/10.1088/1612-202X/ad21eb},
year = {2024},
month = {feb},
publisher = {IOP Publishing},
volume = {21},
number = {3},
pages = {035501},
author = {Kumar Debnath, Pankaj and Chakrabarti, Barnali and Leslie Lekala, Mantile},
title = {Quench dynamics of a Tonks-Girardeau gas in one dimensional anharmonic trap},
journal = {Laser Physics Letters}
}

@article{Dutta:2023,
    author = {S. Dutta and A. U. J. Lode and O. Alon},
    title = {Fragmentation and correlations in a rotating Bose–Einstein condensate undergoing breakup},
    journal = {Sci Rep},
    volume = {13},
    pages = {3343},
    year = {2023},
    doi = {10.1038/s41598-023-29516-w},
    url = {https://doi.org/10.1038/s41598-023-29516-w}
}

@article{Roy:2023,
    author = {Rhombik Roy and Barnali Chakrabarti and Arnaldo Gammal},
    title = {Out of equilibrium many-body expansion dynamics of strongly interacting bosons},
    journal = {SciPost Phys. Core},
    volume = {6},
    pages = {073},
    year = {2023},
    doi = {10.21468/SciPostPhysCore.6.4.073},
    url = {https://doi.org/10.21468/SciPostPhysCore.6.4.073}
}

@article{Aloqali:2024,
    author = {Amer D Al-Oqali and Roger R Sakhel and Asaad R Sakhel},
    title = {Effect of zero-point motion on properties of quantum particles adsorbed on a substrate},
    journal = {Journal of Physics: Condensed Matter},
    volume = {36},
    pages = {245401},
    year = {2024},
    doi = {10.1088/1361-648X/ad3095},
    url = {https://doi.org/10.1088/1361-648X/ad3095}
}

@article{Dutta:2024,
    author = {Sunayana Dutta and Axel U. J. Lode and Ofir E. Alon},
    title = {Condensates breaking up under rotation},
    journal = {J. Phys.: Conf. Ser.},
    volume = {2894},
    pages = {012014},
    year = {2024},
    doi = {10.1088/1742-6596/2894/1/012014},
    url = {https://doi.org/10.1088/1742-6596/2894/1/012014}
}

@article{Haldar:2024,
    author = {Sudip Kumar Haldar and Anal Bhowmik},
    title = {Many-Body Effects in a Composite Bosonic Josephson Junction},
    journal = {Atoms},
    volume = {12},
    pages = {66},
    year = {2024},
    doi = {10.3390/atoms12120066},
    url = {https://doi.org/10.3390/atoms12120066}
}

@article{Chakrabarti:2024,
    author = {Barnali Chakrabarti and Arnaldo Gammal and Luca Salasnich},
    title = {Strongly interacting bosons in a one-dimensional disordered lattice: Phase coherence of distorted Mott phases},
    journal = {Phys. Rev. B},
    volume = {110},
    pages = {184202},
    year = {2024},
    doi = {10.1103/PhysRevB.110.184202},
    url = {https://doi.org/10.1103/PhysRevB.110.184202}
}

@article{Bhowmik:2025,
    author = {Anal Bhowmik and Ofir E Alon},
    title = {Interference of longitudinal and transversal fragmentations in the Josephson tunneling dynamics of Bose–Einstein condensates},
    journal = {New J. Phys.},
    volume = {26},
    pages = {123035},
    year = {2025},
    doi = {10.1088/1367-2630/ada0d3},
    url = {https://doi.org/10.1088/1367-2630/ada0d3}
}

@misc{Chakrabarti:2025-2,
      title={Localization versus incommemsurability for finite boson system in one-dimensional disordered lattice}, 
      author={Barnali Chakrabarti and Arnaldo Gammal},
      year={2025},
      eprint={2502.14440},
      archivePrefix={arXiv},
      primaryClass={cond-mat.quant-gas},
      url={https://arxiv.org/abs/2502.14440}, 
}

@article{Dutta:2025,
    author = {Sunayana Dutta and Ofir Alon},
    title = {Rotation-mediated bosonic Josephson junctions in position and momentum spaces
},
    journal = {arXiv:2503.10153},
    url = {https://doi.org/10.48550/arXiv.2503.10153},
    year = {2025}
}

@article{Roy:2025-7,
    author = {Rhombik Roy and Sunayana Dutta and Ofir E. Alon},
    title = {Rotation quenches in trapped bosonic systems},
    journal = {Scientific Reports},
    volume = {15},
    pages = {27193},
    year = {2025},
    doi = {https://doi.org/10.1038/s41598-025-07144-w}
}

@article{Roy:2025,
    author = {Rhombik Roy and Ofir E. Alon},
    title = {Dynamics and transport of Bose–Einstein condensates in bent potentials},
    doi = {10.1063/5.0301304},
    journal = {J. Chem. Phys.},
    volume = {163},
    pages = {224306},
    year = {2025}
}

@article{Roy:2025-4,
    author = {Rhombik Roy and Ofir E. Alon},
    title = {Assessing small accelerations using a bosonic Josephson junction},
    journal = {Phys. Rev. A},
    volume = {111},
    pages = {043307},
    year = {2025},
    doi = {10.1103/PhysRevA.111.043307},
    url = {https://doi.org/10.1103/PhysRevA.111.043307}
}

@article{Roy:2026,
    author = {Rhombik Roy and Ofir E. Alon},
    title = {Inferring rotations using a bosonic Josephson junction},
    doi = {10.1103/r5wq-lhcr},
    url = {https://doi.org/10.1103/r5wq-lhcr},
    journal = {Phys. Rev. A},
    volume = {114},
    pages = {033305},
    year = {2026}
}

@article{Fischer:2015,
	author = {Uwe R. Fischer and Axel U. J. Lode and Budhaditya Chatterjee},
	title = {Condensate fragmentation as a sensitive measure of the quantum many-body behavior of bosons with long-range interactions},
        journal = {Phys. Rev. A},
	volume = {91},
        pages = {063621},
	year = {2015},
	publisher = {American Physical Society},
	doi = {10.1103/PhysRevA.91.063621},
	url = {https://link.aps.org/doi/10.1103/PhysRevA.91.063621},
}

@article{Chatterjee:2018,
	author = {Chatterjee, Budhaditya and Lode, Axel U. J.},
	doi = {10.1103/PhysRevA.98.053624},
	issue = {5},
	journal = {Phys. Rev. A},
	month = {Nov},
	numpages = {8},
	pages = {053624},
	publisher = {American Physical Society},
	title = {Order parameter and detection for a finite ensemble of crystallized one-dimensional dipolar bosons in optical lattices},
	url = {https://link.aps.org/doi/10.1103/PhysRevA.98.053624},
	volume = {98},
	year = {2018}
}

@article{Bera:2019,
    author = {S. Bera and B. Chakrabarti and A. Gammal and M. C. Tsatsos and M. L. Lekala and B. Chatterjee and C. Lévêque and A. U. J. Lode},
    title = {Sorting Fermionization from Crystallization in Many-Boson Wavefunctions},
    journal = {Scientific Reports},
    volume = {9},
    pages = {17873},
    year = {2019},
    doi = {10.1038/s41598-019-53179-1},
    url = {https://doi.org/10.1038/s41598-019-53179-1},
}

@article{Chatterjee:2019,
	author = {Budhaditya Chatterjee and Marios C Tsatsos and Axel U J Lode},
	doi = {10.1088/1367-2630/aafa93},
	journal = {New J. Phys.},
	month = {mar},
	number = {3},
	pages = {033030},
	publisher = {{IOP} Publishing},
	title = {Correlations of strongly interacting one-dimensional ultracold dipolar few-boson systems in optical lattices},
	url = {https://doi.org/10.1088/1367-2630/aafa93},
	volume = {21},
	year = 2019
}

@article{Bera:2019-symm,
    title = {Correlation Dynamics of Dipolar Bosons in 1D Triple Well Optical Lattice},
    author = {Sangita Bera and Luca Salasnich and Barnali Chakrabarti},
    journal = {Symmetry},
    volume = {11},
    pages = {909},
    year = {2019},
    doi = {10.3390/sym11070909},
    url = {https://doi.org/10.3390/sym11070909}
}

@article{Chatterjee:2020,
	author = {Chatterjee, Budhaditya and L\'ev\^eque, Camille and Schmiedmayer, J\"org and Lode, Axel U. J.},
	doi = {10.1103/PhysRevLett.125.093602},
	issue = {9},
	journal = {Phys. Rev. Lett.},
	month = {Aug},
	numpages = {7},
	pages = {093602},
	publisher = {American Physical Society},
	title = {Detecting One-Dimensional Dipolar Bosonic Crystal Orders via Full Distribution Functions},
	url = {https://link.aps.org/doi/10.1103/PhysRevLett.125.093602},
	volume = {125},
	year = {2020}
}

@article{Roy:2022,
    author = {Rhombik Roy and Barnali Chakrabarti and Andrea Trombettoni},
    title = {Quantum dynamics of few dipolar bosons in a double-well potential},
    journal = {Eur. Phys. J. D},
    volume = {76},
    pages = {24},
    year = {2022},
    doi = {10.1140/epjd/s10053-022-00345-2},
    url = {https://doi.org/10.1140/epjd/s10053-022-00345-2}
}

@article{Hughes:2023,
    author = {Michael Hughes and  Axel U. J. Lode and  Dieter Jaksch and Paolo Molignini},
    title = {Accuracy of quantum simulators with ultracold dipolar molecules: A quantitative comparison between continuum and lattice descriptions},
    journal = {Phys. Rev. A},
    volume = {107},
    pages={033323}, 
    year = {2023},
    url = {https://doi.org/10.1103/PhysRevA.107.033323},
    doi = {10.1103/PhysRevA.107.033323}
}

@article{Bilinskaya:2024,
    author = {Yuliya Bilinskaya and Michael Hughes and Paolo Molignini},
    doi = {10.1103/PhysRevResearch.6.L042024},
    journal = {Phys. Rev. Research},
    pages = {L042024},
    title = {Realizing multiband states with ultracold dipolar quantum simulators},
    url = {https://doi.org/10.1103/PhysRevResearch.6.L042024},
    volume = {6},
    year = {2024}
}

@article{Molignini:2025-3,
    author = {P. Molignini and B. Chakrabarti},
    doi = {10.1103/PhysRevResearch.7.013257},
    journal = {Phys. Rev. Research},
    pages = {013257},
    title = {Interaction quench of dipolar bosons in a one-dimensional optical lattice},
    url = {https://doi.org/10.1103/PhysRevResearch.7.013257},
    volume = {7},
    year = {2025},
}

@article{Molignini:2024-2,
    author = {P. Molignini and B. Chakrabarti},
    doi = {10.1088/1367-2630/ad80b8},
    journal = {New J. Phys.},
    pages = {103030},
    title = {Unbounded entropy production and violent fragmentation for repulsive-to-attractive interaction quench in long-range interacting systems},
    url = {https://doi.org/10.1088/1367-2630/ad80b8},
    volume = {26},
    year = {2024}
}

@article{Roy:2024-annals,
    title = {Quasi-superfluid and Quasi-Mott phases of strongly interacting bosons in shallow optical lattice},
    author = {Subhrajyoti Roy and Rhombik Roy and Arnaldo Gammal and Barnali Chakrabarti and Budhaditya Chatterjee},
    journal = {Ann. Phys.},
    volume = {470}, 
    pages = {169807},
    year = {2024},
    doi = {10.1016/j.aop.2024.169807},
    url = {https://doi.org/10.1016/j.aop.2024.169807}
}

@article{Roy:2024-epjp,
    title = {Expansion of strongly interacting dipolar bosons in 1D optical lattices},
    author = {Rhombik Roy and Andrea Trombettoni and Barnali Chakrabarti},
    journal = {Eur. Phys. J. Plus},
    volume = {139}, 
    pages = {831},
    year = {2024},
    doi = {10.1140/epjp/s13360-024-05651-9},
    url = {https://doi.org/10.1140/epjp/s13360-024-05651-9}
}

@misc{Chakrabarti:2025,
    title = {Transport of ultracold dipolar fermions in one-dimensional optical lattices}, 
    author = {Barnali Chakrabarti and N D Chavda and Andrea Trombettoni and Arnaldo Gammal},
    year = {2025},
    eprint = {2502.06437},
    url = {https://doi.org/10.48550/arXiv.2502.06437},
    doi = {10.48550/arXiv.2502.06437}
}

@article{Molignini:2025-JPCM,
    author = {Paolo Molignini},
    title = {Beyond-mean-field phases of rotating dipolar condensates},
    journal = {J. Phys.: Condens. Matter},
    volume = {37},
    pages = {445401},
    year = {2025},
    url = {https://doi.org/10.1088/1361-648X/ae0fd3},
    doi = {10.1088/1361-648X/ae0fd3},
}

@article{Lode:2017,
    title = {Fragmented Superradiance of a Bose-Einstein Condensate in an Optical Cavity},
    author = {Lode, Axel U.J. and Bruder, Christoph},
    journal = {Phys. Rev. Lett.},
    volume = {118},
    pages = {013603},
    year = {2017},
    doi = {10.1103/PhysRevLett.118.013603},
    url = {https://doi.org/10.1103/PhysRevLett.118.013603}
}

@article{Lode:2018,
    author = {Lode, Axel U.J. and Diorico, Fritz S. and Wu, Rugway and Molignini, Paolo and Papariello, Luca and Lin, Rui and {L{\'{e}}v{\^{e}} Que}, Camille and Exl, Lukas and Tsatsos, Marios C. and Chitra, R. and Mauser, Norbert J.},
    title = {Many-body physics in two-component Bose-Einstein condensates in a cavity: Fragmented superradiance and polarization},
    journal = {New J. Phys.},
    volume = {20},
    pages = {055006},
    year = {2018},
    url = {https://doi.org/10.1088/1367-2630/aabc3a},
    doi = {10.1088/1367-2630/aabc3a}
}

@article{Molignini:2018,
    author = {Molignini, Paolo and Papariello, Luca and Lode, Axel U.J. and Chitra, R.},
    journal = {Phys. Rev. A},
    title = {Superlattice switching from parametric instabilities in a driven-dissipative Bose-Einstein condensate in a cavity},
    volume = {98},
    pages = {053620},
    year = {2018},
    url = {https://doi.org/10.1103/PhysRevA.98.053620},
    doi = {10.1103/PhysRevA.98.053620}
}

@article{Lin:2019,
    author = {Lin, Rui and Papariello, Luca and Molignini, Paolo and Chitra, R. and Lode, Axel U. J.},
    journal = {Phys. Rev. A},
    pages = {013611},
    title = {Superfluid--Mott-insulator transition of ultracold superradiant bosons in a cavity},
    volume = {100},
    year = {2019},
    url = {https://doi.org/10.1103/physreva.100.013611},
    doi = {10.1103/physreva.100.013611},
}

@article{Lin:2020-PRA,
	author = {Rui Lin and Paolo Molignini and Axel U. J. Lode and R. Chitra},
	journal = {Phys. Rev. A},
	pages = {061602(R)},
	title = {Pathway to chaos through hierarchical superfluidity in blue-detuned cavity-BEC systems},
	volume = {101},
	year = {2020},
	url = {https://doi.org/10.1103/PhysRevA.101.061602},
	doi = {10.1103/PhysRevA.101.061602}
}

@article{Lin:2021,
    title = {Mott transition in a cavity-boson system: A quantitative comparison between theory and experiment},
    author = {Rui Lin and Christoph Georges and Jens Klinder and Paolo Molignini and Miriam Buettner and Axel U. J. Lode and R. Chitra and Andreas Hemmerich and Hans Kessler},
    journal = {SciPost Phys.},
    volume = {11},
    pages = {030},
    year = {2021},
    url = {https://doi.org/10.21468/SciPostPhys.11.2.030},
    doi = {10.21468/SciPostPhys.11.2.030}
}

@article{Molignini:2022,
    title = {Crystallization via cavity-assisted infinite-range interactions},
    author = {Paolo Molignini and Camille L\'{e}v\^{e}que and Hans Kessler and Dieter Jaksch and R. Chitra and Axel U. J. Lode},
    journal = {Phys. Rev. A},
    volume = {106},
    pages = {L011701},
    year = {2022},
    doi = {10.1103/PhysRevA.106.L011701},
    url = {https://doi.org/10.1103/PhysRevA.106.L011701}
}

@article{Rosa-Medina:2022,
    author = {Rodrigo Rosa-Medina and Francesco Ferri and Fabian Finger and Nishant Dogra and Katrin Kroeger and Rui Lin and R. Chitra and Tobias Donner and Tilman Esslinger},
    title = {Observing Dynamical Currents in a Non-Hermitian Momentum Lattice},
    journal = {Phys. Rev. Lett.},
    volume = {128}, 
    pages = {143602},
    year = {2022},
    url = {https://doi.org/10.1103/PhysRevLett.128.143602},
    doi = {10.1103/PhysRevLett.128.143602}
}

@article{Ortuno-Gonzalez:2025,
    author = {Daniel Ortuño-Gonzalez and Rui Lin and Justyna Stefaniak and Alexander Baumgärtner and Gabriele Natale and Tobias Donner and R. Chitra},
    title = {Pauli crystal superradiance},
    journal = {Phys. Rev. Lett.},
    volume = {136},
    pages = {083405},
    year = {2026},
    url = {https://doi.org/10.1103/g5fp-ws7z},
    doi = {10.1103/g5fp-ws7z},
}

@article{Marzari:2012,
  title = {Maximally localized Wannier functions: Theory and applications},
  author = {Marzari, Nicola and Mostofi, Arash A. and Yates, Jonathan R. and Souza, Ivo and Vanderbilt, David},
  journal = {Rev. Mod. Phys.},
  volume = {84},
  pages = {1419},
  year = {2012},
  doi = {10.1103/RevModPhys.84.1419},
  url = {https://link.aps.org/doi/10.1103/RevModPhys.84.1419}
}

@article{Dutta:2022,
  title = {Density Matrix Renormalization Group for Continuous Quantum Systems},
  author = {Dutta, Shovan and Buyskikh, Anton and Daley, Andrew J. and Mueller, Erich J.},
  journal = {Phys. Rev. Lett.},
  volume = {128},
  pages = {230401},
  year = {2022},
  doi = {10.1103/PhysRevLett.128.230401},
  url = {https://link.aps.org/doi/10.1103/PhysRevLett.128.230401}
}

@article{Marzari:1997,
  title = {Maximally localized generalized Wannier functions for composite energy bands},
  author = {Marzari, Nicola and Vanderbilt, David},
  journal = {Phys. Rev. B},
  volume = {56},
  issue = {20},
  pages = {12847--12865},
  numpages = {0},
  year = {1997},
  month = {Nov},
  publisher = {American Physical Society},
  doi = {10.1103/PhysRevB.56.12847},
  url = {https://link.aps.org/doi/10.1103/PhysRevB.56.12847}
}

@article{NakajimaNP2021,
  author  = {Nakajima, Shuta and Takei, Nobuyuki and Sakuma, Keita and Kuno, Yoshihito and Marra, Pasquale and Takahashi, Yoshiro},
  title   = {Competition and interplay between topology and quasi-periodic disorder in {Thouless} pumping of ultracold atoms},
  journal = {Nat. Phys.},
  volume  = {17},
  number  = {7},
  pages   = {844--849},
  year    = {2021},
  doi     = {10.1038/s41567-021-01229-9},
}

@article{GottlobPRXQ2025,
  author  = {Gottlob, Emmanuel and Borgnia, Dan S. and Slager, Robert-Jan and Schneider, Ulrich},
  title   = {Quasiperiodicity Protects Quantized Transport in Disordered Systems Without Gaps},
  journal = {PRX Quantum},
  volume  = {6},
  pages   = {020359},
  year    = {2025},
  doi     = {10.1103/zvng-w46m},
}

@article{PadhanMishraPRB2024,
  author  = {Padhan, Ashirbad and Mishra, Tapan},
  title   = {Quantized {Thouless} charge pumping in a system with onsite quasiperiodic disorder},
  journal = {Phys. Rev. B},
  volume  = {109},
  pages   = {174206},
  year    = {2024},
  doi     = {10.1103/PhysRevB.109.174206}
}

@article{YangPNAS2024,
  author  = {Yang, Kai and Fu, Qidong and Prates, Henrique C. and Wang, Peng and Kartashov, Yaroslav V. and Konotop, Vladimir V. and Ye, Fangwei},
  title   = {Observation of {Thouless} pumping of light in quasiperiodic photonic crystals},
  journal = {Proc. Natl. Acad. Sci. U.S.A.},
  volume  = {121},
  number  = {47},
  pages   = {e2411793121},
  year    = {2024},
  doi     = {10.1073/pnas.2411793121},
}

@article{PengELight2025,
  author  = {Peng, Ruihan and Yang, Kai and Fu, Qidong and Chen, Yanli and Wang, Peng and Kartashov, Yaroslav V. and Konotop, Vladimir V. and Ye, Fangwei},
  title   = {Topological pumping of light governed by {Fibonacci} numbers},
  journal = {eLight},
  volume  = {5},
  number  = {1},
  pages   = {16},
  year    = {2025},
  doi     = {10.1186/s43593-025-00095-9},
}

@article{ChiaracanePRB2021,
  title = {Quantum dynamics in the interacting Fibonacci chain},
  author = {Chiaracane, Cecilia and Pietracaprina, Francesca and Purkayastha, Archak and Goold, John},
  journal = {Phys. Rev. B},
  volume = {103},
  issue = {18},
  pages = {184205},
  numpages = {11},
  year = {2021},
  month = {May},
  publisher = {American Physical Society},
  doi = {10.1103/PhysRevB.103.184205},
  url = {https://link.aps.org/doi/10.1103/PhysRevB.103.184205}
}

@article{MaceSciPost2019,
	title = {Many-body localization in a quasiperiodic Fibonacci chain},
	pages = {050},
	author = {Macé, Nicolas and Laflorencie, Nicolas and Alet, Fabien},
	journal = {SciPost Phys.},
	volume = {6},
	year = {2019},
	publisher = {SciPost},
	doi = {10.21468/SciPostPhys.6.4.050},
	url = {https://scipost.org/10.21468/SciPostPhys.6.4.050}
}

@misc{Mathieu,
         key = "{\relax DLMF}",
       title = "{\it NIST Digital Library of Mathematical Functions}",
howpublished = "\url{https://dlmf.nist.gov/}, Release 1.2.7 of 2026-06-15",
         url = "https://dlmf.nist.gov/",
        note = "F.~W.~J. Olver, A.~B. {Olde Daalhuis}, D.~W. Lozier, B.~I. Schneider,
                R.~F. Boisvert, C.~W. Clark, B.~R. Miller, B.~V. Saunders,
                H.~S. Cohl, and M.~A. McClain, eds."}

@article{ghosh2025quantum,
  title = {Quantum state transfer and maximal entanglement between distant qubits using a minimal quasicrystal pump},
  author = {Ghosh, Arnob Kumar and Souto, Rub\'en Seoane and Azimi-Mousolou, Vahid and Black-Schaffer, Annica M. and Holmvall, Patric},
  journal = {Phys. Rev. B},
  volume = {112},
  issue = {20},
  pages = {205427},
  numpages = {14},
  year = {2025},
  month = {Nov},
  publisher = {American Physical Society},
  doi = {10.1103/8zys-w2v4},
  url = {https://link.aps.org/doi/10.1103/8zys-w2v4}
}

@misc{ghosh2026FCwaveguide,
      title={Observation of end-to-end pumping in a quasiperiodic Fibonacci-type photonic chain}, 
      author={Arnob Kumar Ghosh and Ang Chen and Ashraf El Hassan and Patric Holmvall and Mohamed Bourennane and Annica M. Black-Schaffer},
      year={2026},
      eprint={2605.13116},
      archivePrefix={arXiv},
      primaryClass={cond-mat.mes-hall},
      url={https://arxiv.org/abs/2605.13116}, 
}

@article{ChaohuaPRA2023,
  title = {Observation of topological pumping of a defect state in a Fock photonic lattice},
  author = {Wu, Chaohua and Liu, Weijie and Jia, Yuechen and Chen, Gang and Chen, Feng},
  journal = {Phys. Rev. A},
  volume = {107},
  issue = {3},
  pages = {033501},
  numpages = {8},
  year = {2023},
  month = {Mar},
  publisher = {American Physical Society},
  doi = {10.1103/PhysRevA.107.033501},
  url = {https://link.aps.org/doi/10.1103/PhysRevA.107.033501}
}

@article{ThoulessPRB1983,
  title = {Quantization of particle transport},
  author = {Thouless, D. J.},
  journal = {Phys. Rev. B},
  volume = {27},
  issue = {10},
  pages = {6083--6087},
  numpages = {0},
  year = {1983},
  month = {May},
  publisher = {American Physical Society},
  doi = {10.1103/PhysRevB.27.6083},
  url = {https://link.aps.org/doi/10.1103/PhysRevB.27.6083}
}

@article{PhysRevLett.129.053201,
  title = {Topological Pumping in a Floquet-Bloch Band},
  author = {Minguzzi, Joaqu\'{\i}n and Zhu, Zijie and Sandholzer, Kilian and Walter, Anne-Sophie and Viebahn, Konrad and Esslinger, Tilman},
  journal = {Phys. Rev. Lett.},
  volume = {129},
  issue = {5},
  pages = {053201},
  numpages = {6},
  year = {2022},
  month = {Jul},
  publisher = {American Physical Society},
  doi = {10.1103/PhysRevLett.129.053201},
  url = {https://link.aps.org/doi/10.1103/PhysRevLett.129.053201}
}

@article{JagannathanRMP2021,
  title = {The Fibonacci quasicrystal: Case study of hidden dimensions and multifractality},
  author = {Jagannathan, Anuradha},
  journal = {Rev. Mod. Phys.},
  volume = {93},
  issue = {4},
  pages = {045001},
  numpages = {37},
  year = {2021},
  month = {Nov},
  publisher = {American Physical Society},
  doi = {10.1103/RevModPhys.93.045001},
  url = {https://link.aps.org/doi/10.1103/RevModPhys.93.045001}
}

@article{MeyrathPRA2005,
  title   = {Bose-Einstein condensate in a box},
  author  = {Meyrath, T. P. and Schreck, F. and Hanssen, J. L. and Chuu, C.-S. and Raizen, M. G.},
  journal = {Phys. Rev. A},
  volume  = {71},
  pages   = {041604},
  year    = {2005},
  doi     = {10.1103/PhysRevA.71.041604}
}

@article{GauntPRL2013,
  title   = {Bose-Einstein Condensation of Atoms in a Uniform Potential},
  author  = {Gaunt, Alexander L. and Schmidutz, Tobias F. and Gotlibovych, Igor and Smith, Robert P. and Hadzibabic, Zoran},
  journal = {Phys. Rev. Lett.},
  volume  = {110},
  pages   = {200406},
  year    = {2013},
  doi     = {10.1103/PhysRevLett.110.200406}
}

@article{RaiPRB2021,
  title = {Bulk topological signatures of a quasicrystal},
  author = {Rai, Gautam and Schl\"omer, Henning and Matsumura, Chris and Haas, Stephan and Jagannathan, Anuradha},
  journal = {Phys. Rev. B},
  volume = {104},
  issue = {18},
  pages = {184202},
  numpages = {8},
  year = {2021},
  month = {Nov},
  publisher = {American Physical Society},
  doi = {10.1103/PhysRevB.104.184202},
  url = {https://link.aps.org/doi/10.1103/PhysRevB.104.184202}
}

@article{ElsePRX2021,
  title = {Quantum Many-Body Topology of Quasicrystals},
  author = {Else, Dominic V. and Huang, Sheng-Jie and Prem, Abhinav and Gromov, Andrey},
  journal = {Phys. Rev. X},
  volume = {11},
  issue = {4},
  pages = {041051},
  numpages = {21},
  year = {2021},
  month = {Dec},
  publisher = {American Physical Society},
  doi = {10.1103/PhysRevX.11.041051},
  url = {https://link.aps.org/doi/10.1103/PhysRevX.11.041051}
}

@article{VarjasPRL2019,
  title = {Topological Phases without Crystalline Counterparts},
  author = {Varjas, D\'aniel and Lau, Alexander and P\"oyh\"onen, Kim and Akhmerov, Anton R. and Pikulin, Dmitry I. and Fulga, Ion Cosma},
  journal = {Phys. Rev. Lett.},
  volume = {123},
  issue = {19},
  pages = {196401},
  numpages = {6},
  year = {2019},
  month = {Nov},
  publisher = {American Physical Society},
  doi = {10.1103/PhysRevLett.123.196401},
  url = {https://link.aps.org/doi/10.1103/PhysRevLett.123.196401}
}

@article{MadsenPRB2013,
  title = {Topological equivalence of crystal and quasicrystal band structures},
  author = {Madsen, Kevin A. and Bergholtz, Emil J. and Brouwer, Piet W.},
  journal = {Phys. Rev. B},
  volume = {88},
  issue = {12},
  pages = {125118},
  numpages = {6},
  year = {2013},
  month = {Sep},
  publisher = {American Physical Society},
  doi = {10.1103/PhysRevB.88.125118},
  url = {https://link.aps.org/doi/10.1103/PhysRevB.88.125118}
}

@article{JagannathanPRB20205,
  title = {Missing link between the two-dimensional quantum Hall problem and one-dimensional quasicrystals},
  author = {Jagannathan, Anuradha},
  journal = {Phys. Rev. B},
  volume = {112},
  issue = {10},
  pages = {L100102},
  numpages = {6},
  year = {2025},
  month = {Sep},
  publisher = {American Physical Society},
  doi = {10.1103/stk9-d9vf},
  url = {https://link.aps.org/doi/10.1103/stk9-d9vf}
}

@misc{JagannathanArxiv2026,
      title={Topological connections between the 2D Quantum Hall problem and the 1D quasicrystal}, 
      author={Anuradha Jagannathan},
      year={2026},
      eprint={2601.09432},
      archivePrefix={arXiv},
      primaryClass={cond-mat.mes-hall},
      url={https://arxiv.org/abs/2601.09432}, 
}

@misc{MarsalArxiv2026,
      title={Quantum metric and localization in a quasicrystal}, 
      author={Quentin Marsal and Patric Holmvall and Annica M. Black-Schaffer},
      year={2026},
      eprint={2506.15575},
      archivePrefix={arXiv},
      primaryClass={cond-mat.mes-hall},
      url={https://arxiv.org/abs/2506.15575}, 
}

@Article{MoustajCondMat2025,
AUTHOR = {Moustaj, Anouar and Krebbekx, Julius and Morais Smith, Cristiane},
TITLE = {Anomalous Polarization in One-Dimensional Aperiodic Insulators},
JOURNAL = {Condens. Matter},
VOLUME = {10},
YEAR = {2025},
NUMBER = {1},
ARTICLE-NUMBER = {3},
URL = {https://www.mdpi.com/2410-3896/10/1/3},
ISSN = {2410-3896},
DOI = {10.3390/condmat10010003}
}

@article{BonselQuantum2026,
doi = {10.22331/q-2026-04-23-2081},
url = {https://doi.org/10.22331/q-2026-04-23-2081},
title = {Fibonacci {W}aveguide {Q}uantum {E}lectrodynamics},
author = {B{\"{o}}nsel, Florian and Kunst, Flore K. and Roccati, Federico},
journal = {{Quantum}},
issn = {2521-327X},
publisher = {{Verein zur F{\"{o}}rderung des Open Access Publizierens in den Quantenwissenschaften}},
volume = {10},
pages = {2081},
month = apr,
year = {2026}
}

@Article{FanFP2022,
  author   = {Fan, Jiahao and Huang, Huaqing},
  journal  = {Front. Phys.},
  title    = {Topological states in quasicrystals},
  year     = {2022},
  issn     = {2095-0470},
  number   = {1},
  pages    = {13203},
  volume   = {17},
  doi      = {10.1007/s11467-021-1100-y},
  refid    = {Fan2021},
  url      = {https://doi.org/10.1007/s11467-021-1100-y},
}

@article{WangPRB2024,
  title = {Superconductivity in the Fibonacci chain},
  author = {Wang, Ying and Rai, Gautam and Matsumura, Chris and Jagannathan, Anuradha and Haas, Stephan},
  journal = {Phys. Rev. B},
  volume = {109},
  issue = {21},
  pages = {214507},
  numpages = {11},
  year = {2024},
  month = {Jun},
  publisher = {American Physical Society},
  doi = {10.1103/PhysRevB.109.214507},
  url = {https://link.aps.org/doi/10.1103/PhysRevB.109.214507}
}

@article{SandbergPRB2024,
  title = {Josephson effect in a Fibonacci quasicrystal},
  author = {Sandberg, Anna and Awoga, Oladunjoye A. and Black-Schaffer, Annica M. and Holmvall, Patric},
  journal = {Phys. Rev. B},
  volume = {110},
  issue = {10},
  pages = {104513},
  numpages = {22},
  year = {2024},
  month = {Sep},
  publisher = {American Physical Society},
  doi = {10.1103/PhysRevB.110.104513},
  url = {https://link.aps.org/doi/10.1103/PhysRevB.110.104513}
}

@article{KobialkaPRB2024,
  title = {Topological superconductivity in Fibonacci quasicrystals},
  author = {Kobia\l{}ka, Aksel and Awoga, Oladunjoye A. and Leijnse, Martin and Doma\ifmmode \acute{n}\else \'{n}\fi{}ski, Tadeusz and Holmvall, Patric and Black-Schaffer, Annica M.},
  journal = {Phys. Rev. B},
  volume = {110},
  issue = {13},
  pages = {134508},
  numpages = {20},
  year = {2024},
  month = {Oct},
  publisher = {American Physical Society},
  doi = {10.1103/PhysRevB.110.134508},
  url = {https://link.aps.org/doi/10.1103/PhysRevB.110.134508}
}

@Article{YaoNatComm2013,
    author={Yao, N. Y.
    and Laumann, C. R.
    and Gorshkov, A. V.
    and Weimer, H.
    and Jiang, L.
    and Cirac, J. I.
    and Zoller, P.
    and Lukin, M. D.},
    title={Topologically protected quantum state transfer in a chiral spin liquid},
    journal={Nat Commun},
    year={2013},
    month={Mar},
    day={12},
    volume={4},
    number={1},
    pages={1585},
    issn={2041-1723},
    doi={10.1038/ncomms2531},
    url={https://doi.org/10.1038/ncomms2531}
}

@article{DlaskaQSI2017,
    doi = {10.1088/2058-9565/2/1/015001},
    url = {https://dx.doi.org/10.1088/2058-9565/2/1/015001},
    year = {2017},
    month = {jan},
    publisher = {IOP Publishing},
    volume = {2},
    number = {1},
    pages = {015001},
    author = {Dlaska, C and Vermersch, B and Zoller, P},
    title = {Robust quantum state transfer via topologically protected edge channels in dipolar arrays},
    journal = {Quantum Sci. Technol.},
}

@Article{CitroNRP2023,
    author={Citro, Roberta
    and Aidelsburger, Monika},
    title={Thouless pumping and topology},
    journal={Nat Rev Phys},
    year={2023},
    month={Feb},
    day={01},
    volume={5},
    number={2},
    pages={87-101},
    issn={2522-5820},
    doi={10.1038/s42254-022-00545-0},
    url={https://doi.org/10.1038/s42254-022-00545-0}
}

@article{SinghPRA2015,
  title = {Fibonacci optical lattices for tunable quantum quasicrystals},
  author = {Singh, K. and Saha, K. and Parameswaran, S. A. and Weld, D. M.},
  journal = {Phys. Rev. A},
  volume = {92},
  issue = {6},
  pages = {063426},
  numpages = {8},
  year = {2015},
  month = {Dec},
  publisher = {American Physical Society},
  doi = {10.1103/PhysRevA.92.063426},
  url = {https://link.aps.org/doi/10.1103/PhysRevA.92.063426}
}

@article{VerbinPump2015,
  title = {Topological pumping over a photonic Fibonacci quasicrystal},
  author = {Verbin, Mor and Zilberberg, Oded and Lahini, Yoav and Kraus, Yaacov E. and Silberberg, Yaron},
  journal = {Phys. Rev. B},
  volume = {91},
  issue = {6},
  pages = {064201},
  numpages = {6},
  year = {2015},
  month = {Feb},
  publisher = {American Physical Society},
  doi = {10.1103/PhysRevB.91.064201},
  url = {https://link.aps.org/doi/10.1103/PhysRevB.91.064201}
}

@Article{LangNQI2017,
    author={Lang, Nicolai
    and B{\"u}chler, Hans Peter},
    title={Topological networks for quantum communication between distant qubits},
    journal={npj Quantum Inf},
    year={2017},
    month={Nov},
    day={07},
    volume={3},
    number={1},
    pages={47},
    issn={2056-6387},
    doi={10.1038/s41534-017-0047-x},
    url={https://doi.org/10.1038/s41534-017-0047-x}
}

@article{MeiPRA2018,
  title = {Robust quantum state transfer via topological edge states in superconducting qubit chains},
  author = {Mei, Feng and Chen, Gang and Tian, Lin and Zhu, Shi-Liang and Jia, Suotang},
  journal = {Phys. Rev. A},
  volume = {98},
  issue = {1},
  pages = {012331},
  numpages = {6},
  year = {2018},
  month = {Jul},
  publisher = {American Physical Society},
  doi = {10.1103/PhysRevA.98.012331},
  url = {https://link.aps.org/doi/10.1103/PhysRevA.98.012331}
}

@article{WangPRA2022,
  title = {Arbitrary entangled state transfer via a topological qubit chain},
  author = {Wang, Chong and Li, Linhu and Gong, Jiangbin and Liu, Yu-xi},
  journal = {Phys. Rev. A},
  volume = {106},
  issue = {5},
  pages = {052411},
  numpages = {19},
  year = {2022},
  month = {Nov},
  publisher = {American Physical Society},
  doi = {10.1103/PhysRevA.106.052411},
  url = {https://link.aps.org/doi/10.1103/PhysRevA.106.052411}
}

@article{LonghiPRB2019,
  title = {Topological pumping of edge states via adiabatic passage},
  author = {Longhi, Stefano},
  journal = {Phys. Rev. B},
  volume = {99},
  issue = {15},
  pages = {155150},
  numpages = {9},
  year = {2019},
  month = {Apr},
  publisher = {American Physical Society},
  doi = {10.1103/PhysRevB.99.155150},
  url = {https://link.aps.org/doi/10.1103/PhysRevB.99.155150}
}

@article{LonghiLandauZenerAQT2019,
    author = {Longhi, Stefano and Giorgi, Gian Luca and Zambrini, Roberta},
    title = {Landau–Zener Topological Quantum State Transfer},
    journal = {Adv. Quantum Technol.},
    volume = {2},
    number = {3-4},
    pages = {1800090},
    doi = {https://doi.org/10.1002/qute.201800090},
    url = {https://onlinelibrary.wiley.com/doi/abs/10.1002/qute.201800090},
    year = {2019}
}

@article{RomeroPRApp2024,
  title = {Optimizing edge-state transfer in a Su-Schrieffer-Heeger chain via hybrid analog-digital strategies},
  author = {Romero, Sebasti\'an V. and Chen, Xi and Platero, Gloria and Ban, Yue},
  journal = {Phys. Rev. Appl.},
  volume = {21},
  issue = {3},
  pages = {034033},
  numpages = {19},
  year = {2024},
  month = {Mar},
  publisher = {American Physical Society},
  doi = {10.1103/PhysRevApplied.21.034033},
  url = {https://link.aps.org/doi/10.1103/PhysRevApplied.21.034033}
}

@article{Zurita2023fastquantumtransfer,
  doi = {10.22331/q-2023-06-22-1043},
  url = {https://doi.org/10.22331/q-2023-06-22-1043},
  title = {Fast quantum transfer mediated by topological domain walls},
  author = {Zurita, Juan and Creffield, Charles E. and Platero, Gloria},
  journal = {{Quantum}},
  issn = {2521-327X},
  publisher = {{Verein zur F{\"{o}}rderung des Open Access Publizierens in den Quantenwissenschaften}},
  volume = {7},
  pages = {1043},
  month = jun,
  year = {2023}
}

@article{ZhengCJP2025,
    title = {Robust entangled state transmission and preparation in a trimer-like chain},
    journal = {Chin. J. Phys.},
    volume = {93},
    pages = {471-481},
    year = {2025},
    issn = {0577-9073},
    doi = {https://doi.org/10.1016/j.cjph.2024.12.009},
    url = {https://www.sciencedirect.com/science/article/pii/S0577907324004738},
    author = {Li-Na Zheng and Hong-Fu Wang and Xuexi Yi},
}

@article{WangDaWeiPRA2023,
  title = {Simulating the extended Su-Schrieffer-Heeger model and transferring an entangled state based on a hybrid cavity-magnon array},
  author = {Wang, Da-Wei and Zhao, Chengsong and Yang, Junya and Yan, Ye-Ting and Zhou, Ling},
  journal = {Phys. Rev. A},
  volume = {107},
  issue = {5},
  pages = {053701},
  numpages = {9},
  year = {2023},
  month = {May},
  publisher = {American Physical Society},
  doi = {10.1103/PhysRevA.107.053701},
  url = {https://link.aps.org/doi/10.1103/PhysRevA.107.053701}
}

@article{ZhengLiNaPRA2020,
  title = {Defect-induced controllable quantum state transfer via a topologically protected channel in a flux qubit chain},
  author = {Zheng, Li-Na and Qi, Lu and Cheng, Liu-Yong and Wang, Hong-Fu and Zhang, Shou},
  journal = {Phys. Rev. A},
  volume = {102},
  issue = {1},
  pages = {012606},
  numpages = {9},
  year = {2020},
  month = {Jul},
  publisher = {American Physical Society},
  doi = {10.1103/PhysRevA.102.012606},
  url = {https://link.aps.org/doi/10.1103/PhysRevA.102.012606}
}

@article{QiLuPRA2020,
  title = {Engineering the topological state transfer and topological beam splitter in an even-sized Su-Schrieffer-Heeger chain},
  author = {Qi, Lu and Wang, Guo-Li and Liu, Shutian and Zhang, Shou and Wang, Hong-Fu},
  journal = {Phys. Rev. A},
  volume = {102},
  issue = {2},
  pages = {022404},
  numpages = {10},
  year = {2020},
  month = {Aug},
  publisher = {American Physical Society},
  doi = {10.1103/PhysRevA.102.022404},
  url = {https://link.aps.org/doi/10.1103/PhysRevA.102.022404}
}

@article{ZhengLiNaYiPRApp2022,
  title = {Engineering a Phase-Robust Topological Router in a Dimerized Superconducting-Circuit Lattice with Long-Range Hopping and Chiral Symmetry},
  author = {Zheng, Li-Na and Yi, Xuexi and Wang, Hong-Fu},
  journal = {Phys. Rev. Appl.},
  volume = {18},
  issue = {5},
  pages = {054037},
  numpages = {16},
  year = {2022},
  month = {Nov},
  publisher = {American Physical Society},
  doi = {10.1103/PhysRevApplied.18.054037},
  url = {https://link.aps.org/doi/10.1103/PhysRevApplied.18.054037}
}

@article{PalaiodimopoulosPRA2021,
  title = {Fast and robust quantum state transfer via a topological chain},
  author = {Palaiodimopoulos, N. E. and Brouzos, I. and Diakonos, F. K. and Theocharis, G.},
  journal = {Phys. Rev. A},
  volume = {103},
  issue = {5},
  pages = {052409},
  numpages = {9},
  year = {2021},
  month = {May},
  publisher = {American Physical Society},
  doi = {10.1103/PhysRevA.103.052409},
  url = {https://link.aps.org/doi/10.1103/PhysRevA.103.052409}
}

@article{QiPRRTR2021,
  title = {Topological router induced via long-range hopping in a Su-Schrieffer-Heeger chain},
  author = {Qi, Lu and Yan, Yu and Xing, Yan and Zhao, Xue-Dong and Liu, Shutian and Cui, Wen-Xue and Han, Xue and Zhang, Shou and Wang, Hong-Fu},
  journal = {Phys. Rev. Research},
  volume = {3},
  issue = {2},
  pages = {023037},
  numpages = {10},
  year = {2021},
  month = {Apr},
  publisher = {American Physical Society},
  doi = {10.1103/PhysRevResearch.3.023037},
  url = {https://link.aps.org/doi/10.1103/PhysRevResearch.3.023037}
}

@Article{ZhaonppjQuantumInf2023,
    author={Zhao, Xuedong
    and Xing, Yan
    and Cao, Ji
    and Liu, Shutian
    and Cui, Wen-Xue
    and Wang, Hong-Fu},
    title={Engineering quantum diode in one-dimensional time-varying superconducting circuits},
    journal={npj Quantum Inf},
    year={2023},
    month={Jun},
    day={20},
    volume={9},
    number={1},
    pages={59},
    issn={2056-6387},
    doi={10.1038/s41534-023-00729-1},
    url={https://doi.org/10.1038/s41534-023-00729-1}
}

@article{CaoPRA2021,
  title = {Controllable photon-phonon conversion via the topologically protected edge channel in an optomechanical lattice},
  author = {Cao, Ji and Cui, Wen-Xue and Yi, X. X. and Wang, Hong-Fu},
  journal = {Phys. Rev. A},
  volume = {103},
  issue = {2},
  pages = {023504},
  numpages = {7},
  year = {2021},
  month = {Feb},
  publisher = {American Physical Society},
  doi = {10.1103/PhysRevA.103.023504},
  url = {https://link.aps.org/doi/10.1103/PhysRevA.103.023504}
}

@article{WangDaWeiPRA2024,
  title = {Controllable excitation transfer based on the coupling of an atom with a finite-size Su-Schrieffer-Heeger chain},
  author = {Wang, Da-Wei and Zhao, Chengsong and Yang, Junya and Yan, Ye-Ting and Zhou, Ling},
  journal = {Phys. Rev. A},
  volume = {109},
  issue = {3},
  pages = {033708},
  numpages = {10},
  year = {2024},
  month = {Mar},
  publisher = {American Physical Society},
  doi = {10.1103/PhysRevA.109.033708},
  url = {https://link.aps.org/doi/10.1103/PhysRevA.109.033708}
}

@article{HanJinXuanPRApp2024,
  title = {Fast and controllable topological excitation transfers in hybrid magnon-photon systems},
  author = {Han, Jin-Xuan and Wu, Jin-Lei and Yuan, Zhong-Hui and Chen, Yong-Jian and Xia, Yan and Jiang, Yong-Yuan and Song, Jie},
  journal = {Phys. Rev. Appl.},
  volume = {21},
  issue = {1},
  pages = {014057},
  numpages = {21},
  year = {2024},
  month = {Jan},
  publisher = {American Physical Society},
  doi = {10.1103/PhysRevApplied.21.014057},
  url = {https://link.aps.org/doi/10.1103/PhysRevApplied.21.014057}
}

@article{DAngelisPRR2020,
  title = {Fast and robust quantum state transfer in a topological Su-Schrieffer-Heeger chain with next-to-nearest-neighbor interactions},
  author = {D'Angelis, Felippo M. and Pinheiro, Felipe A. and Gu\'ery-Odelin, David and Longhi, Stefano and Impens, Fran\ifmmode \mbox{\c{c}}\else \c{c}\fi{}ois},
  journal = {Phys. Rev. Research},
  volume = {2},
  issue = {3},
  pages = {033475},
  numpages = {11},
  year = {2020},
  month = {Sep},
  publisher = {American Physical Society},
  doi = {10.1103/PhysRevResearch.2.033475},
  url = {https://link.aps.org/doi/10.1103/PhysRevResearch.2.033475}
}

@article{TianPRB2024,
  title = {Nonadiabatic topological transfer in a nanomechanical phononic lattice},
  author = {Tian, Tian and Cai, Han and Zhang, Liang and Zhang, Yichuan and Duan, Chang-Kui and Zhou, Jingwei},
  journal = {Phys. Rev. B},
  volume = {109},
  issue = {12},
  pages = {125123},
  numpages = {10},
  year = {2024},
  month = {Mar},
  publisher = {American Physical Society},
  doi = {10.1103/PhysRevB.109.125123},
  url = {https://link.aps.org/doi/10.1103/PhysRevB.109.125123}
}

@article{YuanAPLPh2021,
    author = {Yuan, Jiale and Xu, Chenran and Cai, Han and Wang, Da-Wei},
    title = {Gap-protected transfer of topological defect states in photonic lattices},
    journal = {APL Photonics},
    volume = {6},
    number = {3},
    pages = {030803},
    year = {2021},
    month = {03},
    issn = {2378-0967},
    doi = {10.1063/5.0037394},
    url = {https://doi.org/10.1063/5.0037394}
}

@article{Fernandez2024,
  title = {Flying Spin Qubits in Quantum Dot Arrays Driven by Spin-Orbit Interaction},
  author = {Fern{\'{a}}ndez-Fern{\'{a}}ndez, D. and Ban, Yue and Platero, G.},
  journal = {Quantum},
  volume = {8},
  pages = {1533},
  year = {2024},
  month = {Nov},
  doi = {10.22331/q-2024-11-21-1533},
  url = {https://doi.org/10.22331/q-2024-11-21-1533}
}

@article{JAKSCHAP2005,
title = {The cold atom Hubbard toolbox},
journal = {Ann. Phys.},
volume = {315},
number = {1},
pages = {52-79},
year = {2005},
note = {Special Issue},
issn = {0003-4916},
doi = {https://doi.org/10.1016/j.aop.2004.09.010},
url = {https://www.sciencedirect.com/science/article/pii/S0003491604001782},
author = {D. Jaksch and P. Zoller}
}

@article{Lewenstein2007,
author = {Maciej Lewenstein and Anna Sanpera and Veronica Ahufinger and Bogdan Damski and Aditi Sen(De) and Ujjwal Sen},
title = {Ultracold atomic gases in optical lattices: mimicking condensed matter physics and beyond},
journal = {Adv. Phys.},
volume = {56},
number = {2},
pages = {243--379},
year = {2007},
publisher = {Taylor \& Francis},
doi = {10.1080/00018730701223200}
}

@Article{CiracNP2012,
    author={Cirac, J. Ignacio
    and Zoller, Peter},
    title={Goals and opportunities in quantum simulation},
    journal={Nature Phys},
    year={2012},
    month={Apr},
    day={01},
    volume={8},
    number={4},
    pages={264-266},
    issn={1745-2481},
    doi={10.1038/nphys2275},
    url={https://doi.org/10.1038/nphys2275}
}

@Article{BlochNP2012,
    author={Bloch, Immanuel
    and Dalibard, Jean
    and Nascimb{\`e}ne, Sylvain},
    title={Quantum simulations with ultracold quantum gases},
    journal={Nature Phys},
    year={2012},
    month={Apr},
    day={01},
    volume={8},
    number={4},
    pages={267-276},
    issn={1745-2481},
    doi={10.1038/nphys2259},
    url={https://doi.org/10.1038/nphys2259}
}

@Article{LangenNP2024,
author={Langen, Tim
and Valtolina, Giacomo
and Wang, Dajun
and Ye, Jun},
title={Quantum state manipulation and cooling of ultracold molecules},
journal={Nature Phys.},
year={2024},
month={May},
day={01},
volume={20},
number={5},
pages={702-712},
issn={1745-2481},
doi={10.1038/s41567-024-02423-1},
url={https://doi.org/10.1038/s41567-024-02423-1}
}

@Article{NakajimaNP2016,
    author={Nakajima, Shuta
    and Tomita, Takafumi
    and Taie, Shintaro
    and Ichinose, Tomohiro
    and Ozawa, Hideki
    and Wang, Lei
    and Troyer, Matthias
    and Takahashi, Yoshiro},
    title={Topological Thouless pumping of ultracold fermions},
    journal={Nature Phys},
    year={2016},
    month={Apr},
    day={01},
    volume={12},
    number={4},
    pages={296-300},
    issn={1745-2481},
    doi={10.1038/nphys3622},
    url={https://doi.org/10.1038/nphys3622}
}

@Article{LohseNP2016,
    author={Lohse, M.
    and Schweizer, C.
    and Zilberberg, O.
    and Aidelsburger, M.
    and Bloch, I.},
    title={A Thouless quantum pump with ultracold bosonic atoms in an optical superlattice},
    journal={Nature Phys},
    year={2016},
    month={Apr},
    day={01},
    volume={12},
    number={4},
    pages={350-354},
    issn={1745-2481},
    doi={10.1038/nphys3584},
    url={https://doi.org/10.1038/nphys3584}
}

@article{ChristianScience2017,
    author = {Christian Gross  and Immanuel Bloch },
    title = {Quantum simulations with ultracold atoms in optical lattices},
    journal = {Science},
    volume = {357},
    number = {6355},
    pages = {995-1001},
    year = {2017},
    doi = {10.1126/science.aal3837},
    URL = {https://www.science.org/doi/abs/10.1126/science.aal3837}
}

@article{HollandScience2023,
    author = {Connor M. Holland  and Yukai Lu  and Lawrence W. Cheuk },
    title = {On-demand entanglement of molecules in a reconfigurable optical tweezer array},
    journal = {Science},
    volume = {382},
    number = {6675},
    pages = {1143-1147},
    year = {2023},
    doi = {10.1126/science.adf4272},
    URL = {https://www.science.org/doi/abs/10.1126/science.adf4272}
}

@article{YichengScience2023,
    author = {Yicheng Bao  and Scarlett S. Yu  and Loïc Anderegg  and Eunmi Chae  and Wolfgang Ketterle  and Kang-Kuen Ni  and John M. Doyle },
    title = {Dipolar spin-exchange and entanglement between molecules in an optical tweezer array},
    journal = {Science},
    volume = {382},
    number = {6675},
    pages = {1138-1143},
    year = {2023},
    doi = {10.1126/science.adf8999},
    URL = {https://www.science.org/doi/abs/10.1126/science.adf8999}
}

\end{document}